\documentclass[printer]{aa}  

\usepackage{graphicx}
\usepackage{txfonts}
\usepackage{bm}
\usepackage[]{hyperref}%draft
\usepackage{xcolor}
\usepackage{tikz}
\usepackage[labelformat=simple]{subfig}

\renewcommand{\thesubfigure}{\relax}

\renewcommand*{\vec}[1]{\bm{#1}}
\newcommand*{\mat}[1]{\mathbf{#1}}

\DeclareMathOperator*{\pr}{Pr}
\DeclareMathOperator{\e}{e}
\DeclareMathOperator{\dd}{d}

\DeclareFontFamily{U}{wncy}{}
\DeclareFontShape{U}{wncy}{m}{n}{<->wncyr10}{}
\DeclareSymbolFont{mcy}{U}{wncy}{m}{n}
\DeclareMathSymbol{\Sh}{\mathord}{mcy}{"58} 
\newcommand{\ch}[1]{{\textcolor{black}{#1}}}
\newcommand{\chd}[1]{{\textcolor{black}{#1}}}

\begin{document}

	\title{Improving exoplanet detection capabilities with the false inclusion probability}
	\subtitle{Comparison with other detection criteria in the context of radial velocities}

	\author{Nathan C. Hara
		\inst{\ref{i:geneve}}\thanks{CHEOPS fellow}
		\and
        Nicolas Unger\inst{\ref{i:geneve}, \ref{i:uba}}
		\and
        Jean-Baptiste Delisle\inst{\ref{i:geneve}}
		\and
		Rodrigo F. D\'iaz\inst{\ref{i:icas},\ref{i:uba},\ref{i:conicet}}
		\and
		Damien Ségransan\inst{\ref{i:geneve}}
	}
	
	\institute{
		Observatoire Astronomique de l’Université de Genève, Chemin de Pegasi 51 b, 1290 Versoix, Switzerland\label{i:geneve}  \email{nathan.hara@unige.ch}
		\and
		International Center for Advanced Studies (ICAS) and ICIFI (CONICET), ECyT-UNSAM, Campus Miguelete, 25 de Mayo y Francia, (1650) Buenos Aires, Argentina.\label{i:icas}
        \and
		Universidad de Buenos Aires, Facultad de Ciencias Exactas y Naturales. Buenos Aires, Argentina\label{i:uba}
		\and
		CONICET - Universidad de Buenos Aires. Instituto de Astronomía y Física del Espacio (IAFE). Buenos Aires, Argentina\label{i:conicet}        
	}

	\abstract
% context heading (optional)
{
It is common practice to claim the detection of a signal if, for a certain statistical significance metric, the signal significance exceeds a certain threshold fixed in advance. 
In the context of exoplanet searches in radial velocity data, the most common statistical significance metrics are the Bayes factor and the false alarm probability (FAP). Both criteria have proved useful, but do not directly address whether an exoplanet detection should be claimed. Furthermore, it is unclear which detection threshold should be taken \ch{and how robust the detections are to model misspecification}. 
} 
{The present work aims at defining a detection criterion which conveys as precisely as possible the information needed to claim an exoplanet detection\ch{, as well as efficient numerical methods to compute it. } We compare this new criterion to existing ones in terms of sensitivity, and robustness to \ch{a change in the model.}
}
% methods heading (mandatory)
{We define a detection criterion based on \ch{the joint posterior distribution of the number of planets and of their orbital elements} called the false inclusion probability (FIP). In the context of exoplanet detections, it provides the probability of presence of a planet with a period in a certain interval. 
Posterior distributions are computed with the nested sampling package \textsc{polychord}. We show that for FIP and Bayes factor calculations, defining priors on linear parameters as Gaussian mixture models allows to significantly speed up computations. The performances of the FAP, Bayes factor and FIP are studied with simulations as well as analytical arguments. We compare the methods assuming the model is correct, then evaluate their sensitivity to the prior and likelihood choices. 
}
% results heading (mandatory)
{Among other properties, the FIP offers ways to test the reliability of the significance levels, it is particularly efficient to account for aliasing and allows to exclude the presence of planets with a certain confidence.   We find that, in our simulations, the FIP outperforms existing detection metrics. We show that planet detections are sensitive to priors on period and semi-amplitude \ch{and that  letting free the noise parameters offers better performances than fixing a noise model based on a fit to ancillary indicators.} 
	 }
{}

	
	\maketitle

	\section{Introduction}
	\label{sec:introduction}
	
	Detection problems arise in many areas of signal processing. Based on a certain dataset, one wants to determine if a detection of a particular signal (pattern, correlations, periodicity...) can be confidently claimed. To do so, one would typically select a statistical significance metric $m$. This one is a function of the data $y$ which retrieves a real number $m(y)$. To claim a detection, $m(y)$ has to be greater than a certain threshold fixed in advance. Ensuring this often is accompanied by an analysis addressing whether the significant signal might be due to other effects than the one looked for. There are several classical significance metrics whose properties have been studied in depth~\citep[e.g.][]{casellaberger2001, lehmannromano2005}.

	In the present work, we focus the discussion on the detection of exoplanets in radial velocity (RV) data. An observer on Earth can measure the velocity of a star in the direction of the line of sight (or radial velocity) thanks to the Doppler effect. If a planet is present, the star has a reflex motion which translates into periodic RV variations. To search for exoplanets, the observer takes a time series of radial velocities and looks for periodic signals that might indicate the presence of planets. Although we focus on the RV analysis, the principles outlined in this work are applicable to a wider range of problems such as other searches for periodic signature in time-series (evenly sampled or not), and more generally any kind of parametric pattern search. 

	The exoplanet detection process usually consists in assessing sequentially whether an additional planet should be included in the model. Planet detections are typically claimed based on one of  two approaches. The first one is the computation of a periodogram, that is a systematic scan for periodicity on a grid of frequencies. This is followed by the computation of a false alarm probability (FAP) to assess the significance of a detection. 
	There are several definitions of the periodogram, corresponding to different assumptions on the data~\citep[e.g.][]{lomb1976,ferrazmello1981,scargle,reegen2007,baluev2008, baluev2009,baluev2013_vonmises, baluev2015, delisle2019a}. The second approach consists in computing the Bayes factor (BF), defined as the ratio of the Bayesian evidences of competing models~\citep{jeffreys1946,kassraftery1995}. Here, the competing models are taken as ones with $k$ and $k+1$ planets~\citep[e.g.][]{gregory2007a, tuomi2011}.  

	The FAP and the BF offer valuable information to determine the number of exoplanets orbiting a given star. However, they might be difficult to interpret. Detections of a $k+1$\textsuperscript{th} planet based on the BF are usually claimed if the BF comparing $k+1$ and $k$ planets model is greater than 150. This value is set based on~\cite{jeffreys1946} and gives reasonable results in practice. However, a given BF threshold does not correspond to an intuitive property. 
	The FAP relies on a $p$-value, whose interpretation is not straightforward, in particular because it measures the probability of an event that has not occurred, as noted in~\citep{jeffreys1961}. Furthermore,
	detections based on FAPs rely on defining null hypothesis models,  \ch{usually} in a sequential manner, as planets are added in the model one at a time. If, at some step, a poor model choice is made, this affects all the following inferences. This happens for instance when the period of the planet added to the model is incorrect due to aliasing~\citep{dawsonfabricky2010, hara2017}. Detections can also be claimed on the explicit posterior probability of the number of planets (PNP)~\citep{brewer2014}. This metric has a more straightforward interpretation but concerns the number of planets and, alone, does not provide information on the period of the planet detected. 
	
	In the present article, we define a detection criterion aiming at expressing as clearly as possible whether an exoplanet should be detected, called the false inclusion probability (FIP).  It is based on the Bayesian formalism and can be computed as a by-product of evidence calculations, necessary to compute the BF.
	The FIP is designed to have the following meaning: assuming that the priors and likelihood are correct, when the detection of a planet with a period in $[P_1, P_2]$ is claimed with FIP $\alpha$, then there is a probability $\alpha$ that no planet with period in $[P_1, P_2]$ orbits the star. \ch{This quantity has been considered in \cite{brewerdonovan2015}, we here study its properties when it is used systematically as a detection criterion.}% 
	
	As in other types of Bayesian analysis, we define prior probabilities for the orbital elements and the number of planets in the system as well as a form for the probability of the data knowing the parameters (the likelihood function).
	The property of the FIP described above holds if the priors and likelihood functions used match the true distribution of the parameters in nature. However, the chosen prior might not accurately represent the true distribution of parameters in a population, and the noise models (likelihoods) might be inaccurate. These problems do occur in radial velocity data analysis, where the true distribution of parameters is not known but searched for, and the star introduces complex, correlated patterns in the data, which are not fully characterised~\cite[e.g.,][]{queloz2001, boisse2009, meunier2010b, dumusque2011i, dumusque2014, haywood2014, haywood2016, collier2019}.
	We study the dependency of FIP and other criteria (FAP, Bayes factor and posterior number of planets) to model misspecification.

	 The article is organised as follows. In section~\ref{sec:framework}, we define precisely the RV analysis framework, as well as the existing detection metrics. In section~\ref{sec:newcrit}, we define the FIP, we present its main properties and show how it can be computed. In section~\ref{sec:practicalcomp}, we present a practical numerical method to compute the FIP and validate our algorithm with numerical simulations. In Section~\ref{sec:application}, we give an example of application of the FIP to the HARPS observations of HD 10180.
	 In section~\ref{sec:discussion}, we compare the FIP to other detection criteria and highlight some of its key advantages. We also study the sensitivity of detections to prior and likelihood choices  and we conclude on the best practices in section~\ref{sec:conc}. 

	 \section{Exoplanet detection metrics}
	 \label{sec:framework}
	 
	 \subsection{Model}
	 \label{sec:model}
	 
	 We will be concerned with the detection of planets in radial velocity data.	Let us suppose that we have a time series of $N$ radial velocity measurements at times $\bm{t} = (t_i)_{i=1..N}$,  denoted by $\bm{y} = (y(t_i))_{i=1..N}$.
	 We use an additive noise representation of the data
	 \begin{align}
	 	\bm{y} = \bm{f}(\bm{\theta}) + \bm{\epsilon},
	 	\label{eq:ymodel}
	 \end{align}
	 where  $\bm{f}(\bm{\theta})$ is a deterministic model and $\bm{\epsilon}$ is a  Gaussian noise whose covariance is parametrized by the vector $\vec \beta$. 	 It includes in particular Gaussian process models of the data ~\citep[e.g.][]{haywood2014,rajpaul2015, faria2016,jones2017}. In the following, $\bm{f}(\bm{\theta})$ is a sum of periodic Keplerian functions, such that the parameters $\bm{\theta}$ include the orbital elements of each planet, in particular their period. Precise mathematical expressions are given in Appendix~\ref{app:model}.

	 One wishes to determine how many planets are in the system, that is how many Keplerian functions must be included in the model, as well as their orbital elements.
	 The existing methods to do so are presented in the following sections.

	 \subsection{Periodogram and false alarm probability}
	 \label{sec:fap}
	 
	 In the context of radial velocity data, FAPs are computed on the basis of a periodogram. This one has many variants, which all rely on comparing the maximum likelihoods obtained with a base model $H_0$ and a model containing $H_0$ plus a periodic component $\omega$. The periodogram $P(\omega, \vec y)$ thus depends on the data $\vec y$ as well as a frequency $\omega$, and is computed on a grid of frequencies. 
	 The base model $H_0$ can be white, Gaussian noise~\citep{schuster1898,  lomb1976, scargle}, include a mean~\citep{ferrazmello1981, cumming1999, zechmeister2009},  a general linear model~\citep{baluev2008}, or a model fitte non linearly at each trial frequency~\citep[][]{baluev2013_vonmises, angladaescude2012}. It is also possible to generalize the definition of periodograms to non sinusoidal periodic functions~\citep{cumming2004, otoole2009, zechmeister2009, baluev2013_vonmises, baluev2015}, several periodic components~\citep{baluev2013} or non-white noises~\citep{delisle2019a}.

	 For a given definition of the periodogram $P(\omega, \vec y)$, a grid of frequency $(\omega_k)_{k=1..M}$ and a dataset $\vec y$, the FAP is defined as follows. Let us suppose that the periodogram of the data of interest has been computed, and has a maximum value $P_\mathrm{max}$.  The false alarm probability is defined as 
	 \begin{align}
	 	\mathrm{FAP}  = 	\pr\left\{ \max\limits_{\omega \in \Omega} P(\omega, \vec y) \geqslant P_\mathrm{max} \;  \Big| \;\vec y \sim H_0   \right\}
	 	\label{eq:fap}
	 \end{align}
	 where $\vec y \sim H_0$ means that the data follows the distribution $H_0$ and Pr stands for probability. 
	 
	 Estimating the FAP can be done by generating datasets that follow the distribution $H_0$ and  computing the empirical distribution of the maxima of periodograms. This method requires extensive computations, especially to estimate very low levels of FAP. Alternatively, one can use sharp analytical approximations, which are very   accurate in the low FAP regime. Analytical approximations are provided in~\cite{baluev2008, baluev2009, baluev2013, baluev2015, delisle2019a}. There are also semi-analytical approaches, where a generalised extreme value distribution is fitted onto the maxima of simulated periodograms~\citep{suveges2014}.
	  The number of planets is then estimated starting at $k=0$ planets. One computes the periodogram and the associated FAP. If the FAP is lower than a certain threshold, typically 0.1\%, the $k+1$ signal model is validated. The planetary origin of the signal must also be discussed. The orbital elements are fitted through a non linear least square minimisation, where the period of the $k+1$-th planet is initialised at the maximum of the peridogram~\citep[e. g.][]{wrighthoward2009}. Then $k$ is incremented and the process is repeated until no detection is found. This method is adopted for instance in \cite{lovis2006, udry2019}.

	 \subsection{Bayes factor}
	 \label{sec:bf}
	 
	 The Bayes factor is a metric comparing two alternative models and relies on the choice of two quantities. First, one must define the likelihood function, that is the probability of data $\vec y$  knowing the model parameters $\vec \theta$, $ p(\bm{y} |  \bm{\theta})$, and secondly, the prior distribution $p(\vec \theta)$, which is the distribution of orbital elements expected before seeing the data (it can also be viewed as a subjective measure of belief~\citep[e.g.][]{cox1946, jain2011}.  %
	 We here call a model a couple prior - likelihood defined on a certain parameter space.
	 The evidence, or marginal likelihood of a model $\mathcal{M}$  is defined as
	 \begin{align}
	 	\label{eq:evidence}
	 	\mathrm{Pr} \{\bm{y} |  \mathcal{M}\} = \int_{\theta \in \mathcal{M}} p(\bm{y} |  \bm{\theta}) p(\bm{\theta}) \dd \bm{\theta} .
	 \end{align}
	 The Bayes factor is then defined as the ratio of the Bayesian evidence of two models~\citep{jeffreys1946,kassraftery1995}. 
	  In the context of exoplanets, one compares models  with $k$ and $k+1$ planets, the  evidences of which  are denoted by $\mathrm{Pr} \{\bm{y} |  {k} \}$ and $\mathrm{Pr} \{\bm{y} |  {k+1} \}$. The model selection is made by computing the Bayes factor, 
	 \begin{equation}
	 	B_{k+1} = \frac{\mathrm{Pr} \{\bm{y} |  {k+1} \}}{\mathrm{Pr} \{\bm{y} |  {k} \}}.
	 	\label{eq:bf}
	 \end{equation}
	 The number of planets is selected as follows. Starting at $k=0$, if the Bayes factor is greater than a certain threshold, typically 150, the $k+1$ model is validated, and $k$ is incremented until no detection is found. Here also, the planetary origin of the signals must be discussed.
	 The evidences of the models with $k$ planets are estimated numerically, typically with Monte-Carlo Markov chains (MCMC) or Nested sampling algorithms~\citep{nelson2018}.
	 The validation of a planet is in general coupled to a periodogram analysis~\citep[e. g.][]{haywood2014} or an analysis of the posterior distribution of periods~\citep[e. g.][]{gregory2007a}, to check that planet candidates have a well defined period. 
	 The computation of~\eqref{eq:evidence} is known to be a difficult numerical problem and evidence estimates must be provided with uncertainties~\citep[e. g.][]{gregory2005,gregory2007a, nelson2018}.

	 \subsection{Posterior number of planets}
	 \label{sec:pnp}
	 
	 One can also compute the posterior number of planets (PNP), that is the probability to have $k$ planets knowing the data. With the notations of the precedent section, for a number of planets $k$, we define the PNP as
	 \begin{align}
	 	\mathrm{Pr}\{k | \bm{y} \}  =  \frac{\mathrm{Pr}\{ \bm{y} | k\} \mathrm{Pr}\{k\}  }{    \sum\limits_{i=0}^{k_{\mathrm{max}} }   \mathrm{Pr}\{ \bm{y} | i \} \mathrm{Pr}\{i\}    }.
	 	\label{eq:jointpost}
	 \end{align}
	 This criterion is suggested by~\cite{brewer2014, brewerdonovan2015} and used in~\cite{faria2018, faria2020}, which uses a nested sampler qualified as trans-dimensional, that is it can explore parameter spaces with sub-spaces of different dimensions. Here, this means that the sampler can jump between models with different number of planets.  
	 In that case, the validation of a planet is in general coupled to a periodogram analysis, an analysis of the posterior distribution of periods \ch{or a composite distribution of the posterior densities defined in \cite{brewerdonovan2015}}, to check that planet candidates have a well defined period. Note that the PNP can also be evaluated with non trans-dimensional samplers, one can evaluate separately the terms of Eq.~\eqref{eq:jointpost}.

	 \subsection{Others}

	 The periodogram, FAP and BF are the most used tools for exoplanets detection, but other approaches have been proposed. 
	 The $\ell_1$ periodogram, as defined in~\citep{hara2017}, has been used in several works~\citep[e.g][]{hobson2018, hobson2019,santerne2019, hara2020, leleu2021}. 
	 This  tool is based on a sparse recovery technique called the basis pursuit algorithm~\citep{chen1998}. The $\ell_1$ periodogram takes in a frequency grid and an assumed covariance matrix of the noise as input. It aims to find a representation of the RV time series as a sum of a small number of sinusoids whose frequencies are in the input grid. It outputs a figure which has a similar aspect as a regular periodogram, but with fewer peaks due to aliasing. 
	 The $\ell_1$ periodogram can be used to select the periods, whose significance is then assessed with a FAP or an approximation of the Bayes factor~\citep{nelson2018}.

	 There are several variations of periodograms relying on the marginalisation of parameters other than period, such as~\cite{mortier2015, feng2017}.
	 Other methods exist, which are more agnostic to the shape of the signal. \cite{mortier2017} suggests to compute periodogram adding one point at a time to check whether the evolution of a peak amplitude is compatible with a purely sinusoidal origin.  \cite{gregory2016} suggests to include in the model a so-called apodization term, that is a multiplicative factor of the Keplerian signals which determines whether they are consistent through time or transient. \cite{zucker2015,zucker2016} suggests to use a Hoeffding test, based on the phase folded-data. One can also look for statistical dependency of the data with an angle variable  (phase correlation periodograms~\citep{zucker2018}.

	 \section{The FIP as a detection criterion: definition and main properties}
	 \label{sec:newcrit}
	 
	 \subsection{Motivation}
	 
	 The detection criteria described in section~\ref{sec:fap},~\ref{sec:bf} and~\ref{sec:pnp} have several shortcomings. 
	 First, except the PNP which is an actual probability, the meaning of a given detection threshold for the FAP and BF is unclear. The scale of the BF is empirical~\citep{jeffreys1946}, and does not have an easy interpretation. The FAP is not a probability of an observed event, but of a hypothetical one. Though in practice useful, it is not as easy to interpret as the probability of a certain event knowing the data. Secondly, it might happen that the FAP, BF or PNP support the detection of an additional planet, while not giving a clear indication of its period, but one would not claim the detection of a planet without being confident in its period. Thirdly, the FAP ignores the potential underlying population of orbital elements, such that it cannot distinguish planets with very rare characteristics, for which a high likelihood is required for detection, and common ones.  \chd{The Bayes factor is asymptotically consistent \citep{chib2016}, but in finite sample comparing models only two by two might be problematic if only sequential comparisons are made (1 planet versus 0, 2 planets vs 1 planet and so on)}. As noted in \cite{brewerdonovan2015}, the Bayes factor does not marginalise over possible models. For instance if two planets are presents, the Bayes factor of the models with 1 planet versus 0 ignores the possibility of a second planet, which potentially results in incorrect decisions.  On the contrary, for instance the PNP is not necessary limited to comparing models two by two.

	 Our goal is to define a detection criterion which combines the information on the number of planets and their orbital elements, especially the period, and is put on scale where the detection threshold has a clear meaning. Validating a planet would essentially come down to the following statement: the data cannot be explained without a planet with period in interval $I$ (or more generally with orbital elements in a certain region of the parameter space).

	 \subsection{The false inclusion probability (FIP)}
	 \label{sec:fip}
	 \subsubsection{Definition}
 
	 We define a new detection criterion based on the joint posterior distribution of the orbital elements and the number of planets. We define the true inclusion probability (TIP) as  the posterior probability of the following event: for a given range of periods $I$, there is at least one planet with period $P \in I$. By analogy with the FAP, we also define the false inclusion probability (FIP), which is the probability that there is no planet with period in interval $I$; Formally, the TIP and FIP are 
	 \begin{align}
	 	\label{eq:criterion}
	 	\mathrm{TIP}_I &= \mathrm{Pr}\{ \exists P, P \in I| \bm{y}  \}. \\
	 	\mathrm{FIP}_I &= 1 - \mathrm{TIP}_I .
	 	\label{eq:fip}
	 \end{align}
	 Unlike the Bayes factor and  the FAP, Eq.~\eqref{eq:criterion} is not computed iteratively when comparing models with $k$ and $k+1$ planets, but by ``averaging'' the detection of the planet over the possible number of planets. 
	 
	 The TIP and FIP can be defined in other contexts. The TIP is simply the explicit expression of the feature to be detected. Assuming the data has a model parametrized by $\theta$ in a parameter space $\Theta$ and a the searched feature corresponds to a subspace of parameters $\Theta'$, the TIP can be defined as the probability that  $\mathrm{Pr}\{ \theta \in \Theta' |y\}$. For any periodic signal detection, for instance in the context of planetary transits, the space $\Theta'$ can be defined as a period interval as in eq.~\eqref{eq:criterion}. \ch{ The definition of the TIP is close to the posterior inclusion probability (PIP), defined in the context of linear regression \citep{barbieriberger2004}, and a similar quantity is defined in \citep{brewerdonovan2015}. We further discuss the relationship of the methods of the present work with existing works in Section~\ref{sec:previouswork}.  }

	 \subsubsection{Computation from classical samplers}
	 \label{sec:comp1}
	 
	 In practice, one can evaluate Eq.~\eqref{eq:criterion}  as follows.
	 We suppose that there is a maximum number of Keplerian signals in the data $k_{\mathrm{max}}$. We denote by $p(\vec \theta|k)$ the prior probability of model parameters knowing there are $k$ planets, and $p(\bm{y}|\vec \theta,k)$ the likelihood of the data $\bm{y}$ knowing the number of planets $k$ and the orbital parameters. We suppose a prior probability for the model with $k$ planets $\mathrm{Pr}\{k\}$. Then, as defined in Eq.~\eqref{eq:criterion}, the TIP is
	 \begin{align}
	 	\label{eq:calc0}
	 	\mathrm{TIP}_I  = \sum\limits_{k=0}^{k_{\mathrm{max}}}    \mathrm{Pr}\{ \exists i \in [1..k] , P_i \in I | \bm{y}, k  \} \mathrm{Pr}\{k | \bm{y} \} % \\
	 \end{align}
	 where $(P_i )_i={1..k}$ are the periods of the $k$ planets in the model. 
	 The terms appearing in Eq.~\eqref{eq:calc0} are computed as follows. The expression of  $\mathrm{Pr}\{k | \bm{y} \}$ is given in Eq.~\eqref{eq:jointpost}. Its computation necessitates to evaluate the marginal likelihood, which can be computed via importance sampling or nested sampling, as described in~\cite{nelson2018}.
 	 The quantity  $ \mathrm{Pr}\{ \exists i \in [1..k] , P_i \in I | \bm{y}, k  \} $ can easily be estimated from samples of the posterior distribution of the parameters for a number of planets fixed to $k$, $p(\vec\theta | \vec y, k)$.   One only needs to compute the number of samples for which there is at least a period of one of the planetary signals in $I$, divided by the total number of samples. There are several ways to sample the distribution $p(\vec \theta | \vec y, k)$, such as Monte-Carlo Markov chains with parallel tempering~\citep[e.g.][]{gregory2007} and nested sampling~\citep[e.g.][]{buchner2021}. \ch{Note that Eq.~\eqref{eq:calc0} can be straightforwardly generalised to average over models of the data, for instance over different noise models. } 
	 
	 \subsubsection{Computation from trans-dimensional samplers}
	 \label{sec:comp2}
	 
	 Eq.~\eqref{eq:criterion} can also be computed from the joint posterior probability of the number of planets and the orbital elements $p(k,\vec \theta|\bm{y}$). In that case, we simply need to compute how many samples are such that at least one planet has a period in $I$. 
	 Several samplers can handle parameter spaces with different numbers of dimensions, such as reversible jump MCMC~\citep{green1995} and trans-dimensional nested samplers~\citep{brewer2014}.

	 \subsubsection{Practical use: the FIP periodogram}
	 \label{sec:practicaluse}

\begin{figure*}
	 	\centering
	 	\includegraphics[width=0.7\linewidth]{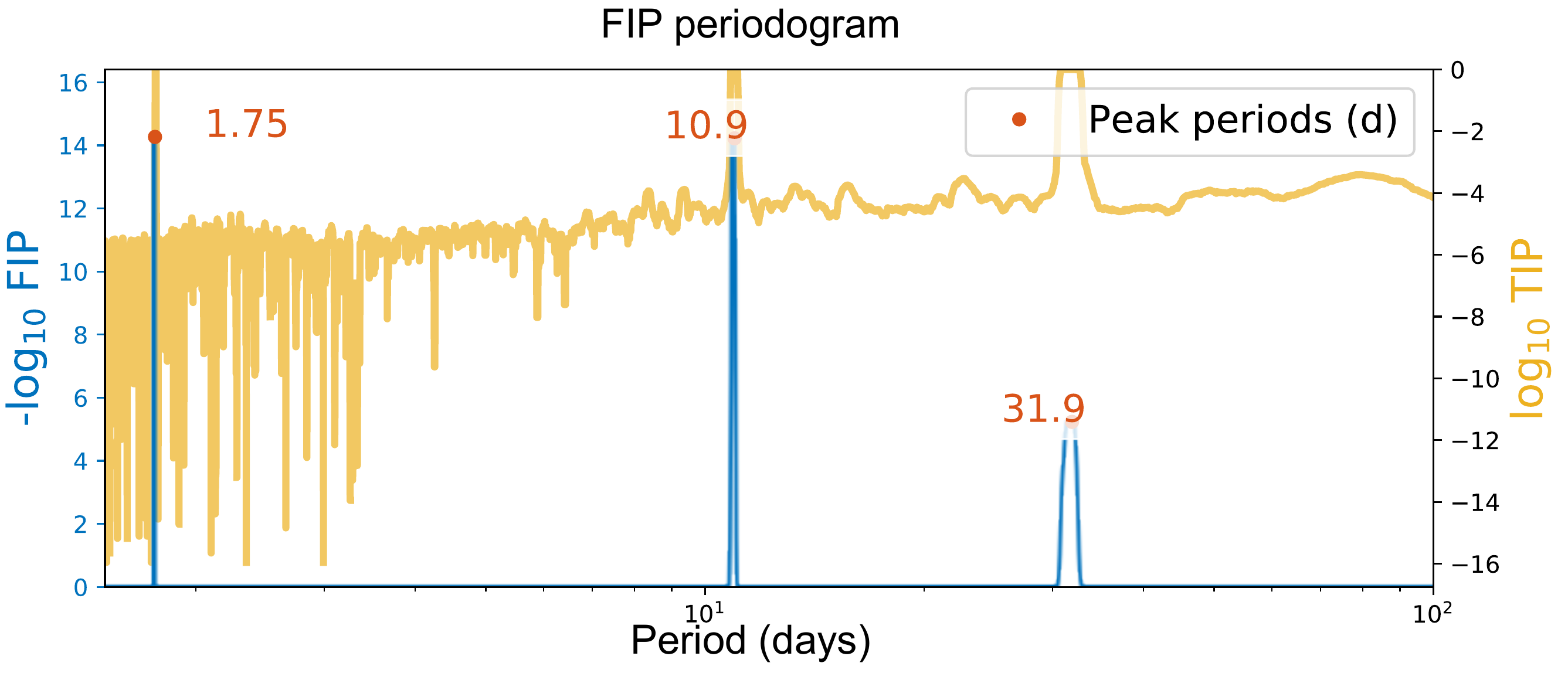}
	 	\caption{FIP periodogram of a Simulated system with three injected planets. The periods of the peaks are indicated in red points, the -$\log_{10} $FIP and $\log_{10} $TIP are represented as a function of the center of the period interval considered, in blue and yellow respectively. }
	 	\label{fig:ex1}
	 \end{figure*}

	 The periods where to search the planets are unknown \textit{a priori}. We compute the FIP as a function of period as follows. 
	 We consider a grid of frequency intervals with a fixed length. The element $k$ of the grid is $I_k$ is defined as $[\omega_k - \Delta \omega/2, \omega_k + \Delta \omega/2]$ where $ \Delta \omega = 2\pi/T_\mathrm{obs}$, $T_\mathrm{obs}$ is the total observation timespan and  $ \omega_k = k\Delta \omega / N_\mathrm{oversampling}$. We take $N_\mathrm{oversampling} = 5$. The rationale of this choice is that the resolution in frequency is approximately constant, and of width $\approx \Delta \omega$. We call the resulting figure a FIP-periodogram.
	 In Fig~\ref{fig:ex1}, we show an example of such a calculation for a simulated system. This one is generated the 80 first measurements of HD 69830~\citep{lovis2006} \ch{(this dataset is presented in Appendix \ref{app:datasets})}, it contains three circular planets whose randomly selected periods are 1.75, 10.9 and 31.9 days. The noise is white and generated according to the nominal uncertainties ($\approx 1$ m/s). The semi-amplitude of planets are generated from a Rayleigh distribution with $\sigma = 1.5$ m/s.
	 The $x$ axis represents the period and the $y$ axis represents the FIP on a grid of intervals $I_k$ as defined above. To emulate the aspect of a classical periodogram, we represent on the $y$ axis, in blue, $-\log_{10} \mathrm{FIP}$, so that high peaks correspond to confident detections. We also represent the $\log_{10} \mathrm{TIP}$ (TIP = 1-FIP) in yellow, in order to spot peaks with low significance. The scale is given on the right $y$ axis. 
	  The signals injected are confidently recovered with a FIP of $10^{-14.2}$,  $10^{-14.2}$ and $10^{-5.2}$.
	 In the following, we refer to figures such as  Fig~\ref{fig:ex1} as FIP periodograms.
	 
	 The FIP is marginalised on the number of planets of the system. However, in practice, one does not know the maximum number of planets. To decide when to stop searching for additional planets, we proceed as follows. For a given maximum number of planets $k$, we compute the FIP-periodogram as defined above. We then compute the difference between the FIP-periodogram with $k$ and $k-1$ planets. If the maximum of the absolute difference between the two is such that the decision about which planets are detected does not change, then one can stop the calculations. As an order of magnitude, a maximum difference below 1 corresponds to a change of FIP of at most a factor 10, which is often sufficiently precise to conclude. 
	 One can also use as a convergence criterion that both the difference between FIP-periodograms of $k+2$ and $k+1$ planets and  $k+1$ and $k$ planets are below a fixed threshold. This criterion is more robust, but also more computationally costly. 
	 For the comparison of FIP periodograms with $k$ and $k+1$ planets to be meaningful, it must be checked that each of those FIP periodograms are accurate. The reliability of the FIP periodogram with $k$ planets can be checked by computing it with different runs of the algorithm, and check that the variation of the FIP periodograms values between runs is below a certain threshold. Examples of application of the convergence tests are given in Section~\ref{sec:application}.

\subsubsection{Relation with existing works}
\label{sec:previouswork}

\ch{It is apparent in Eq.~\eqref{eq:calc0} that the TIP is a particular case of Bayesian model averaging~\citep[e.g.][]{hoeting1999}, since we estimate the probability of a quantity of interest weighted by the posterior distribution of the alternative models defined. \cite{barbieriberger2004} introduce a quantity similar to the TIP, the posterior inclusion probability (PIP). This one is defined in the context of linear regression where the data $\vec y$ (vector of size $N$) has a model $\vec y = \mat X \beta + \vec \epsilon$, $\mat X$ being a $N\times p$, $p<N$, $\vec \epsilon$ is a random noise and $\vec \beta$ a vector of $p$ parameters. One defines alternative models, corresponding to subsets of $\{1..p\}$. The PIP of index $ i$, $1\leqslant i \leqslant N$ is defined as the sum of posterior probabilities of models that include indices $i$. \cite{barbieriberger2004} show that the Median posterior model (MPM), that is the model corresponding to indices with PIP>0.5, under certain conditions generalised in \cite{barbieri2021}, has the optimal prediction error (quadratic penalty). The threshold of 0.5 simply means that it is more likely than not that $i$ is non zero. The TIP can here be seen as the prolongation of the PIP to the continuous parameter case.   }

The FIP periodogram shares with the Keplerian periodogram~\citep{gregory2007a, gregory2007} as well as AGATHA periodograms~\citep{feng2017}  that the period selection is made by marginalising over parameters other than period. Here we marginalise, in addition, over the number of planetary signals. Therefore, the FIP provides a single detection metric, which, furthermore, can directly be interpreted as a probability. \ch{The definition of the FIP periodogram in Section \ref{sec:practicaluse} is especially close, though not equivalent, to the quantity defined in Eq. (9) of \citet{brewerdonovan2015}, which is the sum of the posterior densities of the $\log$ periods of the planets in the model. The probability that there is at least one planet in a certain period interval $I$ is equal to the sum on $i$ of the probability of events $E_i$: ``the period of planet $i$ is in $I$'' provided the $E_i$ are disjoint, that is the probability of having two different planets in $i$ is zero. In practice this probability is very small but not strictly zero. If the quantity defined in Eq. (9) of \cite{brewerdonovan2015} was binned,  then this would be close to a FIP periodogram with a bin of constant size in $\log$-period. Finally, \cite{brewerdonovan2015} suggest to use the posterior probability of the event Q := ``a planet exists with period between 35 and 37 d'' which is what we suggest to do. In the following Sections, we examine the properties of using systematically the probability of such events (the TIP) as a detection criterion. In particular in the next section, we highlight a property of the FIP which can be seen as a Bayesian false discovery rate \citep[][]{benjamini1995}. }

	 \subsection{Properties}
	 \subsubsection{Fundamental property}
	 \label{sec:fundprop}
	 
	 \paragraph{Property} One of the advantages of the quantity~\eqref{eq:fip} is that it is easy to interpret: if the likelihood and prior accurately represent the data, and a series of statistically independent detections are made with \ch{FIP $= \alpha$ (or TIP $= 1-\alpha$}, then, on average, a fraction $1-\alpha$ are true detections. More precisely, the number of true detections among $M$ detections follows a binomial distribution $B(M,1-\alpha)$. 
	 
	 In practice, this has the following meaning.
	 Let us consider a collection of intervals $I_j$ and of RV datasets $\vec y_j$ $j=1..n$, such that the events ``there exists a planet with period in $I_j$ knowing data $\vec y_j$'' are statistically independent, and such that $\mathrm{TIP}_{I_j} = 1-\alpha$ for a certain $\alpha$ between 0 and 1. The $\vec y_j$ could be the same data set or different ones, we only require independence of the events. Then, provided the likelihood and priors used in the computations of $\mathrm{TIP}_{I_j}$ are correct, 
	 \begin{align}
	 	\lim\limits_{n \rightarrow +\infty} \frac{\#  P \in I_j}{n} = 1-\alpha .
	 	\label{eq:property1}
	 \end{align}  	
	 \ch{We can be even more precise: the number of times a planet with period $P \in I_j$ and FIP = $\alpha$ was indeed present among $M$ statistically independent detections  follows a binomial distribution $B(M,1-\alpha)$}.

	 \paragraph{Consistency test}
	 The property~\eqref{eq:property1} holds if the priors and likelihoods are exactly the same as those that generated the data, which is unlikely to happen in real cases. We suggest to see~\eqref{eq:property1} as a device to calibrate the scale of probability used to detect exoplanets. Indeed, \eqref{eq:property1} is a prediction, which can be used to test if the model used in the analysis is correct. In principle, let us suppose that several data sets (for instance the HARPS data) have been analysed and FIPs are computed at time $t_1$. As more data comes along, at $t_2 > t_1$ the presence of certain planets will be confirmed with very high probability. One can check that statistically independent detections made at $t_1$ with FIP $\alpha$, are such that a fraction $\alpha$ of them are spurious, up to the uncertainties of a Binomial distribution. However, if property~\eqref{eq:property1} is a necessary condition for RV models to be validated, it is not a sufficient one. The fact that it is satisfied does not guarantee that RV models are all correct. 
	 \ch{The property~\eqref{eq:property1} we put forward pertains to a frequency of events, while in Bayesian analyses probabilities are usually interpreted as subjective measures of belief, but as we discuss below, this does not constitute a contradiction. } \chd{The consistency test we suggest is a particular case of Bayesian model calibration \citep[][]{Draper2013}.}

	 \paragraph{Discussion} 	 
 \ch{It has been mathematically established that in any system of quantitative measure of belief satisfying intuitive properties, the update of the belief measure in view of new information has to be made according to Bayes formula~\citep[][ although see comments by \cite{paris1994,halpern1999a}]{cox1946,cox1961}. Probabilites are typically used as subjective measures in the Bayesian context. However, Cox's theorem result does not give prescriptions on how to select the initial belief. In the case of exoplanet detection, it seems to us that, among subjective measures, it seems natural to desire that (1) a probability gives the actual fraction of times you would be wrong when claiming a detection with a certain significance (2) the prior probability represents the distribution of elements in nature, like in a hierarchical Bayes model such as \cite{hogg2010}. 
If so, then \eqref{eq:property1} has to be satisfied. To illustrate our approach, we can consider the context of information theory~\citep{shannon1948}, in which priors reflect the occurrence of a word in a certain context within a given language and the posterior probability of a word at the receiving end the communication channel reflects a frequency of errors. In the context of exoplanets, ``words'' would be the vector of orbital and stellar parameters of a system, and we wish them to be distributed according to the true distribution of planetary systems (the ``language''). A given FIP threshold then corresponds to a concrete, verifiable property, while a Bayes factor scale is harder to interpret. }
 Tying the probabilities to frequencies within Bayesian analysis has been suggested in \cite{vanfraassen1984, shimony1988} from an epistemological point of view.

 \ch{
 One could argue that the influences of the prior vanishes as more data is acquired~\citep[][]{wald1950}, and, as a consequence, it is not necessary to tie the meaning of a prior to an observable, or operate the consistency test we suggest in this section. However, in the context of exoplanets the asymptotic regime ($N\rightarrow \infty$) is not reached: the influence of the prior is rarely negligible. Secondly, results showing the convergence of parameters estimate when $N\rightarrow \infty$ regardless of the prior assume a certain form for the likelihood as a parametrised function, which might not accurately represent the noise properties. In the context of radial velocities, stellar noise models are unreliable, but if they were, \eqref{eq:property1} would be satisfied. } 
 
 \ch{In conclusion, we believe that the property \eqref{eq:property1} offers an intuitive interpretation of the FIP. Furthermore, \chd{it can serve as a model calibration test \citep{Draper2013}}, although it will not give precise indications on whether the prior or the likelihood is faulty.   }

	 \subsubsection{Aliasing}
	 \label{sec:aliasing}
	 
	 Radial velocities have a sampling that is irregular but close to an equispaced sampling with a step of one sidereal day (0.997 day) with missing samples. As such, periodogram signals at frequency $\nu_0$ typically exhibit aliases at $\nu_0 + \nu_s$ and $-\nu_0 + \nu_s$  where $\nu_s = 1/0.997$  day$^{-1}$. As a consequence, it is uncertain whether the signal is at  $\nu_0$ or $  \pm \nu_0 + \nu_s$ is the true signal~\citep{dawsonfabricky2010, robertson2018}. The yearly and monthly repetition of the sampling patterns, other gap in the data potentially create more aliasing problems. 
	 The FIP periodogram provides insight into this problem, since if there is a degeneracy between two periods, the samples will be split between the two in proportion of the probability that they are \ch{supported by the data and assumed likelihood and 
	priors}. 

	 Aliasing can also create problems when several signals are present, and might result in a high periodogram peak which does not correspond to any of the true periods~\citep{ hara2017}. FIP periodograms also address that situation as, by design, several planets are searched at the same time.

	 \subsubsection{Another perspective on error bars}
	 \label{sec:errorbars}	 
	 	 The error bars on the orbital elements of a planetary system often are computed with Monte-Carlo Markov chain methods~\citep[MCMC, e.g.][]{ford2005, ford2006}. Once the planets have been detected, one computes the posterior distribution of the orbital elements. 
	 The FIP can be generalised to express the probability that the model includes a planet with its orbital elements in a certain set. One can define the probability that there is a planet with orbital elements in a set $S$ as
	 \begin{align}
	 	\label{eq:Scriterion}
	 	\mathrm{TIP}_S = \mathrm{Pr}\{ \exists (P, e, \omega, K, M_0) \in S| \bm{y}  \}.
	 \end{align}
	 \ch{Just like in the case of credible regions,} one can define a sequence of probabilities $(p_i)_i$ and a corresponding sequence of sets $S_i$ such that $P_{S_i} = p_i$. In that case, one obtains regions of probability $(p_i)_i$, for which the uncertainties on the number of planets is propagated. This can be useful in particular for population studies. Instead of excluding planets that do not meet a certain detection thresholds, one can take into account marginal detections with a rigorous account on the uncertainty on whether there is a planet.

	 \subsubsection{Excluding planets}
	 \label{sec:nondetec}
	 
	 One of the properties of the formalism we develop is that it can put an easily interpretable condition on the absence of planets. The FIP, by definition, is the probability not to have a planet in a certain range. Excluding the presence of planets with a certain confidence might be of interest to study the architecture of planetary systems. For instance, it can be helpful to put constraints on the total mass of the planets and compare it to the minimum mass Solar nebula~\citep{hayashi1981}.

	 \section{FIP: practical computations}
	 \label{sec:practicalcomp}
	 
	 \subsection{\textsc{polychord}}  
	 \label{sec:polychord}
	 The computation  method suggested in section~\ref{sec:comp1} necessitates to compute the evidence $\mathrm{Pr}\{ \bm{y} | k\}$ of a model with $k$ planets and the posterior of the orbital elements knowing the number of planets, $p(\vec \theta | \bm{y}, k )$.	These two quantities can be computed with a nested sampling algorithm. In the present work, we use \textsc{polychord}~\citep{handley2015,handley2015b}. By default, we set the number of live points as forty times the number of free parameters in the model.

	 \subsection{Marginalising over linear parameters}
	 \label{sec:margin}

	 As described in Section~\ref{sec:model}, our model of the data $\vec y$ is $\bm{y} = vec f(\bm{\theta}) + \bm{\epsilon} $ where $\bm{\epsilon}$   is a random variable whose distribution is parametrized by $\vec \beta$ and $\vec f$ a deterministic function. Let us now separate the model parameters $\vec \theta$ in two categories: the model depends linearly on parameters $\vec x$ and non-linearly on parameters $\vec \eta$.
In appendix~\ref{app:marg}, we show that, provided the likelihood is Gaussian and the prior on the linear parameters $\vec x$ is a Gaussian mixture model, then the integral 
	 \begin{align}
	 	p(\bm{y} |\bm{\eta}, \bm{\beta},k)  = \int p(\bm{y} |\vec x,  \bm{\eta}, \bm{\beta},k)p(\vec x ) \dd \vec x
	 \end{align}
	 has an analytical expression.
	 
	 Marginalising over linear parameters presents the advantage of reducing the number of dimensions to explore with the nested sampling algorithm. To reduce as much as possible the number of non linear parameters, we rewrite the radial velocity due to a planet. Denoting it by $v(t)$, Instead of 
	 \begin{align}
v(t) & = K(\cos\left(\omega + \nu(t)\right) + e \cos \omega)
	 \end{align}
	 where $K$ is the semi-amplitude, $t$ is the time, $\nu$ the true anomaly, $e$ the eccentricity and $\omega$ the argument of periastron, we write 
	 \begin{align}
v(t) & = A\cos\left(\nu(t)\right) + B\sin\left(\nu(t)\right) + C. \label{eq:vexprabbody}
	 \end{align}
	 We obviously only use one offset $C$ in our model, even if several planets are present.

	 The Gaussian mixture components would typically be chosen to represent populations such as Super-Earth, Mini-Neptunes~\citep{fulton2017}, Neptunes, Jupiters etc. 
	 This can be leveraged to speed up computations, by removing dimensions of the parameter space. 
	 When the linear parameters are analytically marginalised, $\bm{\eta}$ and $\bm{\beta}$ are the only free parameters, such that only three parameters are used per planets  instead of five : period, eccentricity $e$ and time of passage at periastron $t_p$ (or equivalently the initial mean anomaly). This idea is similar to the use of a Laplace approximation of the evidence marginalised on certain parameters~\citep[][]{pricewhelan2017}.
	 
	 When the prior on $A$ and $B$ in eq.~\eqref{eq:vexprabbody} is Gaussian with null mean and variance $\sigma^2$, this translates to a Rayleigh prior on $\sqrt{A^2+B^2}$ with parameter $\sigma$. For simplicity, we will refer to this situation as a Rayleigh prior on $K$ with parameter $\sigma$, even though $K \neq \sqrt{A^2+B^2}$ for $e>0$.

	 \subsection{Validation of the FIP computations}
	 \label{sec:validation}
	 
	 To validate our algorithms, we perform several tests on simulated data. Here, our goal is not to evaluate the performances of the FIP as a detection criterion, but to ensure that our numerical methods are retrieving a good approximation of Eq.~\eqref{eq:criterion}. \ch{Claiming a detection is then: ``there is a planet with period in interval $I$''}.
	 We have seen in section~\ref{sec:fundprop} that, when the prior and likelihood are correct, on average, a fraction $\alpha$ of independent detections made with a FIP $\alpha$ are spurious (see Eq.~\eqref{eq:property1}), \ch{and a fraction 1-$\alpha$ is correct}. If our numerical method is correct, then this property must be verified in practice.

	 \ch{We will verify whether the property~\eqref{eq:property1} is true on a thousand generated datasets}. 
	 In the first test, we consider only circular orbits. The model of the signal is 
	 \begin{align}
	 	y(t_j) = C + \sum\limits_{i=1..k} A_i \cos \left(\frac{2\pi}{P_i} t_j\right) + B_i \sin \left(\frac{2\pi}{P_i}  t_j\right)   +  \epsilon_j
	 	\label{eq:circ}
	 \end{align}
	 where $\epsilon_j$ is the noise. We consider $t_i$ from the 80 first measurements of HD 69830~\citep{lovis2006} \ch{(this dataset is presented in Appendix \ref{app:datasets})}. The values of $\vec \theta$ := $k$, $(A_i)_{i=1..k}, (B_i)_{i=1..k}$, $C$ and $P$ \ch{ are generated according to distributions} shown in Table~\ref{tab:priorscirc}. We denote by $G(\mu,\sigma^2)$ a Gaussian distribution of mean $\mu$ and variance $\sigma^2$. Once a value of $\vec \theta$ has been drawn, we create a data set by drawing the $\epsilon_j$ from a Gaussian distribution of null mean and standard \ch{deviation given by} the nominal uncertainties on the 80 first measurements of HD 69830 (\ch{typically 0.45 m/s}). 
	 
	 \ch{We now have a thousand data sets. }For each of them RV, we compute the FIP-periodogram  as defined in section~\ref{sec:practicaluse}   with exactly the same priors and noise model as the ones used to generate the data. \ch{If our calculations of the FIP periodograms have converged, on average a fraction $1-\alpha$ of independent detections made with FIP $\alpha$ should be correct. }
	 
	 We consider a grid of probabilities from 0 to 1, $(\alpha_j)_{j=1..M}$. For each of the probability of the grid, \ch{we search for detections with FIP $\alpha_j$}. \ch{For instance, we fix $\alpha$ = 10\% and search for  intervals $I$ such that the event ``there is a planet with $P \in I$'' has a FIP of 10\%. } If several events  ``presence of a signal in a certain period interval'' have a probability $\alpha_j$ for the same dataset, we select one of them randomly. As a result of this process, for each $\alpha_j$, for each generated system indexed by $n$, we have selected at most one interval $I_j^n$ such that the event ``there is a planet in the interval $I_j^n$'' has probability $\alpha_j$.  If for system $n$  there is no such event, we simply do not include system $n$ in the computation.  
	 
	 Since for a given $\alpha_j$, the events \ch{``there is a planet in the interval $I_j^n$ with FIP $\alpha_j$ ''} we have selected are independent, we expect from section~\ref{sec:fundprop} that for a fraction $\alpha_j$ of them, there is actually no planet in the interval. Equivalently, in a fraction $p_j = 1-\alpha_j$ of them, there will actually  be a planet.

\begin{figure}

        \centering
        \includegraphics[width=8cm]{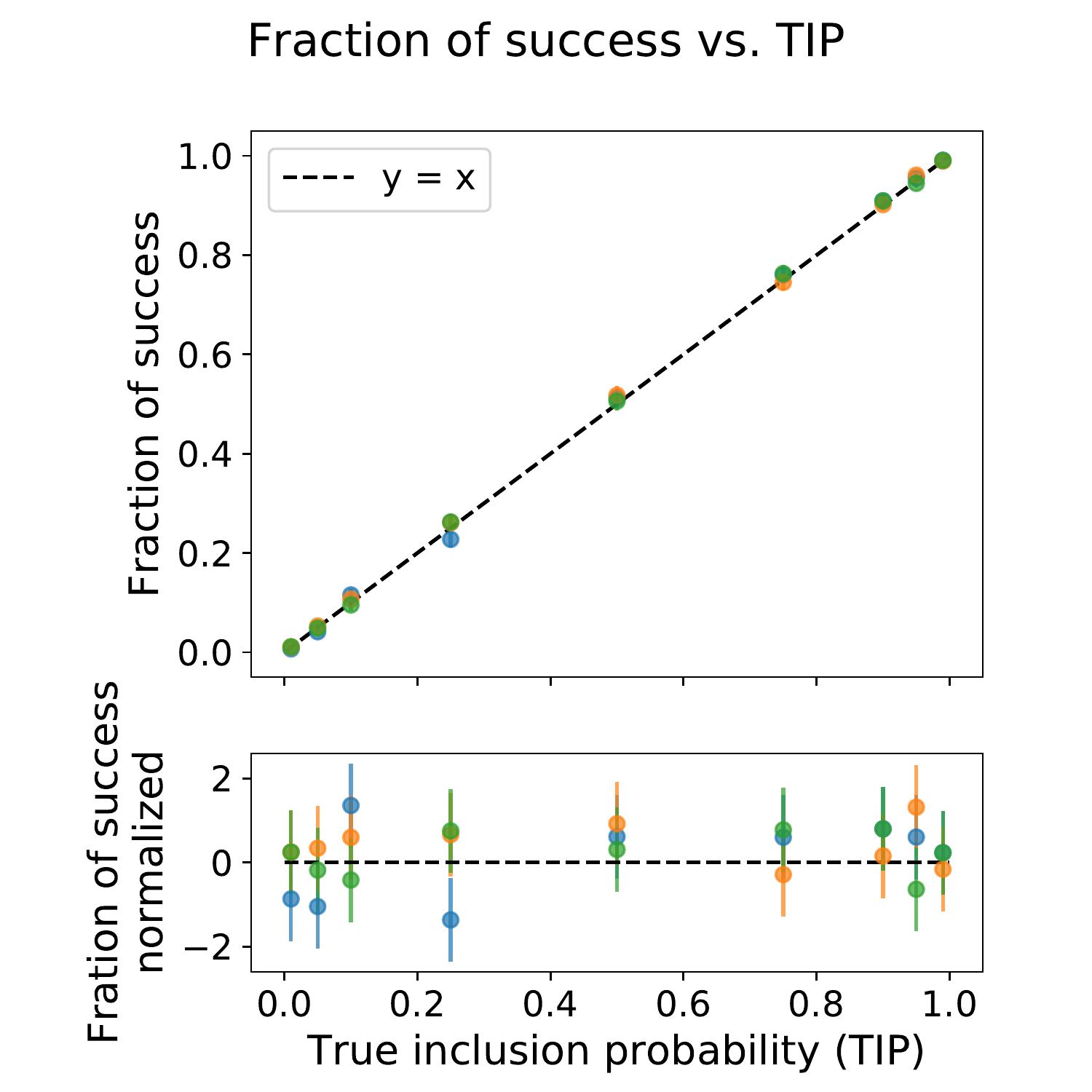}
        \caption{Fraction of events with probability $p_j$ where there actually was a planet injected as a function of $p_j$. The colors blue, orange and green correspond to different, random choices of events with TIP $p_j$.}
        \label{fig:successalpha}
    \end{figure}

	 More precisely, for fixed $j$, the events ``there is a planet in the interval $I_j^n$'' should be independent realizations of a Bernouilli distribution of parameter $p_j $. This means that the number of success (there is indeed a planet) divided by the number of events $N_j$ should be  on average $p_j $  with a standard deviation $ \sqrt{(1-p_j )p_j  /N_j}$. In Fig.~\ref{fig:successalpha}, upper panel, we represent the fraction of success as a function of $p_j$ as well as the error bar $\sigma_j = \sqrt{(1-p_j )p_j  /N_j}$. \ch{Let us recall that for a given dataset $n$, when there are several events ``there is a planet whose period is in the interval $I_j^n$  with probability $\alpha_j$'', we choose one event randomly. Points of different color correspond to different realisation of the random choice.}   In the lower panel, we represent the difference of the fraction of success and the TIP $p_j$, divided by the error bar $\sigma_j$.
	 In the upper panel, we expect a curve which is compatible with $y=x$, which seems to be the case. More precisely, according to our hypotheses, the quantity plotted in the lower panel should be distributed according to a distribution of mean 0 and variance 1, which appears to be the case.  

	 The same test has been repeated in several configurations (0 to 4 planets, red noise, eccentric planets...) described in Appendix~\ref{app:validation}. 	In all cases we find an agreement between the expected and observed distribution of FIPs, with at most two sigmas. We find the highest discrepancy in the simulation allowing highly eccentric orbits. We attribute this to the difficulty to explore the parameter space of highly eccentric orbits, which contains a consequent amount of local minima, as shown in~\cite{baluev2015} and~\cite{hara2019ecc}.

	 \begin{table}
	 	\caption{Priors used to generate and analyse the 1000 systems with circular orbits.}
	 	\begin{tabular}{p{1cm}|p{3.9cm}|p{2.8cm}}
	 		Param- eter & Prior & Values \\ \hline \hline
	 		k & Uniform $[k_\mathrm{min}, k_\mathrm{max}]$ & $k_\mathrm{min} = 0 $, $k_\mathrm{max} = 2$  \\
	 		A & $G(0,\sigma_A^2)$ & $\sigma_A = 1.5$ m/s\\
	 		B & $G(0,\sigma_B^2)$ & $\sigma_B = 1.5$ m/s\\	
	 		C & $G(0,\sigma_C^2)$ & $\sigma_C = 1$ m/s\\		
	 		P & $\log$-uniform on $[P_\mathrm{min}, P_\mathrm{max}]$ & $P_\mathrm{min} = 1.5 $, $P_\mathrm{max} = 100$
	 	\end{tabular}
	 	\label{tab:priorscirc}
	 \end{table}
	 
\section{Example: the FIP periodogram of HD 10180}
		\label{sec:application}
		
In this section, we compute the FIP periodogram of the HARPS data of HD 10180. This system is known to host at least six planets with periods 5.759, 16.35, 49.7, 122, 604, 2205 days and minimum masses ranging from 11 to 65 $M_\oplus$~\citep{lovis2011}. \cite{feroz2011} also find evidence for six planets. Three other unconfirmed planets have been claimed in~\cite{tuomi2012hd10180} at 1.17, 9.65 and 67 days. The first 190 points of the HARPS HD 10180 dataset have been analysed in~\cite{faria2018} (p. 66) with a trans-dimensional nested sampling algorithm~\citep{brewer2014}, in which the number of planets freely varies with the other parameters. In the analysis of~\cite{faria2018}  it is found that, when taking as a detection criterion the Bayes factor and a detection threshold at 150, only  six planets are found. However, it appears that taking a uniform prior on the number of planets, the peak of the posterior number of planets increases monotonically until 19 planets. 

As in~\cite{faria2018}, we analyse the first 190 points of the HARPS dataset\footnote{The data can be downloaded here: \url{https://dace.unige.ch/radialVelocities/?pattern=HD10180}}.  \ch{The radial velocity measurements span on 3.8 years (from BJD 2452948 to 2455376) and have a typical nominal uncertainty of 0.6 m/s. The data are presented in Appendix \ref{app:datasets}. } 
 We here use \textsc{polychord}~\citep{handley2015,handley2015b} to compute the posterior distribution of orbital elements and the Bayesian evidence for models with a fixed number of planets. The FIP is then computed as described in Section~\ref{sec:comp1}, and the FIP-periodogram defined as -$\log_{10} FIP(\omega)$ where $FIP(\omega)$ is the FIP of the event ``there is at least one planet with frequency in the interval $[\omega-\Delta \omega, \omega-\Delta \omega]$'' with $\Delta \omega = 2\pi/T_\mathrm{obs}$, $T_\mathrm{obs}$ being the observation time-span (see Section~\ref{sec:fip} for details). 
 We have defined two convergence tests. In Section~\ref{sec:fip} we suggested to check that the absolute difference of FIP periodogram computed with $k$ and $k+1$ planets, maximised over the period, is such that the decision taken about the data does not change. 
 For a given number of planets, to ensure that the FIP periodogram has converged, as suggested in Section~\ref{sec:fip},  we perform several runs of  \textsc{polychord} (here three) and ensure that the maximum difference of FIP periodogram over all frequencies is below a certain threshold. We computed the FIP periodograms  with priors and likelihood summarised in Table~\ref{tab:priorhd10180}. The prior on semi-amplitude is $\log$-uniform, so that the analytical marginalisation of linear parameters described in Section~\ref{sec:margin} cannot be performed. We set the number of live points to 40 times the number of free parameters, that is 1360 live points for the six planets model. Calculations are made on the DACE cluster (Univ. Geneva) of the LESTA server using 32 cores of the Intel(R)Xeon(R) Gold 5218 CPU @ 2.30GHz.
 
A first calculation of the FIP up to five planets shows that planets at 5.759, 16.35, 49.7, 122, 2205 days have a very low FIP ($10^{-12}$), and are therefore detected with a very high confidence. To improve the convergence of the algorithm, we impose restrictive priors on the periods of the five confidently detected planets. These are centred on the maximum likelihood estimate of these periods and have a width in frequency $\pm 2\pi/T_{obs}$ where $T_{obs}$ is the total observation time-span. This hypothesis changes the marginal likelihood and in turn the PNP and the FIP. To correct for this, we adopt a new prior on the number of planets. Denoting by $p(k)$ the prior on the number of planets $k$ and $p'(k)$ the new one, denoting by $p_B$ the broad prior on period chosen  in Table~\ref{tab:priorhd10180}, by $p_N$ the new narrow prior, and by $P_1, ...P_5$ the periods of the planets confidently detected,
\begin{align}
    p’(k) =  p(k) \prod\limits_{i=1}^5 \frac{p_b(P_i) }{p_N(P_i)}
\end{align}

\begin{figure*}
\centering
    \includegraphics[width=0.7\linewidth]{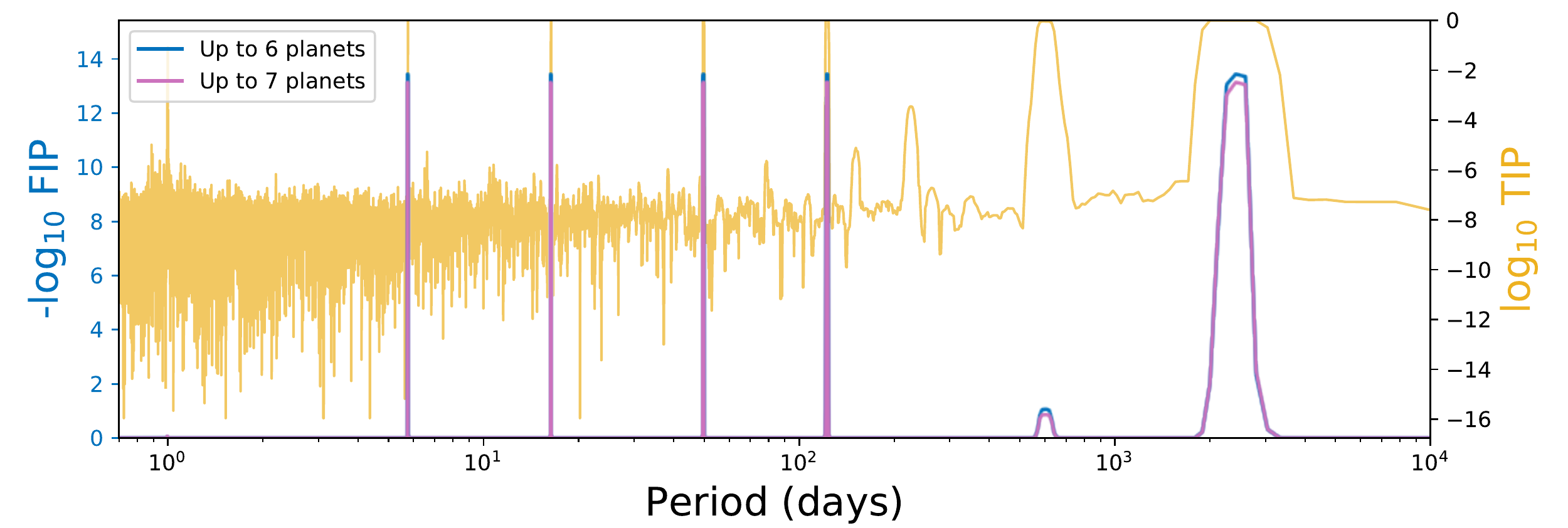}
    \caption{FIP periodogram of 190 HARPS measurements of HD 10180 computed with a $\log$-uniform prior on semi-amplitude on $ [0.1, 20]$ m/s. In blue: FIP periodogram up to six planets, In pink:  FIP periodogram up to seven planets. }
    \label{fig:hd10180log67}

    \includegraphics[width=0.7\linewidth]{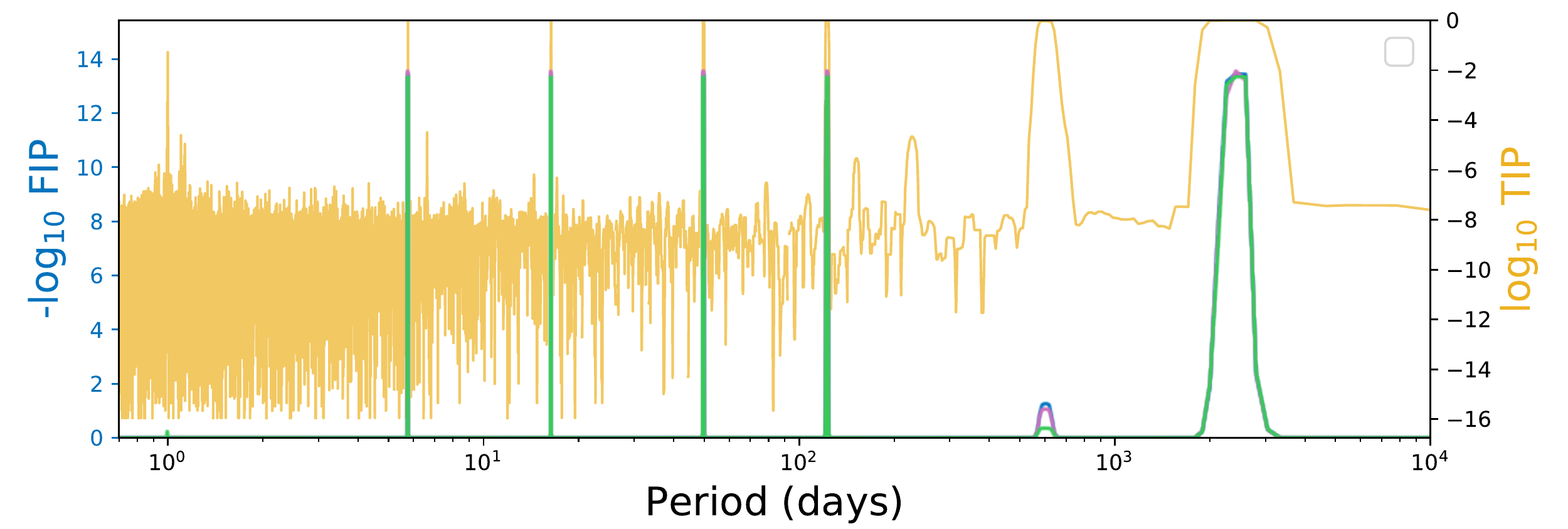}
    \caption{FIP periodogram of 190 HARPS measurements of HD 10180 computed with a $\log$-uniform prior on semi-amplitude on $ [0.1, 20]$ m/s. In blue, pink and green: FIP periodograms corresponding to different runs of calculations of posterior distributions of the parameters of a seven planet model. }
    \label{fig:hd10180logsep}
\end{figure*}

Fig.~\ref{fig:hd10180log67} shows the FIP periodograms obtained with a maximum of six and seven planets (blue and purple, respectively). It appears that the FIP periodogram is essentially unchanged, so that we do not search for an additional planet. In Fig.~\ref{fig:hd10180logsep}, we represent three calculations of the seven planets FIP periodogram obtained with different runs in green, blue and purple. The maximum difference occurs at 600 days, and is below one.
In Table~\ref{tab:run_log}, we summarise the results of our calculation. We provide the $\log$ evidence, its standard deviation across runs, the posterior number of planets and median runtime. It appears that, as in~\cite{faria2018}, the PNP is higher for the seven planets model, however the six planets model is favoured by the FIP. 

\begin{figure*}
    \centering

    \includegraphics[width=0.7\linewidth]{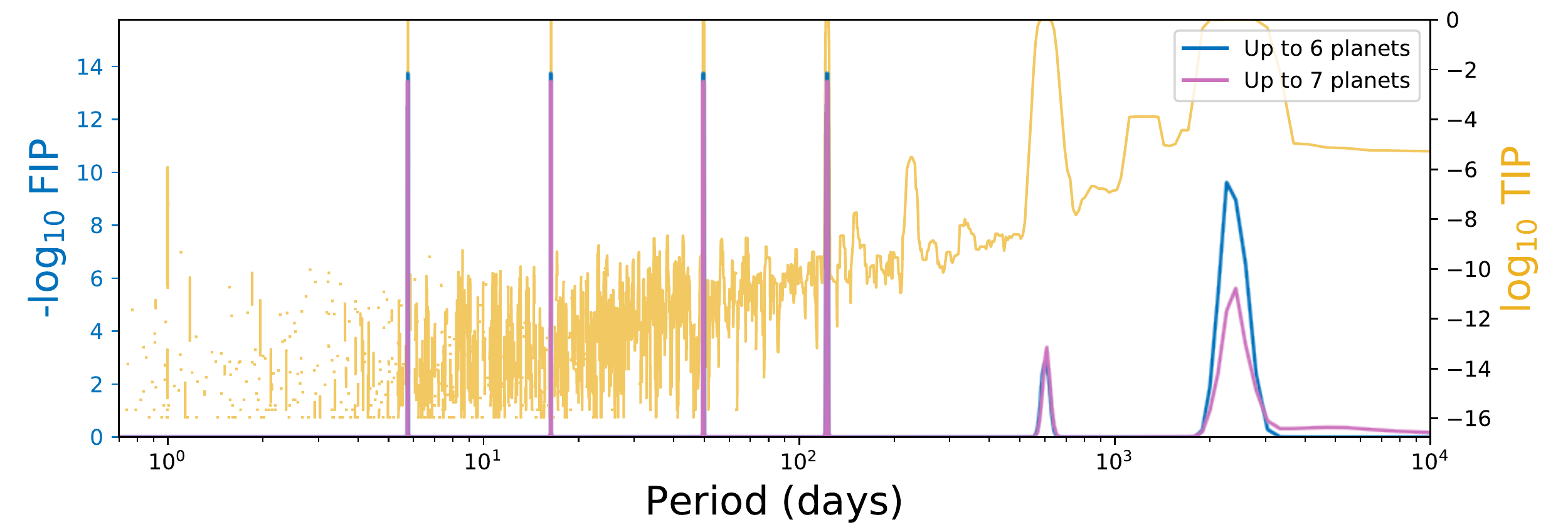}
    \caption{FIP periodogram of 190 HARPS measurements of HD 10180 computed with a Gaussian mixture model linear parameters (two components with $\sigma=1$ and 4 ms). In blue: FIP periodogram up to six planets, In pink:  FIP periodogram up to seven planets. }
    \label{fig:hd10180marg67}
    
    \includegraphics[width=0.7\linewidth]{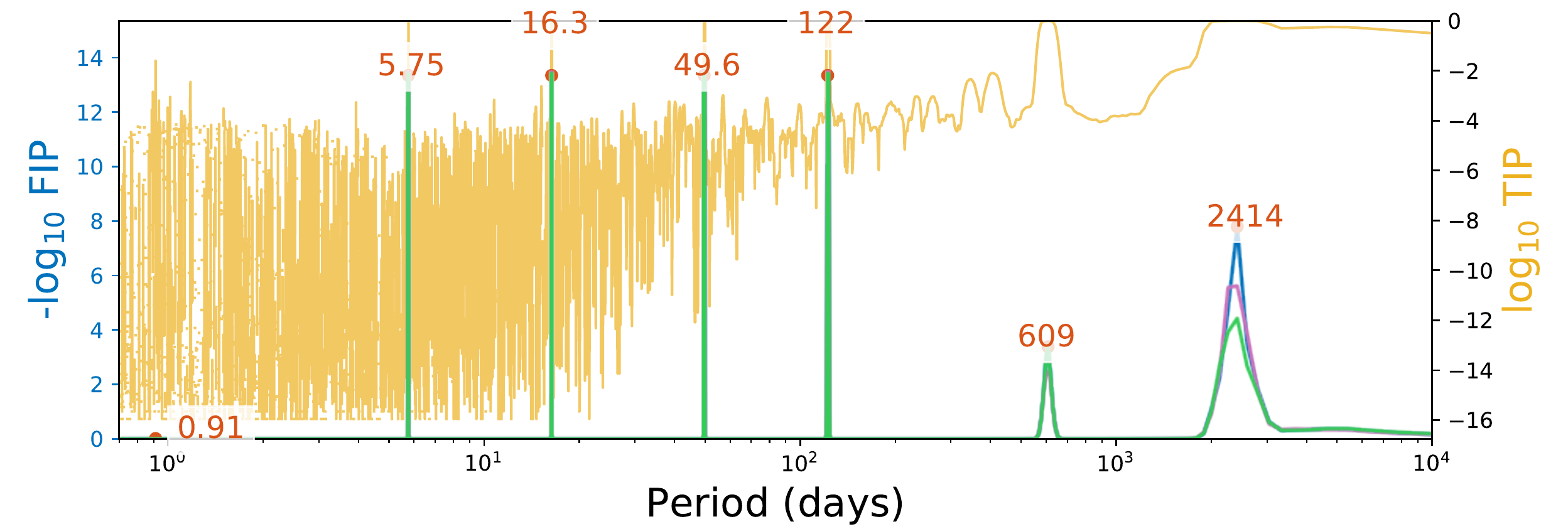}
    \caption{FIP periodogram of 190 HARPS measurements of HD 10180 computed with a Gaussian mixture model linear parameters (two components with $\sigma=1$ and 4 ms). In blue, pink and green: FIP periodograms corresponding to different runs of calculations of posterior distributions of the parameters of a seven planet model. }
    \label{fig:hd10180margsep}
    
\end{figure*}

In Section~\ref{sec:margin}, we stated that when defining the prior on linear parameters as a Gaussian mixture model, calculations can be sped up. We perform the same calculations as above but the priors defined on the linear parameters of Keplerian is a Gaussian mixture with two components of mean 0 and standard deviation 1 and 4 m/s.  Note that if we wanted to define Super-Earth/Mini-Neptunes and Neptnune population more closely, we would need to make the standard deviation of the two components of the Gaussian mixture model depend on the period of the planets as in~\cite{fordgregory2007}, which is not done here for the sake of simplicity. 
We here use a number of live points equal to fifty times the number of free parameters, which are two less by planet because of the analytical marginalisation. For the six planets model, there are 1100 live points. 
Fig.~\ref{fig:hd10180marg67} shows the FIP periodograms obtained with a maximum of six and seven planets (blue and purple, respectively). In Fig.~\ref{fig:hd10180margsep}, we represent three calculations of the seven planets FIP periodogram obtained with different runs in green, blue and purple. In that case, the difference across runs is more important and the significance of the 600 days signal is much higher. This last point illustrates that the prior on semi-amplitude can have a non negligible effect on the significance of signals. In this case the run-to-run difference is more important as well as the 6 vs. 7 planets test. It appears that in the seven planet models, there is a degeneracy between the 2400 days planet and longer periods. However, as shown in Table~\ref{tab:run_marg}, the runtime for six and seven planets is 4h46min and 14h59 min as opposed to 9h42min and 34h14min for the log uniform prior. 

For both prior choices, it appears that even though the 
Bayes factor and PNP slightly favour a seven planet model, the FIP provides the detection of six planets. As discussed further in Section~\ref{sec:prior}, the choice of the semi amplitude prior has an important effect on the significance of small amplitude signals.

\begin{table}[]
\centering
\begin{tabular}{ccc}
\hline
Parameter & Units & Prior                     \\ \hline
$P$       & days  & log-Uniform: [0.7, 10\,000] \\
$K$       & m/s   & log-Uniform: [0.1, 20]    \\
$e$       &       & Beta: [0.867, 3.03]$^{\dagger}$       \\
$\omega$  & rad   & Uniform: [0, 2$\pi$]      \\
$M_0$     & rad   & Uniform: [0, 2$\pi$]      \\ \hline
\end{tabular}
\caption{Priors used for the computation of the FIP periodogram of HD 69830. $^{\dagger}$ \cite{kipping2014}}
\label{tab:priorhd10180}
\end{table}

\begin{table}[]
\caption{Parameters of the different runs of \textsc{polychord} on the HD 10180 dataset, when using a $\log$-uniform prior on semi-amplitude. For a given number of planets in the model, we perform three runs. We give the $\log$ Bayesian evidence ($\log(Z)$, median), the standard deviation of $\log(Z)$ amongst runs ($\sigma_{\log(Z)}$), the PNP and runtime.  }
\label{tab:run_log}
\begin{tabular}{c|c|c|c|c}
Planets & $\log(Z)$  &  $\sigma_{\log(Z)}$  &   PNP &     Runtime  \\ \hline \hline
0   &     -882.52  &  0.32  &  1.71e-108 & 12s \\
1   &     -837.39  &  0.11  &  3.69e-93  & 2 min 10 s \\
2   &     -785.31  &  0.15  &  7.26e-75  & 9 min 36 s  \\
3   &     -731.86  &  0.09  &  9.15e-56  & 28 min 11 s \\
4   &     -670.76  &  0.16  &  3.08e-33  & 1h 06 min 20 s \\
5   &     -595.12  &  0.26  &  4.15e-25 &  2h 22 min 37 s \\
6   &     -587.60  &  0.97  &  0.396   &   9h 42 min 25 s\\
7   &     -587.18  &  2.00  &  0.603   &   34 h 14 min  9 s
\end{tabular}

\end{table}

\begin{table}[]
\caption{Parameters of the different runs of \textsc{polychord} on the HD 10180 dataset, when using a Gaussian mixture prior on semi-amplitude. For a given number of planets in the model, we perform three runs. We give the $\log$ Bayesian evidence ($\log(Z)$, median), the standard deviation of $\log(Z)$ amongst runs ($\sigma_{\log(Z)}$), the PNP and runtime.  }
\label{tab:run_marg}
\begin{tabular}{c|c|c|c|c}
Planets & $\log(Z)$  & $\sigma_{\log(Z)}$  &   PNP &     Runtime  \\ \hline \hline     
0     &   -882.45 & 0.23 &  3.82e-108 & 17s   \\
1     &   -839.36 & 0.19 &  1.08e-93 & 1 min 56 s \\
2     &   -789.03 & 0.24 &  3.72e-76 & 6 min 57 s   \\
3     &   -736.95 & 0.04 &  1.19e-57 &  17 min 39 s \\
4     &   -677.56 & 0.15 &  7.27e-36 & 38 min 41 s \\
5     &   -603.42 & 0.13 &  1.12e-07 & 1 h 33 min 31 s\\
6     &   -590.14 & 0.30 &  6.60e-02 &  4 h 46 min 46 s\\
7     &   -587.49 & 0.22 &  9.34e-01 &  14 h 59 min 57 s \\
\end{tabular}
\end{table}

	 \section{Comparison of the metrics}
	 \label{sec:discussion}
	 
	 \subsection{Outline}
	 
	 In this section, we discuss the properties of the FIP and other detection criteria. First, in section~\ref{sec:comparison}, we compare their performances when the model is known. We then study whether the detection criteria are sensitive to prior and likelihood choices, respectively in sections~\ref{sec:prior} and~\ref{sec:likelihood}.

	 \subsection{Performance comparison of the different metrics when true model is known}
	 \label{sec:comparison}
	 
	 \label{sec:comparisonsimu}
	 
	 \subsubsection{Simulation}

	 To compare the different methods, we consider a set of a thousand simulated data sets, with zero to two injected circular planets. The planet signals are generated with a $\log$-uniform distribution in period on 1.5 to 100 days, uniform phase and a Rayleigh distribution in amplitude with $\sigma = 1.5$ m/s. This allows us to use the analytical marginalisation on linear parameters described in Section~\ref{sec:margin}, which speeds up computations.   The distributions of elements are summarised in Table~\ref{tab:priorscirc}. The time stamps are taken from the first 80 HARPS measurements of HD 69830~\citep{lovis2006} \ch{(this dataset is presented in Appendix \ref{app:datasets})} and the noise is generated according to the nominal error bars, which are typically of 0.54 $\pm$ 0.24 m/s. 
	 
	 We then generate another set of a thousand systems with a lower signal-to-noise ratio. The simulation is made with identical parameters except that a correlated Gaussian noise is added. This one has an exponential kernel with a 1 m/s amplitude and a time scale $\tau = 4$ days. %Finally, the analysis is made with Keplerian orbits.
	 These simulations are intentionally simple, to enable the identification of the hypotheses driving the results. 
	 
	 \subsubsection{Analysis}
	 \label{sec:analysis}
	 
	 We analyse the data with different methods. In all cases, correct likelihood and priors are assumed. In particular, we search only up to two planets, according to the input data. 
	 Except the PNP, we evaluate the methods with a grid of detection thresholds. For a given detection threshold, the methods proceed as follows. 
	 \begin{itemize}
	 	\item Periodogram + FAP:  We compute a  general linear periodogram as defined in~\cite{delisle2019a} with the same grid of frequencies as the one used to generate the data (from 1.5 to 100 d) and the correct covariance matrix. If the FAP (as defined section~\ref{sec:fap}) is below a certain threshold fixed in advance, we add a cosine and sine function at the period of the maximum peak to the linear base model and recompute the periodogram. The planet is added if the FAP is below the FAP threshold. We do not search for a third planet.  
	 	\item Periodogram + Bayes factor: same as above, but here the criterion to add a planet is that the Bayes factor (as defined section~\ref{sec:bf}) is above a certain threshold.   \ch{The evidences (Eq. \eqref{eq:evidence}) are computed with the  distributions used in the simulations, in particular the period is left free between 1.5 and 100 days. } 
	 	\item  $\ell_1$-periodogram\footnote{The $\ell_1$ periodogram code is available at \url{https://github.com/nathanchara/l1periodogram}} + FAP : we compute the $\ell_1$ periodogram~\citep{hara2017} with the same grid of frequencies as the one used to generate the data (from 1.5 to 100 d). If the FAP of the maximum peak is below a certain threshold, it is added to the base model of unpenalized vectors, the $\ell_1$ periodogram is recomputed, the FAP of the maximum peak is assessed. If it is below a certain threshold, the a planet detection is claimed. We do not look for a third planet. 
	 	\item $\ell_1$-periodogram + Bayes factor: same as above, but here the criterion to add a planet is that the Bayes factor (as defined section~\ref{sec:bf}) is above a certain threshold. 
	 	\item FIP: We compute the FIP periodogram as defined in section~\ref{sec:fip} and select the two highest peaks. We select a period if its corresponding FIP is below a certain threshold. 
	 	\item PNP + FIP \ch{periodogram}: here, to select the number of planets we order the peaks of the FIP periodogram with increasing FIP. We select the number of peaks corresponding to the highest PNP, as defined in~\ref{sec:pnp}. 
	 	\item \ch{ FIP  \ch{periodogram} + Bayes factor : the periods are selected as the highest peaks of the FIP periodogram and the number of planets is selected with the Bayes factor. This procedure is very close to \cite{gregory2007, gregory2007a} except that we use the FIP periodogram instead of the marginal distribution of periods for each planets. We do not take the approach of \cite{gregory2007, gregory2007a} to select the periods as nested sampling algorithms such as \textsc{polychord} tend to swap the periods of planets, such that marginal distributions are typically multi-modal. } 
	 	\item \ch{ FIP  \ch{periodogram} + FAP : the periods are selected as the highest peaks of the FIP periodogram and the number of planets is selected with the false alarm probability. }
	 \end{itemize}
	For the computation of Bayes factor, FIP and PNP, the number of live points in the nested sampling algorithm is equal to 200 times the number of planets in the model.

	 \subsubsection{Performance evaluation}
	 \label{sec:performances}
	 
	 To evaluate the performance of the different analysis methods, we use two criteria: the ability of the methods to retrieve the correct number of planets, and their ability to retrieve the planets with the correct periods. 
	 To assess the correct retrieval of the number of planets, we simply count how many planets are detected. We measure the difference between the number of planets claimed and the true one. If this difference is strictly positive or negative, we count the respectively as a false positive and a false negative. For instance if there are two detections while no planet is present, we count two false positives. The total number of mistakes is given by the sum of false negatives and false positives on the 1000 systems analysed. 
	 
	 To verify that periods are appropriately retrieved, we check whether a frequency found is less than $1 /T_\mathrm{obs}$  away from the true frequency, $T_\mathrm{obs}$ being the observation time span. For a given detection threshold and a given simulated system, we consider the planets detected with decreasing significance. If a planet is claimed, 
	  but does not correspond to a true planet with the desired precision on period or no planet is present, it is labelled as a false detection. If the claimed planet corresponds to a true planet, we label it as a correct inference and remove its period from the set of true periods, so that a planet cannot be detected twice. The situation where no planet is claimed but there is a planet in the data, is labelled as a missed detection. The total number of mistakes is here the sum of missed and false detections on the 1000 systems.
	 
	 	\ch{The rationale of evaluating the different methods presented in Section \ref{sec:analysis} is to determine whether the performance of a given method comes from the period selection or the scale of the significant metrics. Periods are typically refined by a MCMC, but this changes the estimate of the frequency by a small fraction of 1/$T_\mathrm{obs}$. }

	 \subsubsection{Simulation  1: white noise}
	 \label{sec:simulation1}
	 
	 In the first simulation, we have only white noise with a typical ratio of semi amplitude and noise standard deviation of 3.4. The number of mistakes for the different detection metrics are shown in Fig.~\ref{fig:wnoise}. The plot on the left (blue/red, a1 to e1) and right (purple/yellow, a2 to e2) represent respectively the performances in terms of retrieval of the number of planets, and periods of planets. Each row corresponds to a different detection metric: periodogram + FAP (a), periodogram + Bayes factor (b), $\ell_1$ periodogram + FAP (c), $\ell_1$ periodogram + Bayes factor (d), FIP (e). 
	 The total number of false positives, false negatives, false detections and missed detections on the 1000 systems are represented in red, blue, yellow and purple shaded area respectively. 
      The solid black lines indicate the minimum number of mistakes as well as the minimum and maximum thresholds at which this minimum is attained. The gray plain and dashed lines represent the number of mistakes obtained by taking the maximum PNP and maximum PNP + FIP, respectively. In all cases, the $x$ axis is oriented such that from left to right the detection criterion is more and more stringent (confidence increases). 
	 
	 As one might expect, for each performance metric we find a U-shaped curve. When the detection criterion is permissive, the total error is dominated by false positives or false detections. Conversely, as the  detection criterion becomes more stringent, detections due to random fluctuation are progressively ruled out and the detection errors are dominated by false negatives or missed detections. We note that the thresholds for which a low false positive rate is expected would typically be chosen, and it is in this range of thresholds that the methods should be compared.
	 
	 We find that in terms of number of planets (left column in Fig.~\ref{fig:wnoise}), all the significance metric exhibit similar behaviours, with a minimum  number of errors of 34 to 46. For comparison, a uniform random guess of the number of planets (0, 1 or 2) would yield a total of 888 errors on average. We find that the maximum PNP leads to the smallest error, as well as a $\log$ Bayes factor close to 0 (see the plain gray line in third plot from the top). This is to be expected, since the maximum PNP has optimality properties and the Bayes factor is designed to compare the number of planets by averaging over all the possible values of the parameters for a given number of planets. In all cases, we see a sharp decrease of the number of false positives as the detection threshold becomes more stringent. 
	 
	 Larger discrepancies in performance happen when the methods are evaluated on their ability to retrieve not only the correct number of planets, but also the correct periods. In that case, the periodogram + FAP and BF (two upper plots) exhibit similar performances. The $\ell_1$-periodogram and FIP (three lower plots) exhibit better performances to find the periods of the planets in two ways: the minimum number of mistakes is smaller and the ratio of false positive to false negative is smaller at the optimum value.

\begin{figure}
	 	\includegraphics[width=0.98\linewidth]{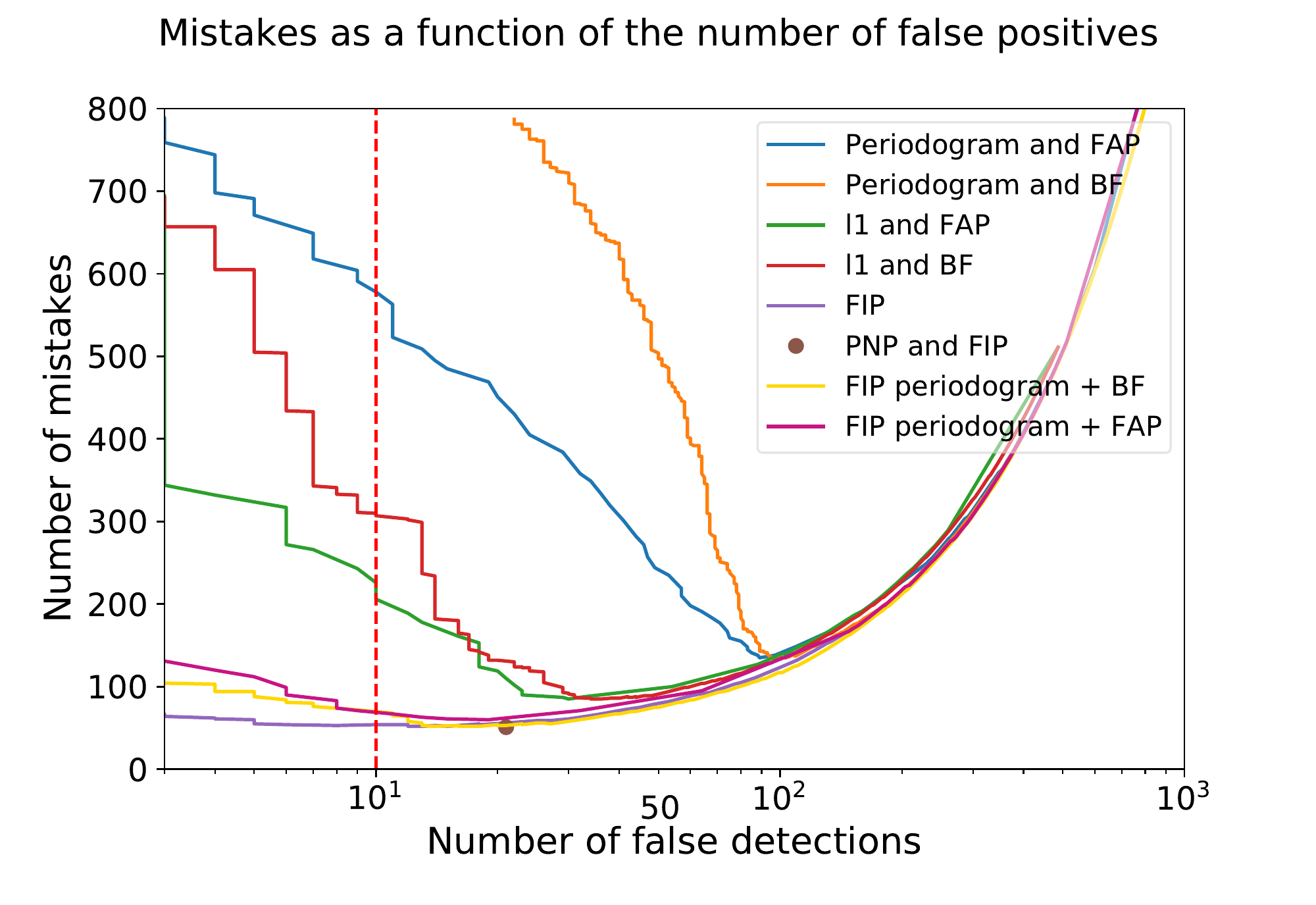}
	 	\caption{Number of mistakes as a function of the number of false detections ($\log$ scale) for the different detection methods. The number of mistakes are defined as the sum of missed true planets and false detections.  The data corresponds to  simulation 1, described in section~\ref{sec:simulation1}: a random number of planets equal to 0, 1 or 2, the noise is white, Gaussian with semi amplitude to noise ratio of 3.5.  }
	 	\label{fig:wmistakes}
	 \end{figure}

	 The scales of thresholds (BF, FAP, FIP) are not in the same units. To further compare the methods, we compute the total number of mistakes in terms of correct retrieval of period and number of planets as a function of the number of false detections. This is plotted in Fig.~\ref{fig:wmistakes}, where we see that in the regime of low number of false positives (stringent detection criterion), there are important differences between the methods. For instance, for 10 false positives, indicated by a red dashed line, the methods exhibit very different performances. We checked that these results are not too dependent on the success criterion, by labelling a detection a false one if the frequency found departs from more than 2/$T_{obs}$ and 3/$T_{obs}$  from a true period, instead of 1/$T_{obs}$. The results are qualitatively identical.  In conclusion, the FIP provides a low number of missed detection even when the number of false detection is small.
	 
	 The performances of the $\ell_1$ periodogram and the FIP comes from the fact that both methods encode the search for several planets simultaneously. On the other hand, since the periodogram searches for one planet at a time, there are cases where the maximum of the periodogram does not occur on any of the true periods~\citep[see][]{hara2017}. Indeed, a detailed analysis shows that at stringent detection criteria for the Periodogram + FAP and Periodogram + BF, almost all of the false detections made by the  methods occur in cases where there are two planets present in the data, and the period selection method selects a spurious peak. \ch{The FIP, FIP periodogram + BF and FIP periodogram + FAP methods, where periods are selected from the FIP periodogram, exhibit much better performances. The performances are similar, a difference is seen only in the region with a low number of false positives (< 12) (lower left, Fig.~\ref{fig:wmistakes}) where the FIP leads to a slightly lower number of false negatives. This suggests that not only the period selection is more efficient with a FIP periodogram, but the significance scale on which the FIP is defined performs well to distinguish true planets from false detections.   }

	 \subsubsection{Simulation  2: lower SNR}
	 \label{sec:simulation2}
	 
\begin{figure}
 	 	\includegraphics[width=1.\linewidth]{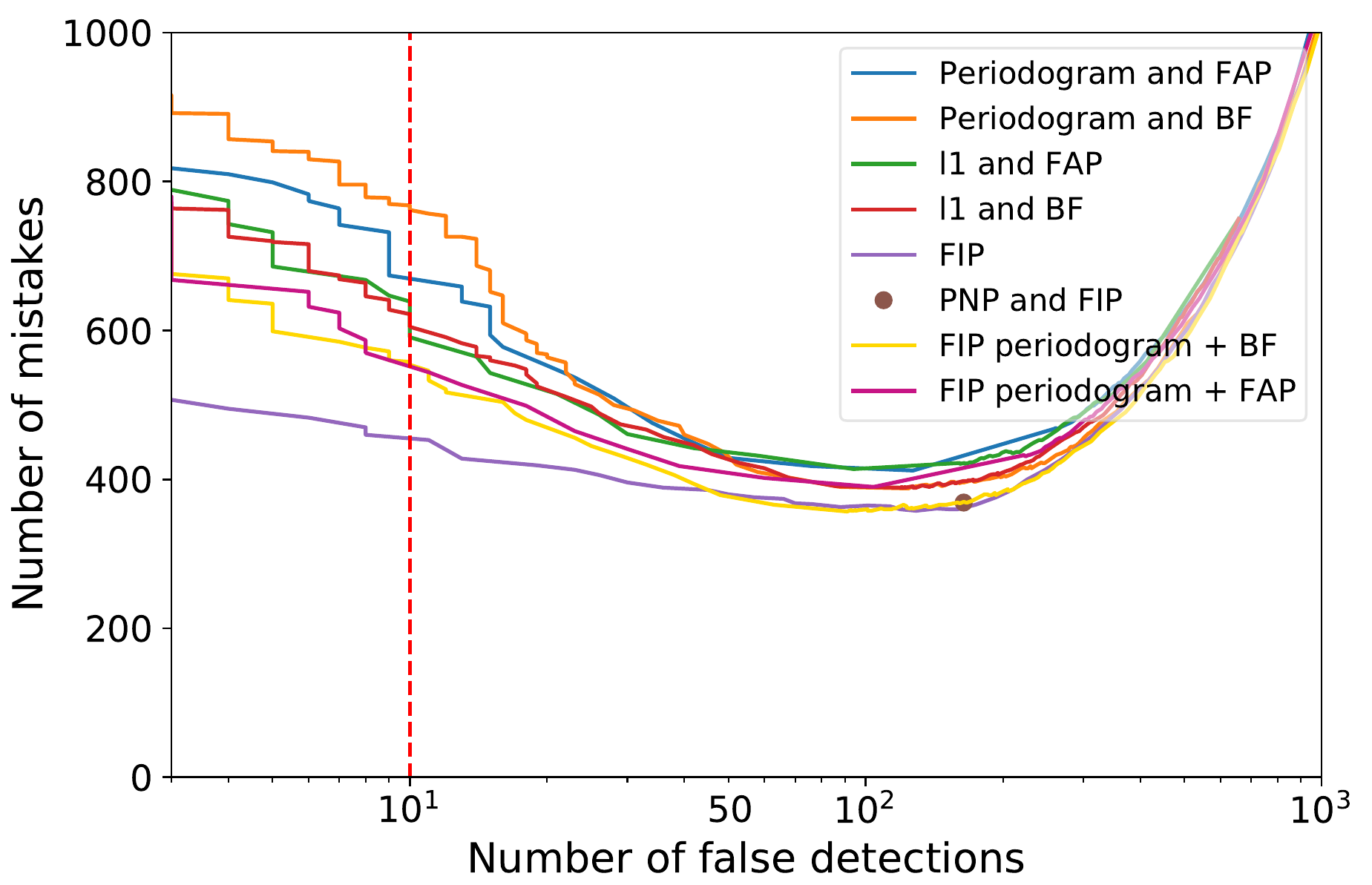}
	 	\caption{Number of mistakes as a function of the number of false detections ($\log$ scale) for the different detection methods.  The data corresponds to simulation 2, described in section~\ref{sec:simulation2}: a random number of planets equal to 0,1 or 2 planets per system, white and correlated noise. This one has a time-scale of 4 days and an amplitude of 1 m/s. The semi amplitude to noise level ratio of 1.7. Mistakes are defined as the sum of missed and false detections. }
        \label{fig:rmistakes}
	 \end{figure}

	 In this simulation, the exact same parameters as simulations 1 are used to generate the data, except that we add a 1 m/s correlated noise with a 4 days exponential decay and a time-scale. In that case, the typical semi amplitude to noise ratio is 1.65, as opposed to 3.48 in the previous simulation. The results are represented in Fig.~\ref{fig:rnoise} with the same conventions as Fig.~\ref{fig:wnoise}. We add the red dashed lines to indicate the detection threshold at which there are only 10 false positive claims. In Fig.~\ref{fig:rmistakes}, we represent the total number of mistakes as a function of the number of false detections (which therefore includes whether the period of planets is appropriately recovered). 
	 
	 Both in terms of optimal thresholds and mistakes at low false positive rates, the differences in performance between different analysis methods are less important. 
	 We attribute this to the fact that the noise level is higher, such that all signals are on average less significant, including spurious peaks. We observe that the FIP still outperforms the other methods. \ch{ Here too, we see a difference in performance in the region with a low number of false detections (< 12), where the FIP leads to a lower number of missed detections. }

	 \color{black}

	 \iffalse
	 \begin{figure}
	 	\centering
	 	\includegraphics[width=0.7\linewidth]{figures/FAPmistakes}
	 	\caption{}
	 	\label{fig:fapmistakes}
	 \end{figure}
	 \begin{figure}
	 	\centering
	 	\includegraphics[width=0.7\linewidth]{figures/BFmistakes}
	 	\caption{}
	 	\label{fig:bfmistakes}
	 \end{figure}
	 \fi

\begin{figure*}
%	\noindent
	\centering
  \subfloat[][]{
	\hspace{-1cm}
	\begin{tikzpicture}
	\path (0,0) node[above right]{\includegraphics[width=0.36\linewidth]{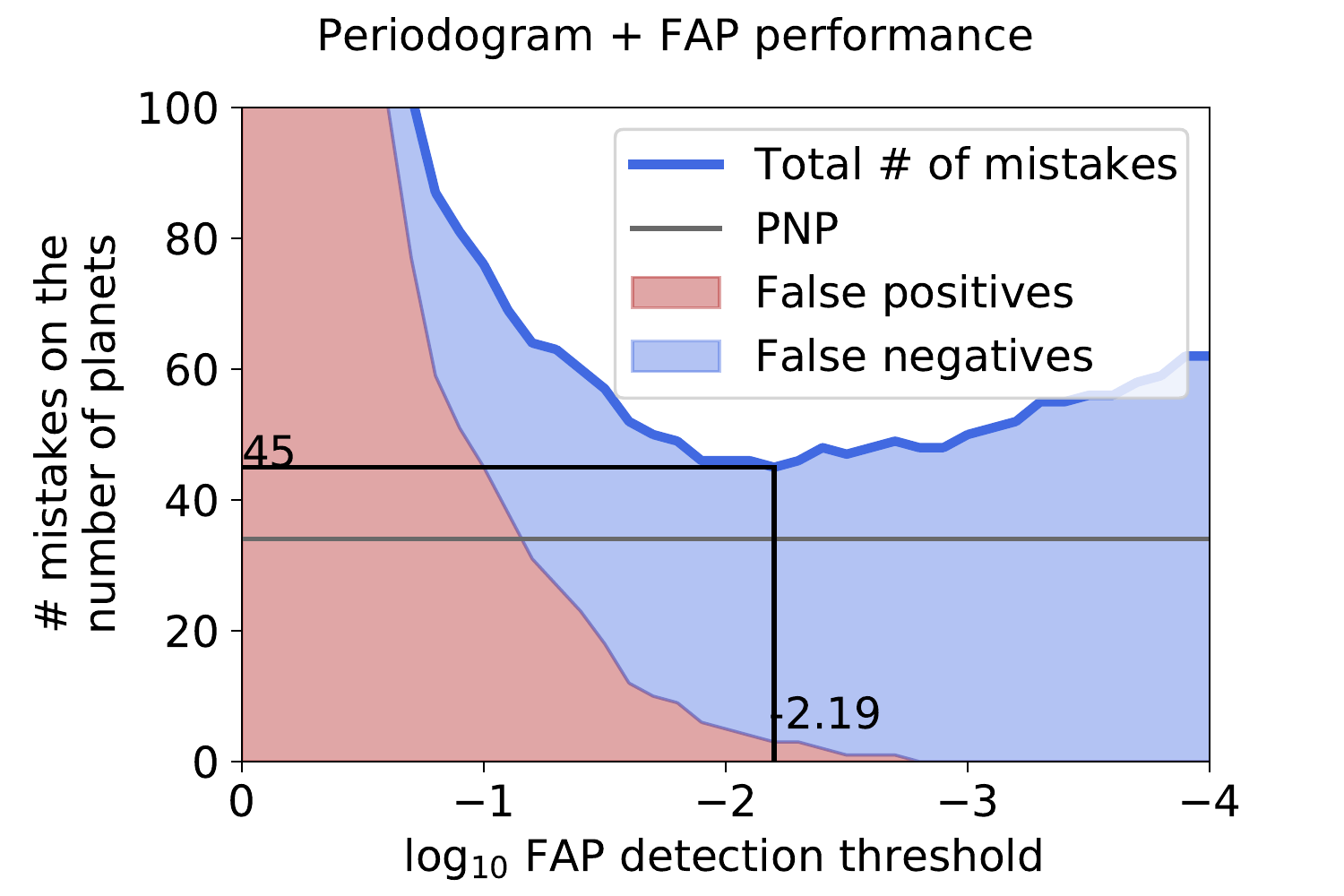}};
	\path (0.15,4) node[above right]{\large(a1)};
	\path (8,0) node[above right]{\includegraphics[width=0.36\linewidth]{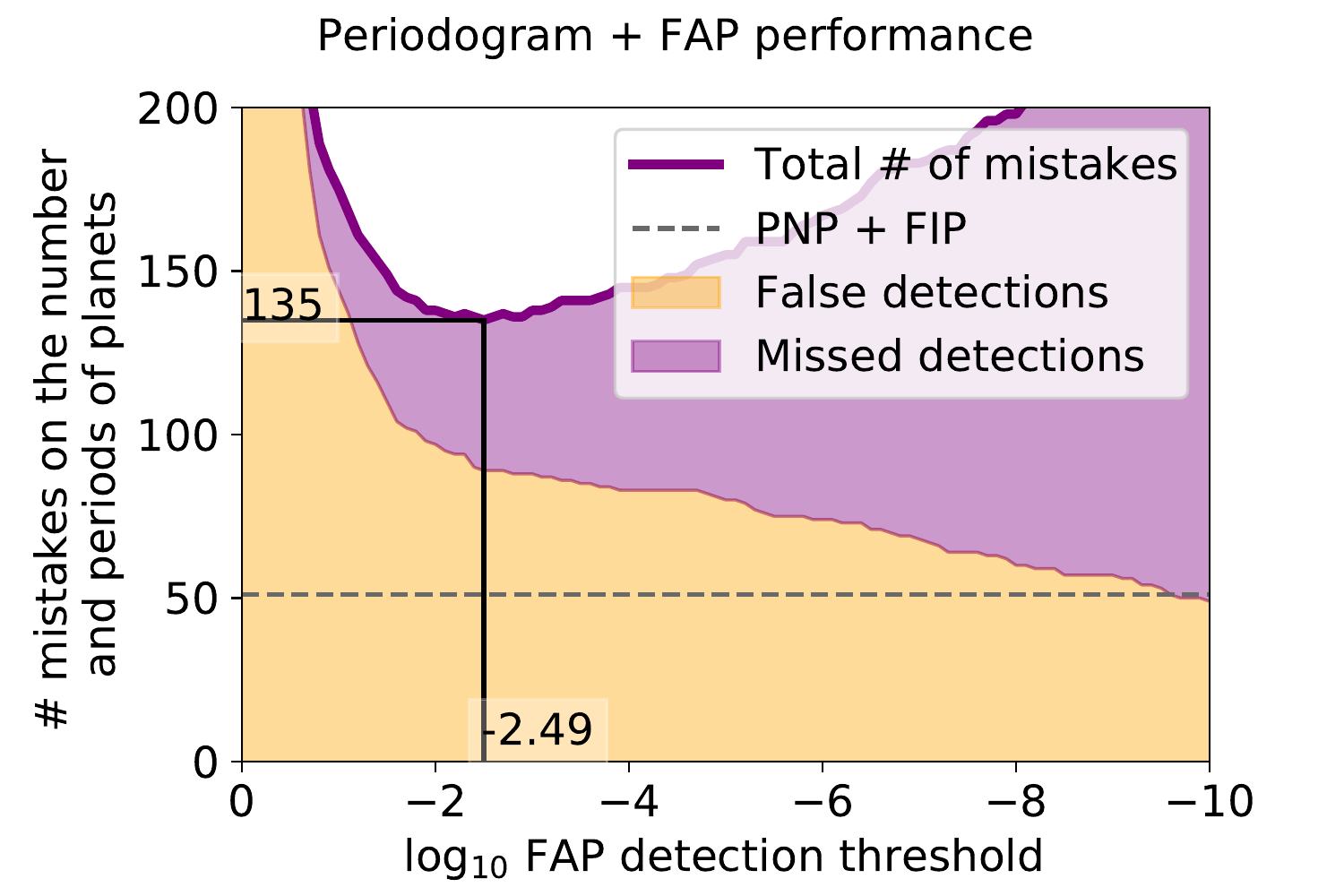}};
	\path (8.15,4) node[above right]{\large(a2)};
	\begin{scope}[yshift=-4.5cm]
	\path (0,0) node[above right]{\includegraphics[width=0.36\linewidth]{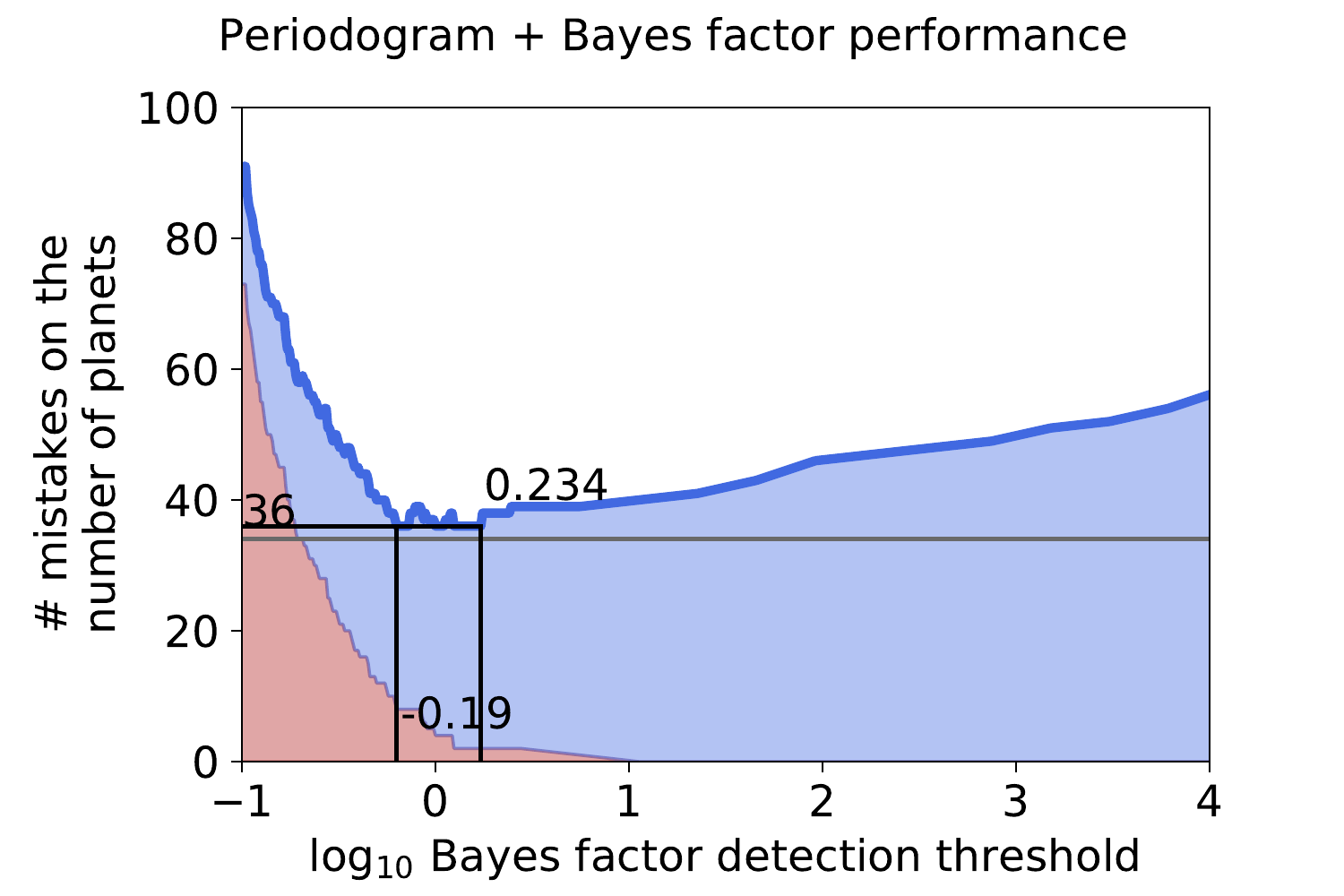}};
	\path (0.15,4) node[above right]{\large(b1)};
	\path (8,0) node[above right]{\includegraphics[width=0.36\linewidth]{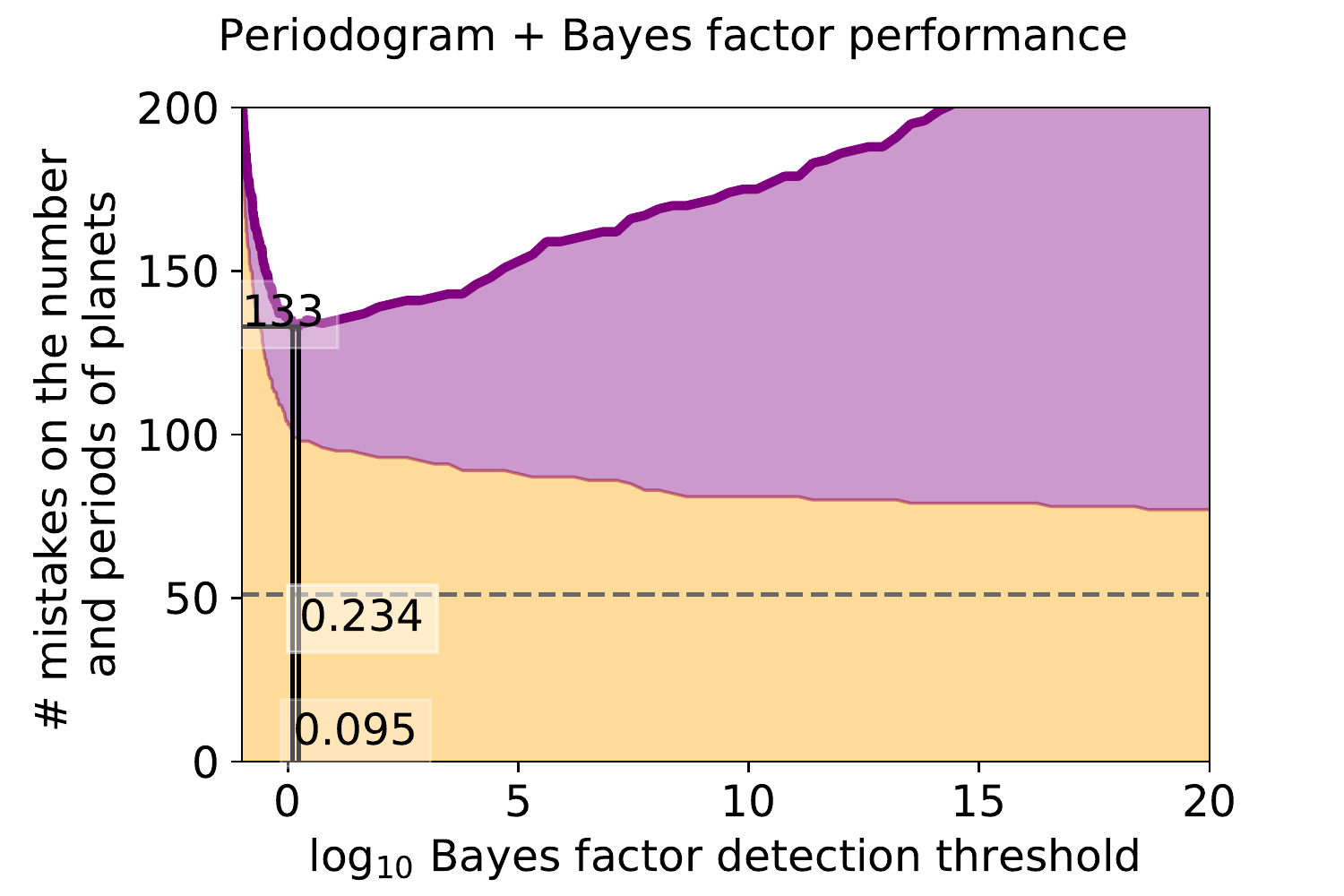}};
	\path (8.15,4) node[above right]{\large(b2)};
	\end{scope}
	\begin{scope}[yshift=-9cm]
	\path (0,0) node[above right]{\includegraphics[width=0.36\linewidth]{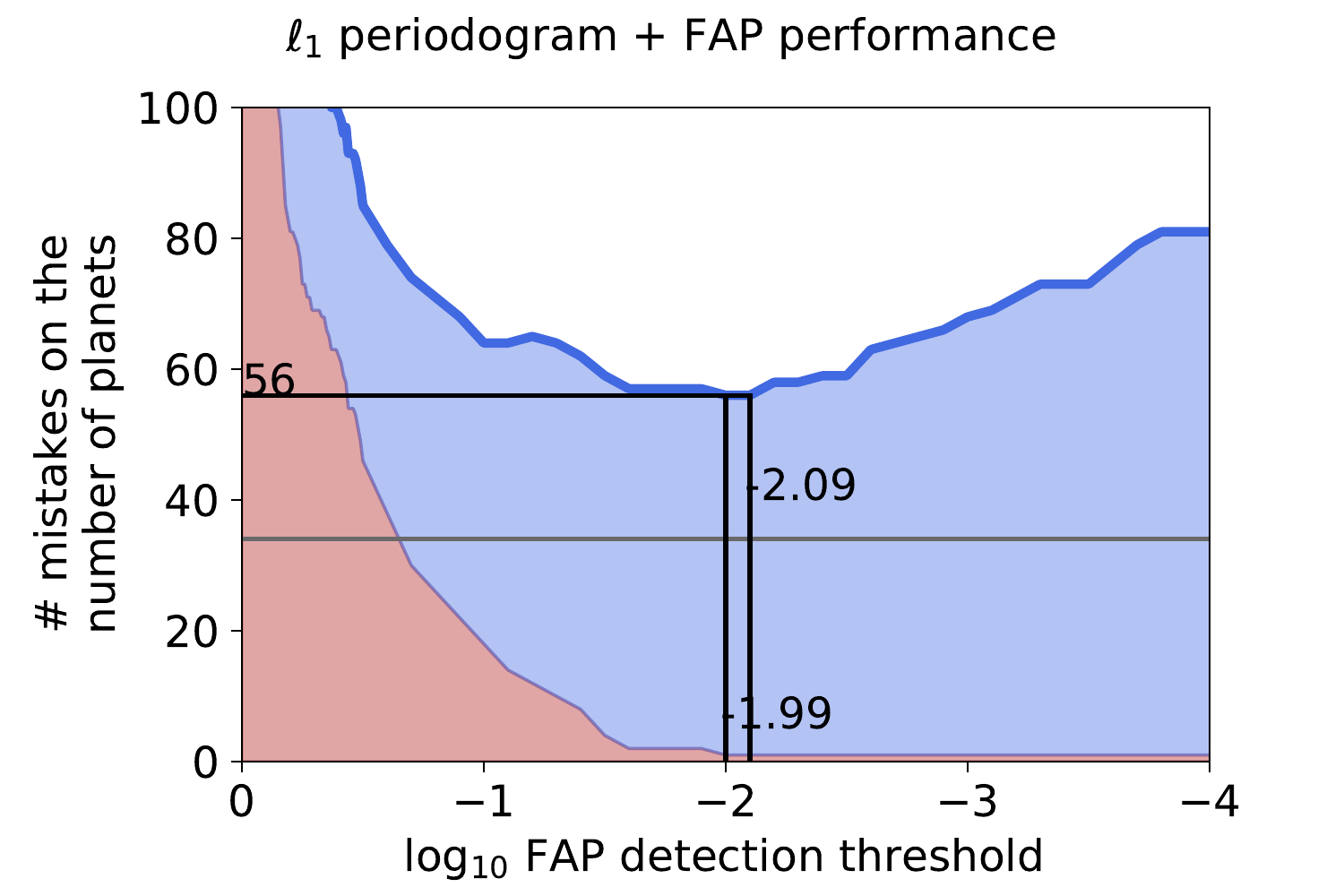}};
	\path (0.15,4) node[above right]{\large(c1)};
	\path (8,0) node[above right]{\includegraphics[width=0.36\linewidth]{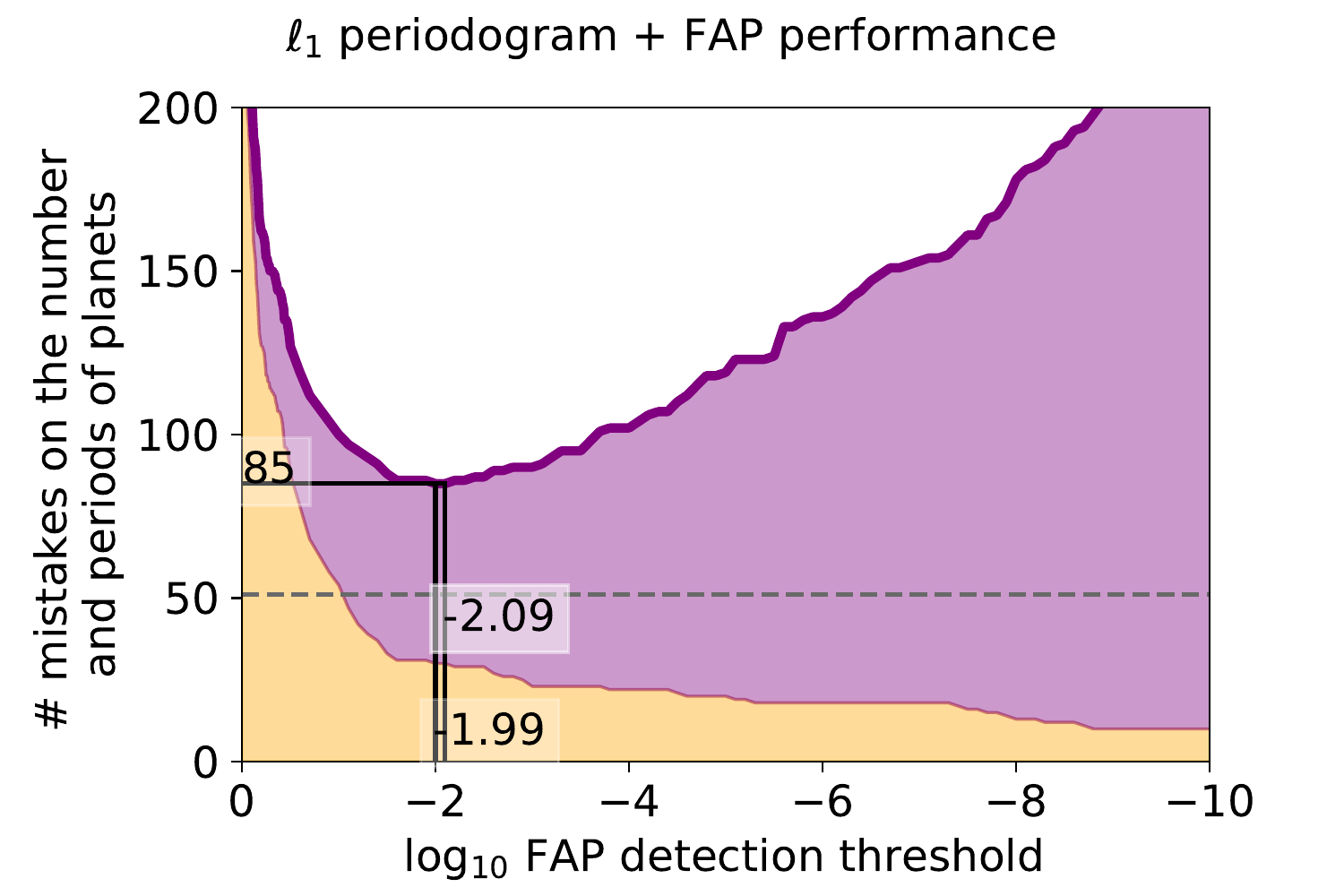}};
	\path (8.15,4) node[above right]{\large(c2)};
	\end{scope}
	\begin{scope}[yshift=-13.5cm]
	\path (0,0) node[above right]{\includegraphics[width=0.36\linewidth]{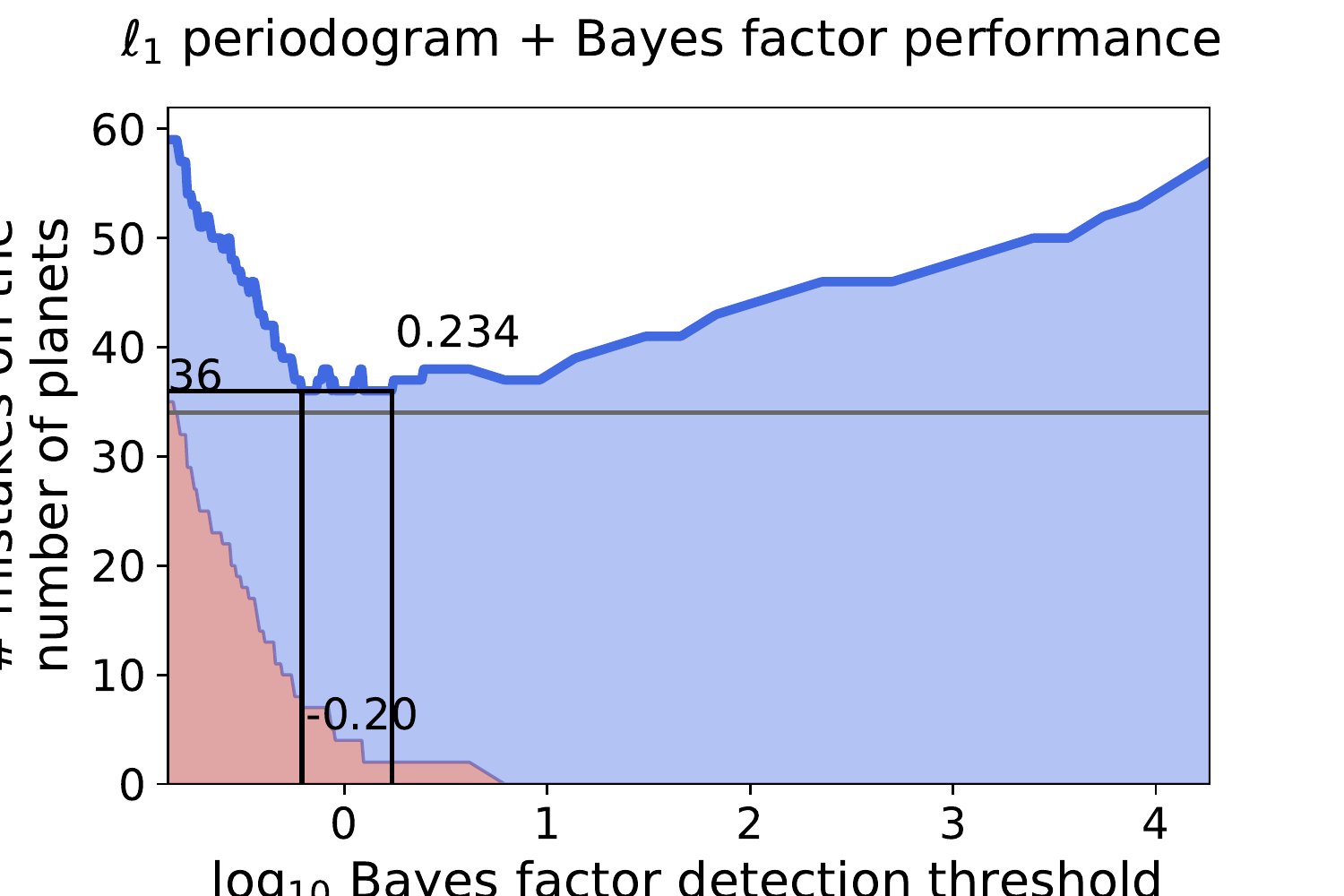}};
	\path (0.15,4) node[above right]{\large(d1)};
	\path (8,0) node[above right]{\includegraphics[width=0.36\linewidth]{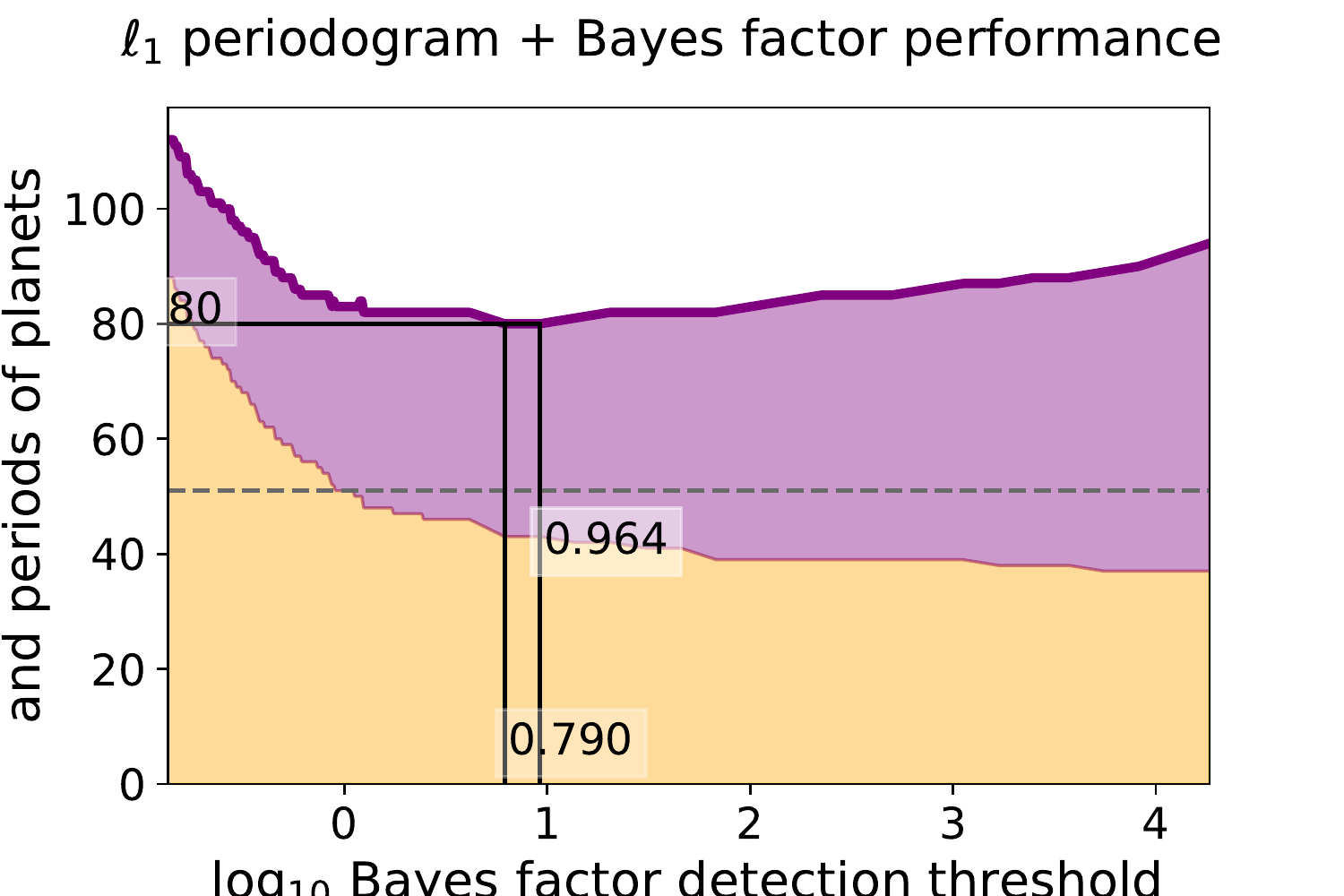}};
	\path (8.15,4) node[above right]{\large(d2)};
	\end{scope}
	\begin{scope}[yshift=-18cm]
	\path (0,0) node[above right]{\includegraphics[width=0.36\linewidth]{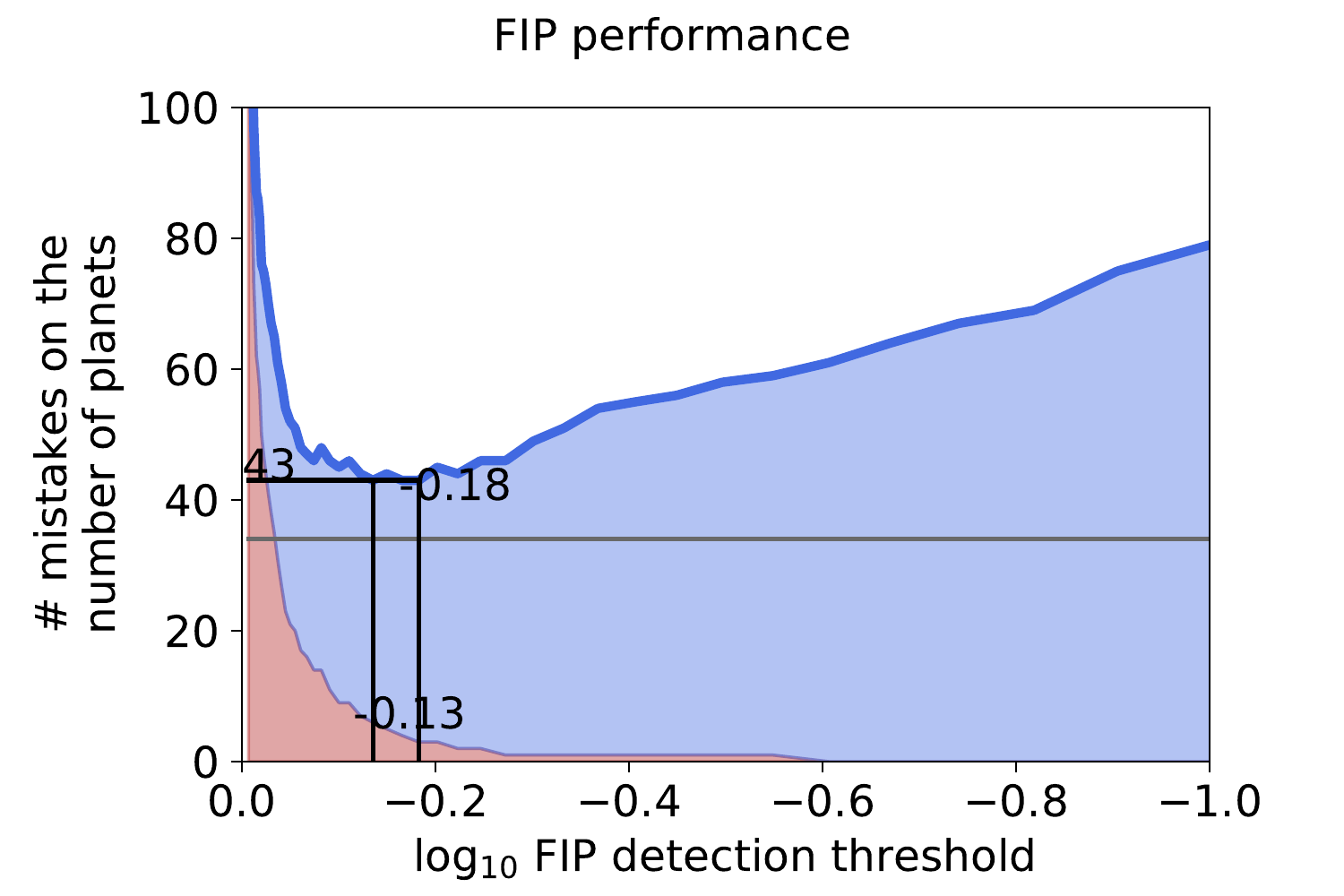}};
	\path (0.15,4) node[above right]{\large(e1)};
	\path (8,0) node[above right]{\includegraphics[width=0.36\linewidth]{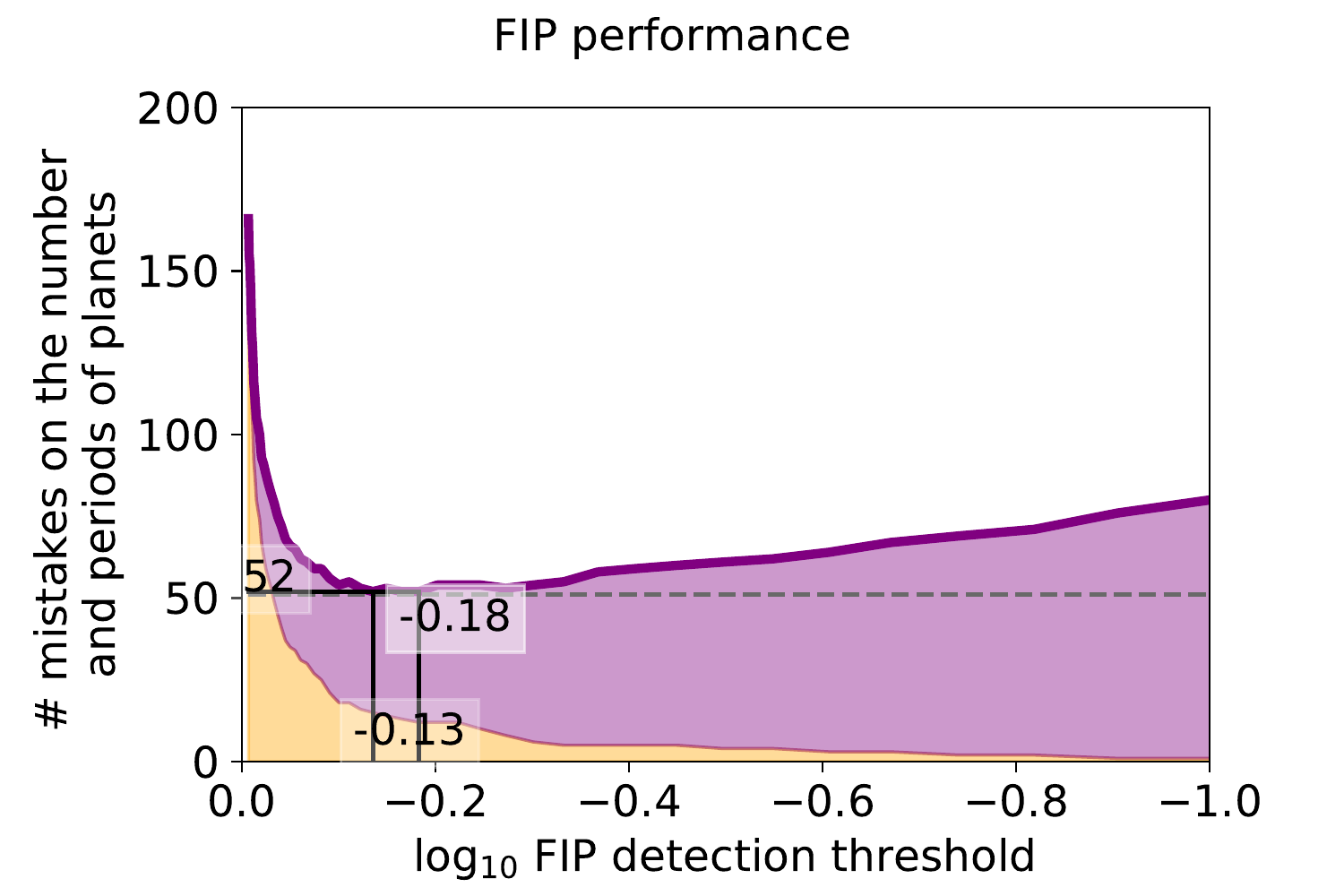}};
	\path (8.15,4) node[above right]{\large(e2)};
	\end{scope}
	\end{tikzpicture}
	}
	\vspace{-0.7cm}
	\caption{ Total number of mistakes on the number of planets (left) and number and period of planets (right) out of a thousand datasets \ch{with white noise (Simulation 1, described in Section~\ref{sec:simulation1})} as a function of the detection threshold for different statistical significance metrics. From top to bottom: Periodogram + FAP, Periodogram + Bayes factor, $\ell_1$ periodogram + FAP and FIP.  The total number of false positives and false negatives are represented in light red and blue \ch{shaded areas}. The false and missed detections are represented in orange and purple \ch{shaded areas}, respectively. The minimum and maximum values of the detection threshold corresponding to the minimum number of mistakes are marked with black lines. \ch{See the continuation of the figure below.} }
	\label{fig:wnoise} 
\end{figure*}

\begin{figure*}
	 \ContinuedFloat 
	 	\centering
	  \subfloat[][]{  	
  \hspace{-1cm}
	\begin{tikzpicture}
	\path (0,0) node[above right]{\includegraphics[width=0.36\linewidth]{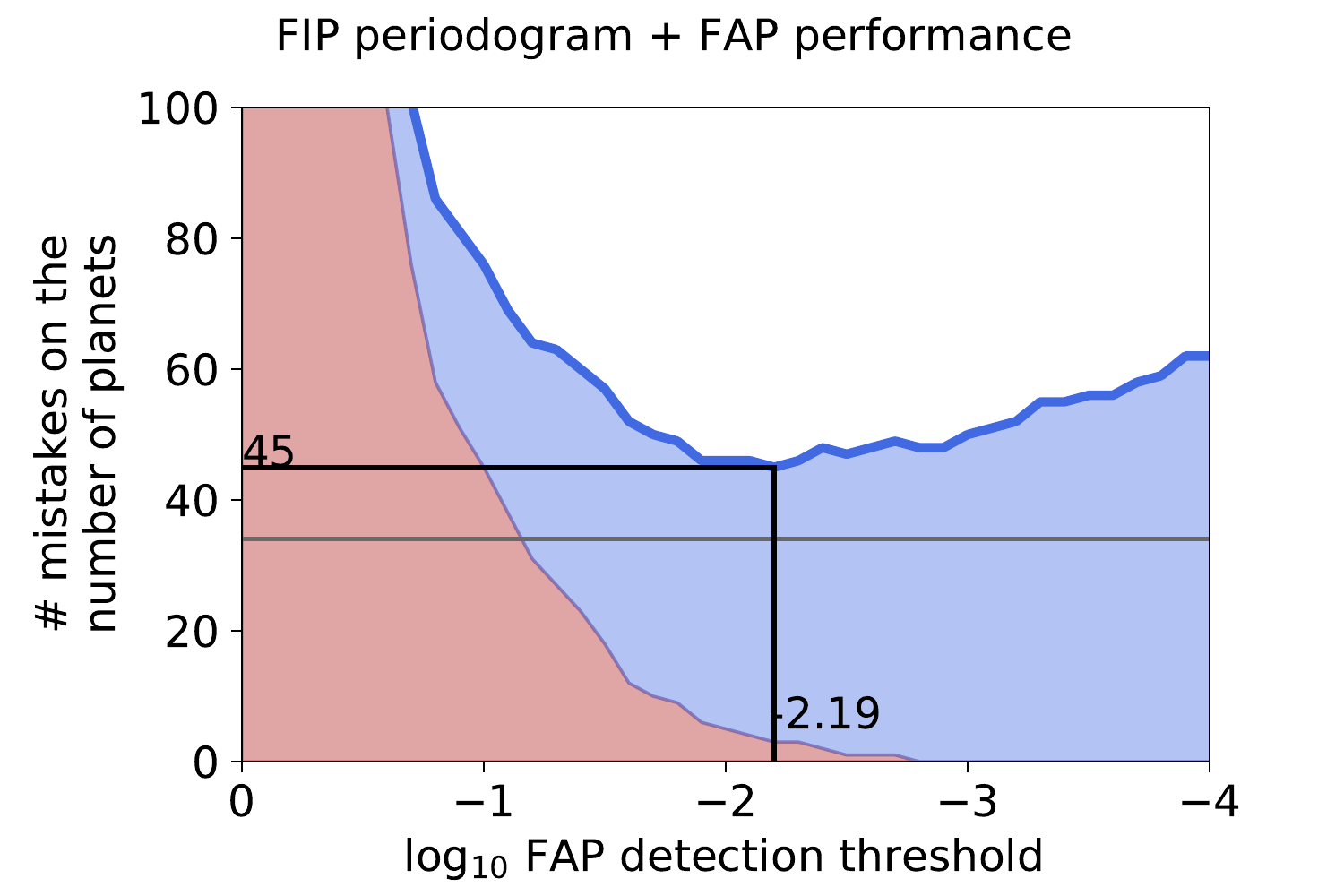}};
	\path (0.15,4) node[above right]{\large(f1)};
	\path (8,0) node[above right]{\includegraphics[width=0.36\linewidth]{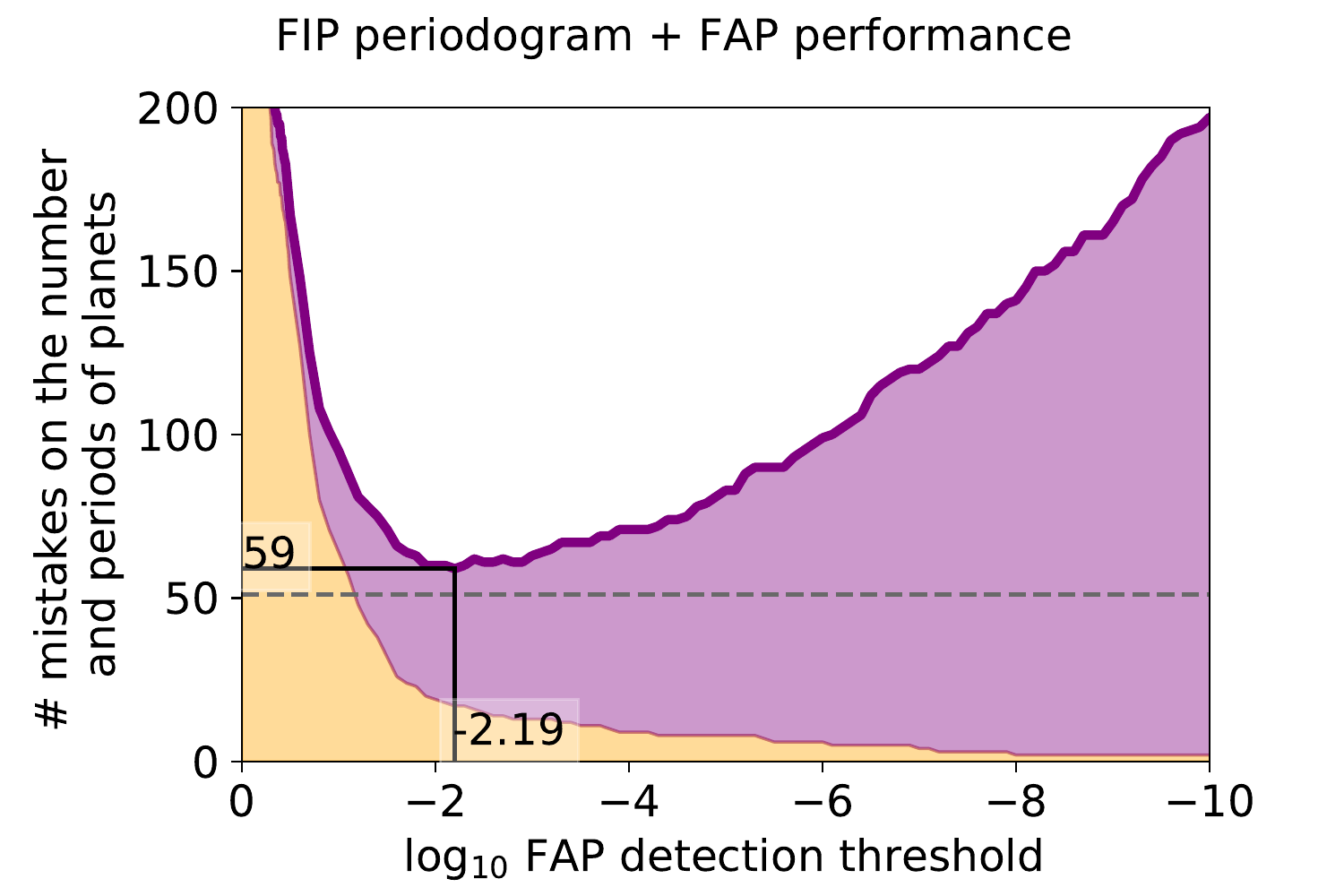}};
	\path (8.15,4) node[above right]{\large(f2)};
	\begin{scope}[yshift=-4.5cm]
	\path (0,0) node[above right]{\includegraphics[width=0.36\linewidth]{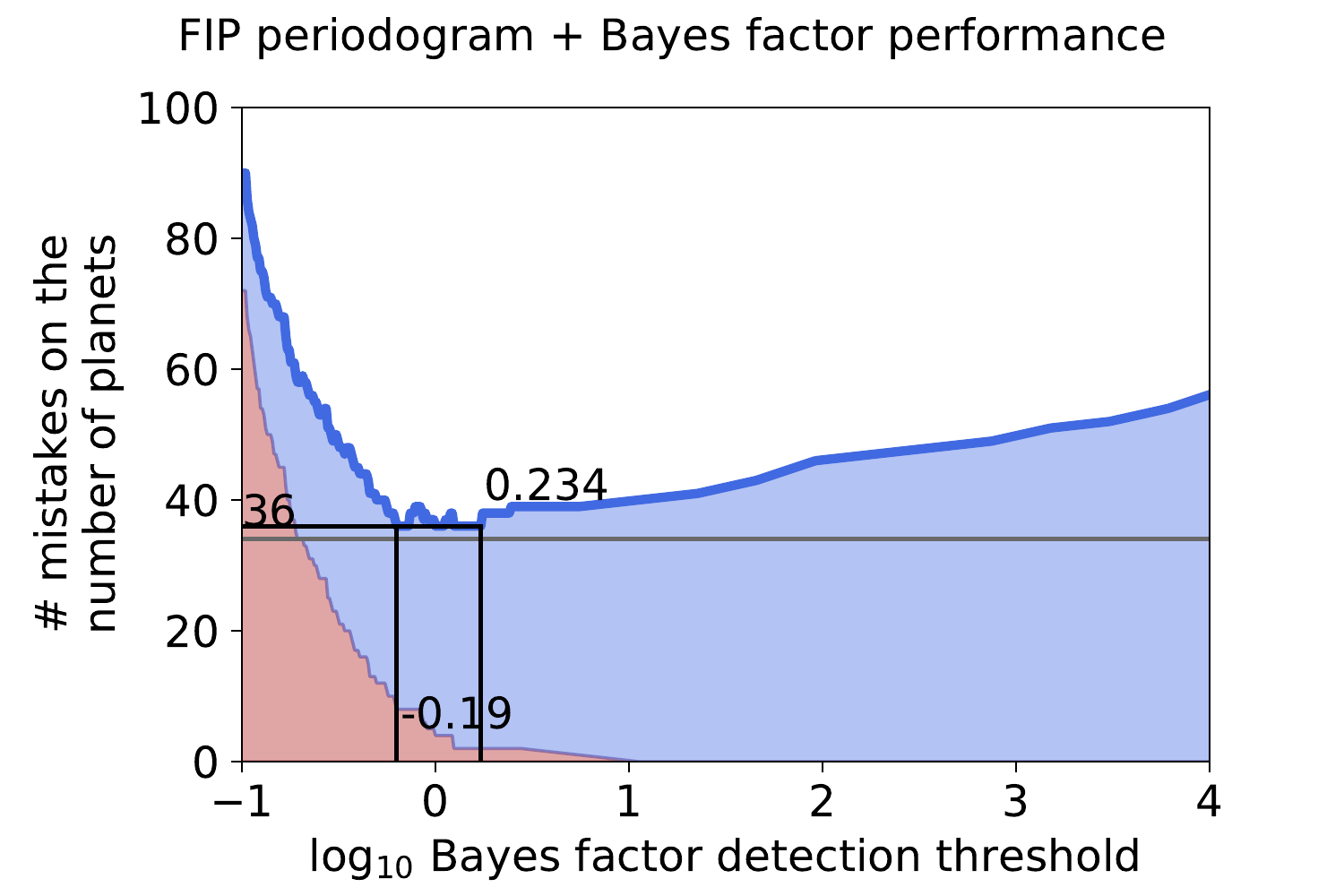}};
	\path (0.15,4) node[above right]{\large(g1)};
	\path (8,0) node[above right]{\includegraphics[width=0.36\linewidth]{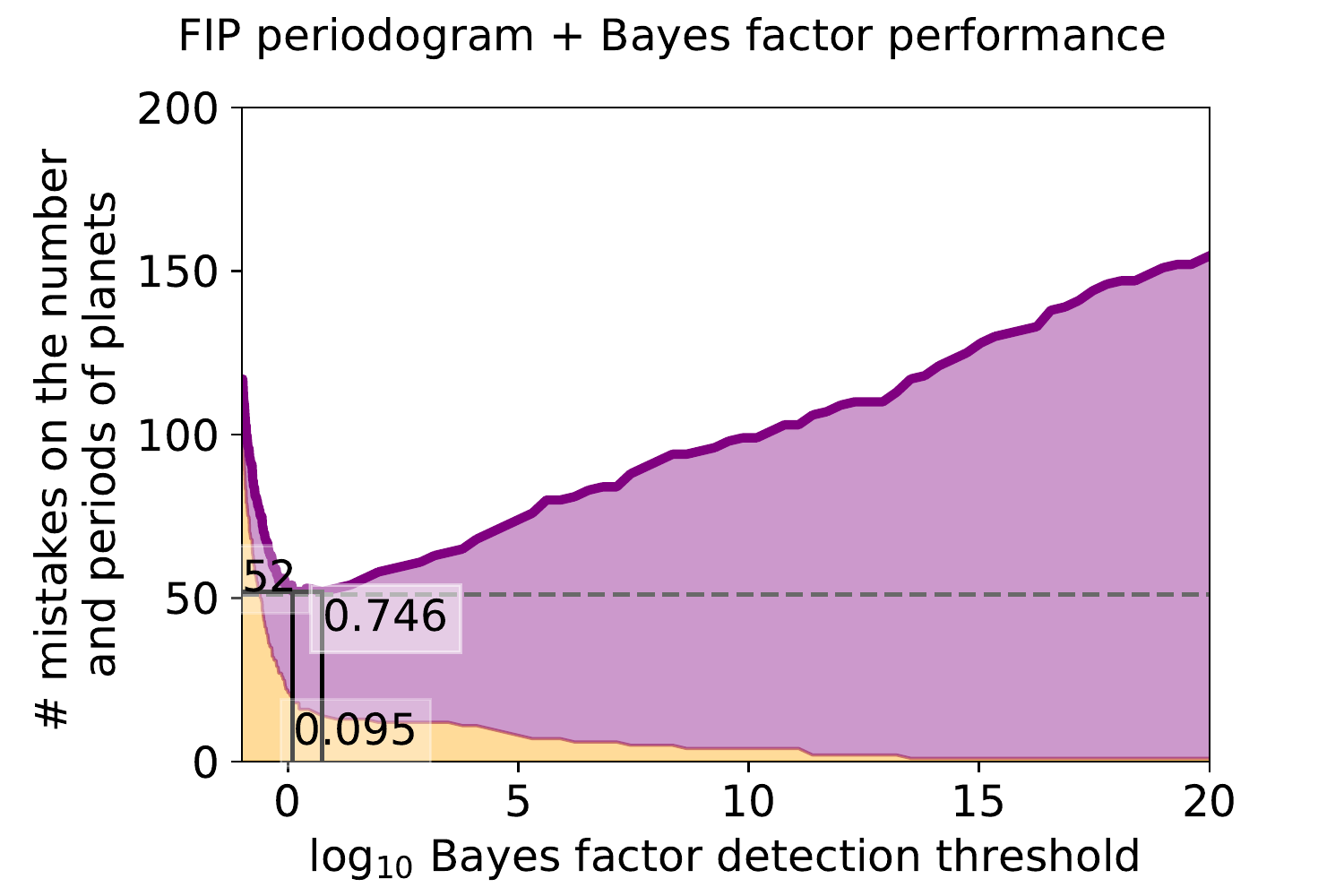}};
	\path (8.15,4) node[above right]{\large(g2)};
	\end{scope}
	\end{tikzpicture}
	}
	\vspace{-0.7cm}
	\caption[]{ \ch{Same quantities as above, for the detection criteria FIP periodogram + FAP (f1 and f2) and FIP periodogram + Bayes factor (g1 and g2). } }
	\label{fig:wnoise} 
\end{figure*}	 	 
	 
\begin{figure*}

	\centering
  \subfloat[][]{
	\hspace{-1cm}
	\begin{tikzpicture}
	\path (0,0) node[above right]{\includegraphics[width=0.36\linewidth]{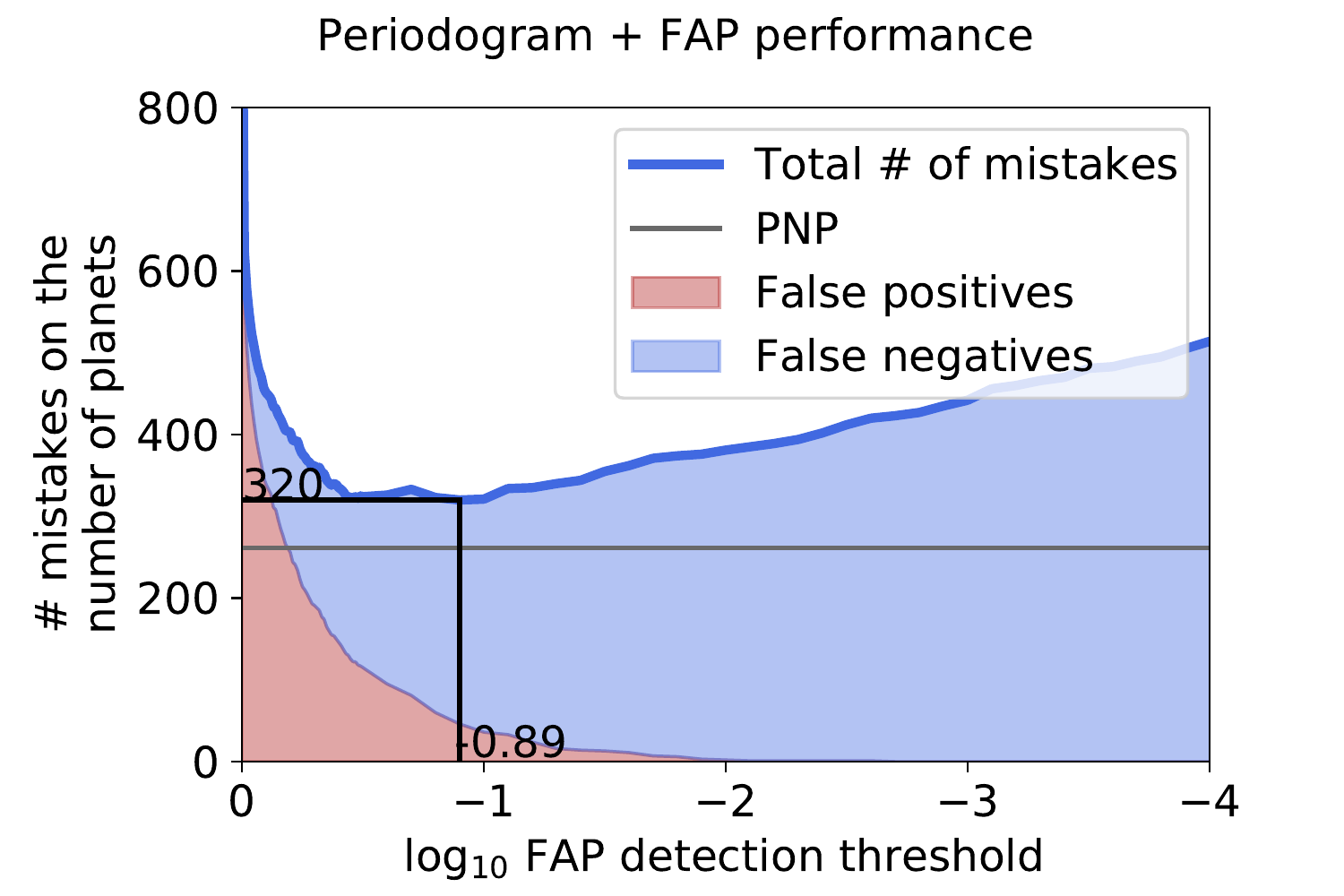}};
	\path (0.15,4) node[above right]{\large(a1)};
	\path (8,0) node[above right]{\includegraphics[width=0.36\linewidth]{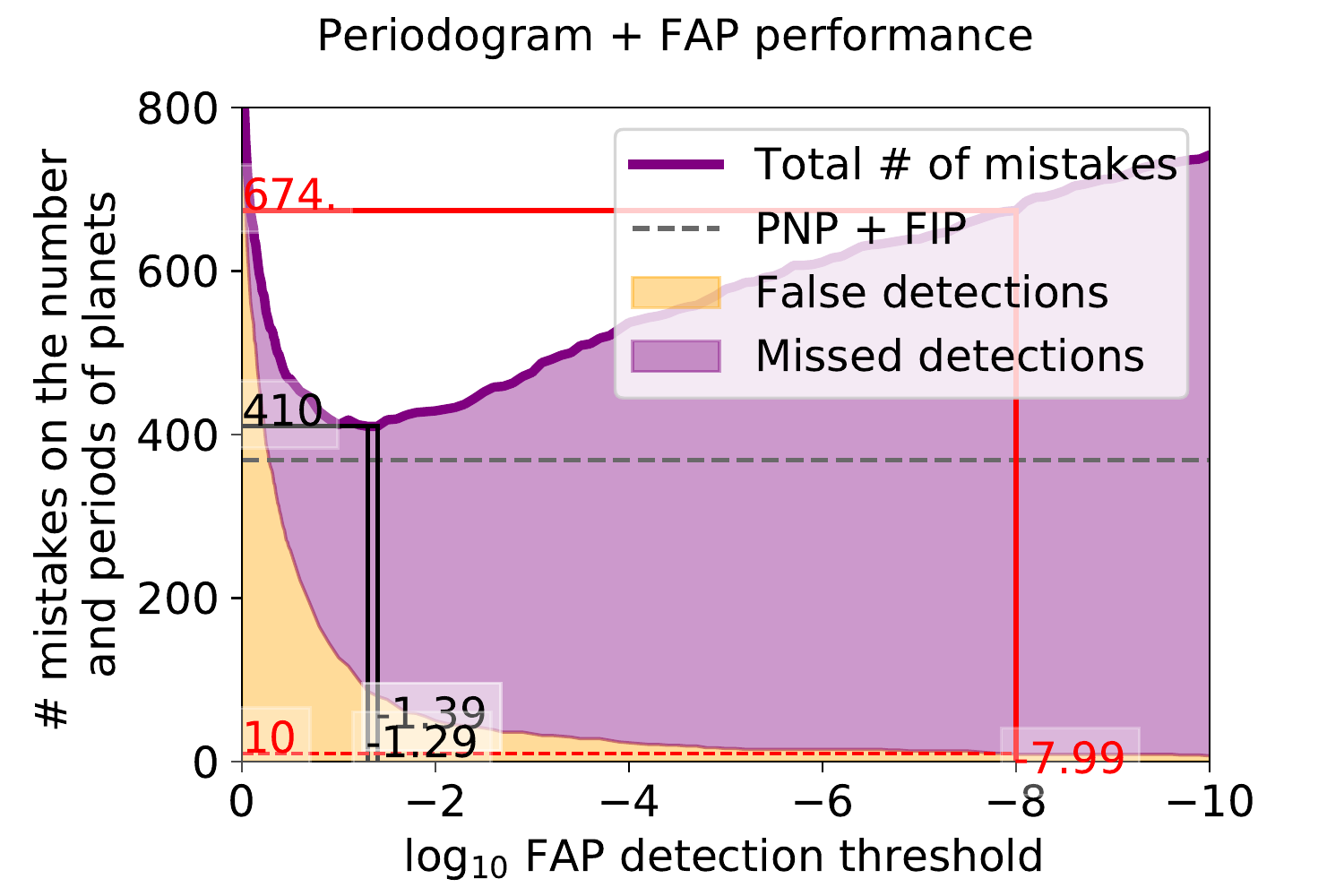}};
	\path (8.15,4) node[above right]{\large(a2)};
	\begin{scope}[yshift=-4.5cm]
	\path (0,0) node[above right]{\includegraphics[width=0.36\linewidth]{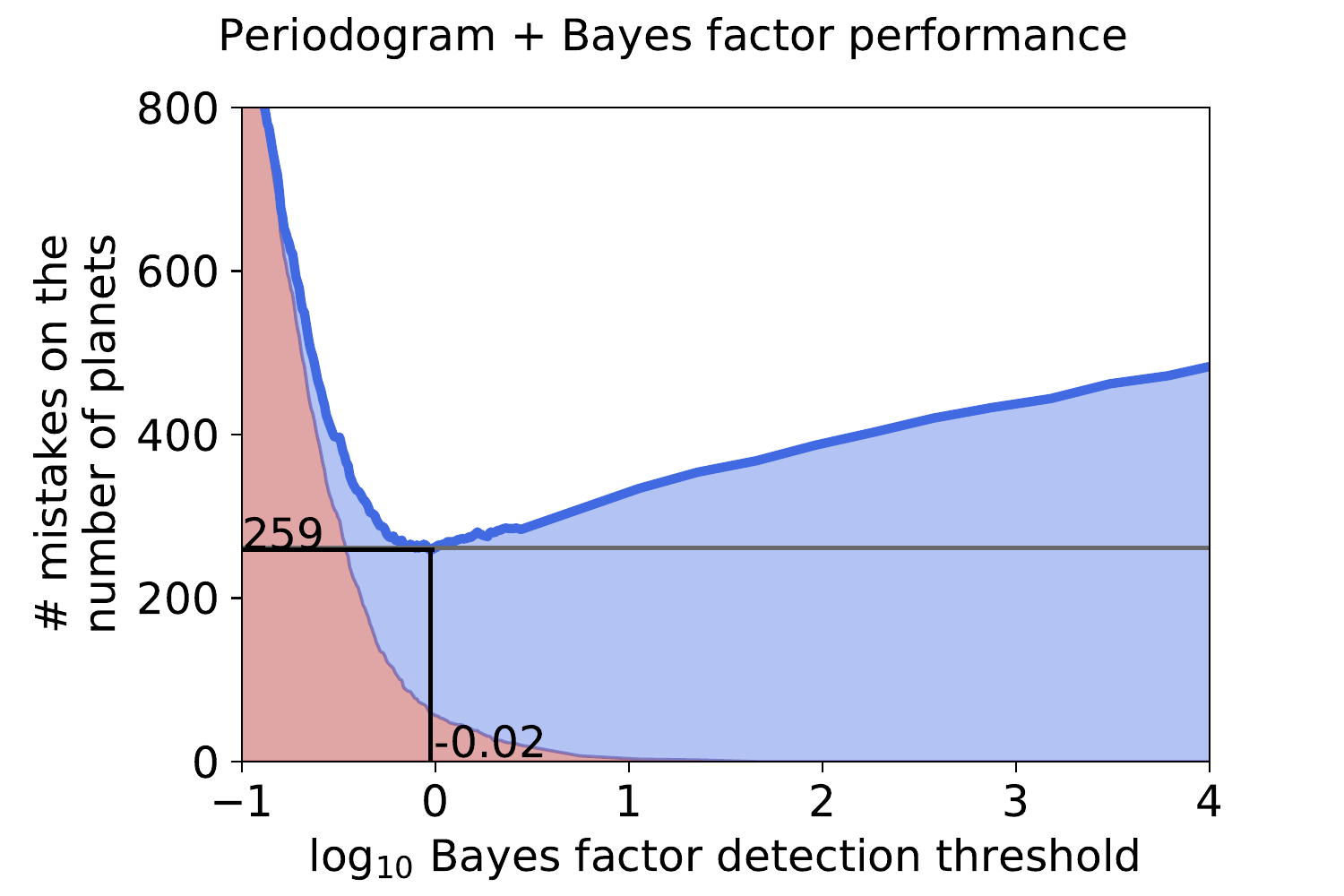}};
	\path (0.15,4) node[above right]{\large(b1)};
	\path (8,0) node[above right]{\includegraphics[width=0.36\linewidth]{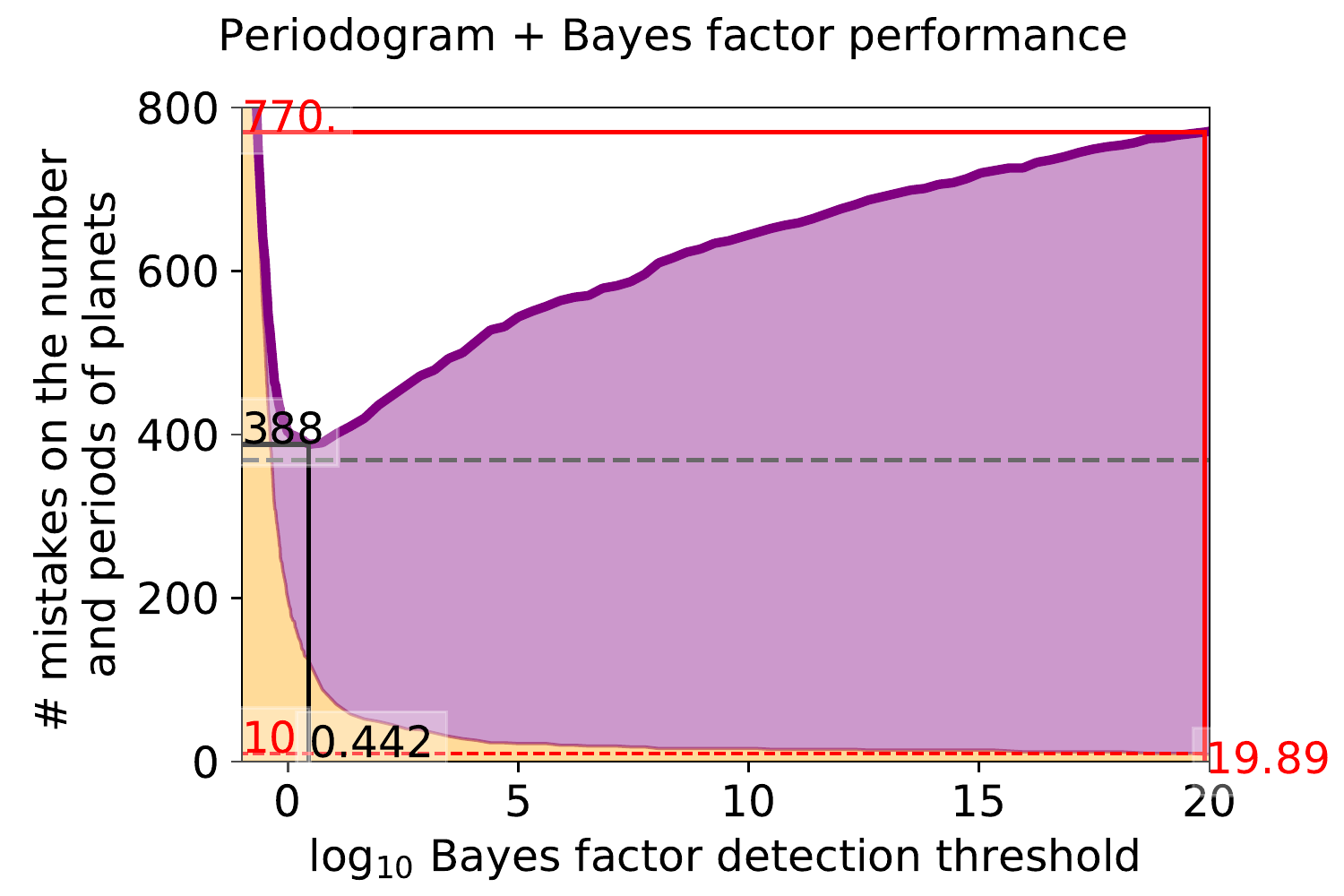}};
	\path (8.15,4) node[above right]{\large(b2)};
	\end{scope}
	\begin{scope}[yshift=-9cm]
	\path (0,0) node[above right]{\includegraphics[width=0.36\linewidth]{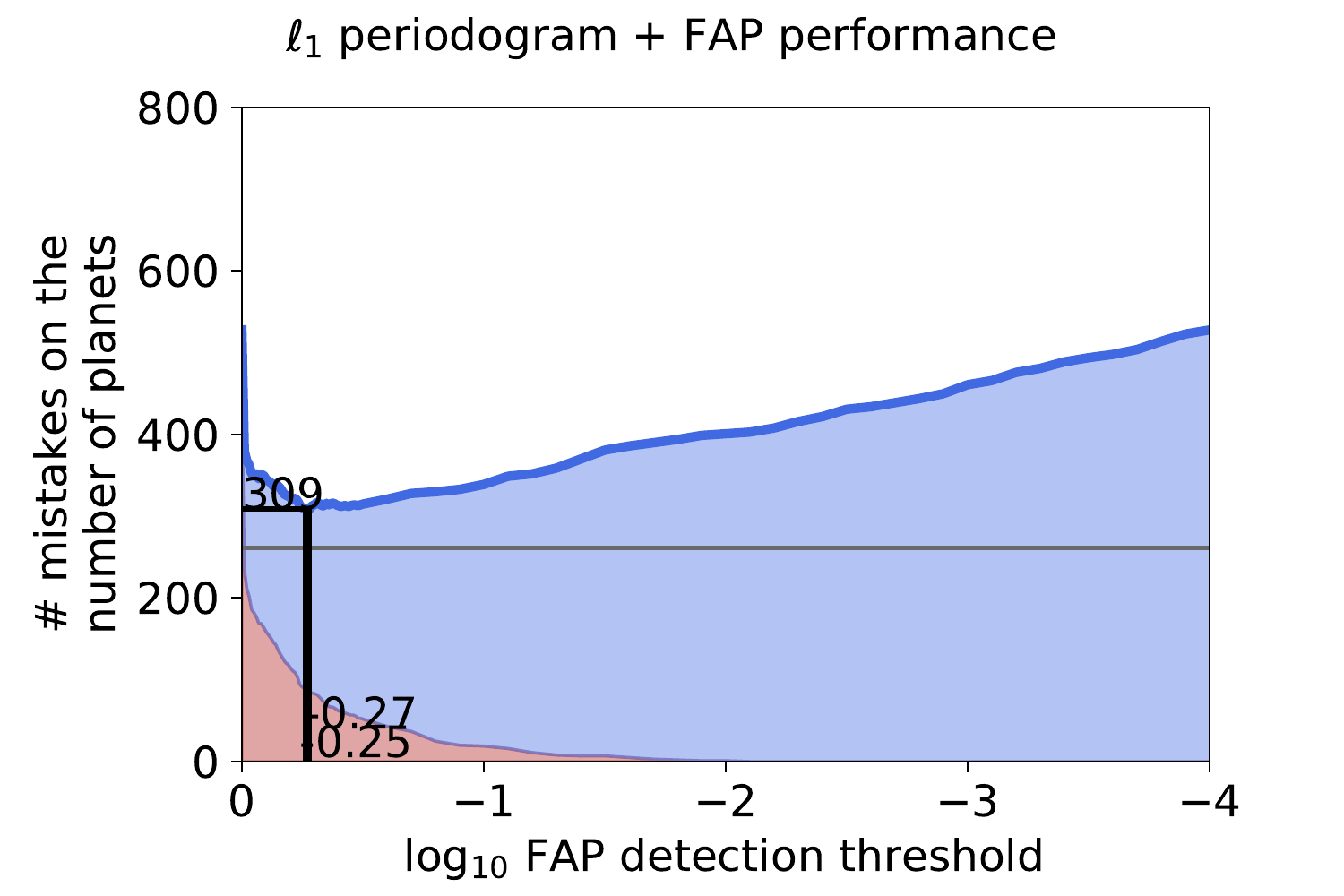}};
	\path (0.15,4) node[above right]{\large(c1)};
	\path (8,0) node[above right]{\includegraphics[width=0.36\linewidth]{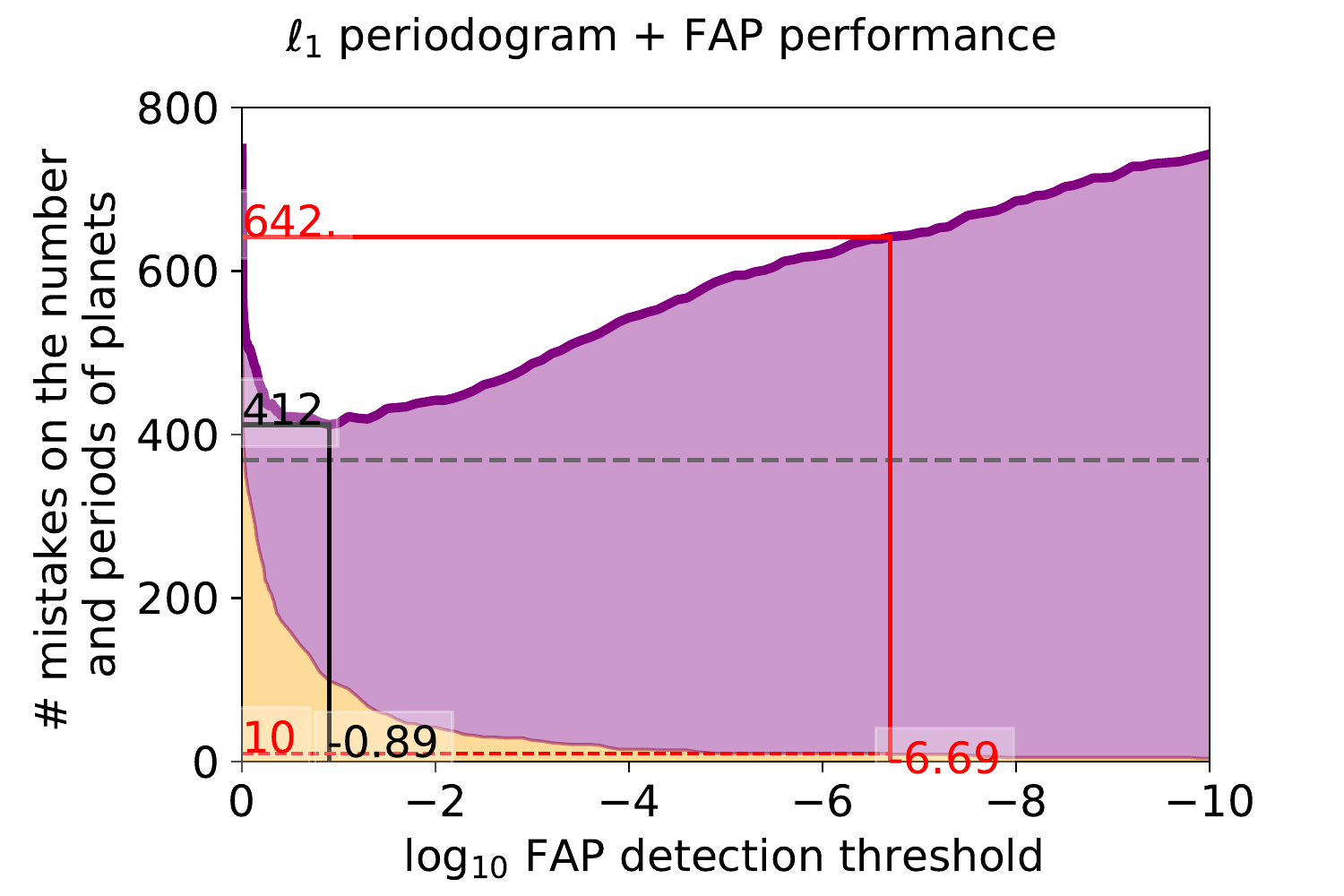}};
	\path (8.15,4) node[above right]{\large(c2)};
	\end{scope}
	\begin{scope}[yshift=-13.5cm]
	\path (0,0) node[above right]{\includegraphics[width=0.36\linewidth]{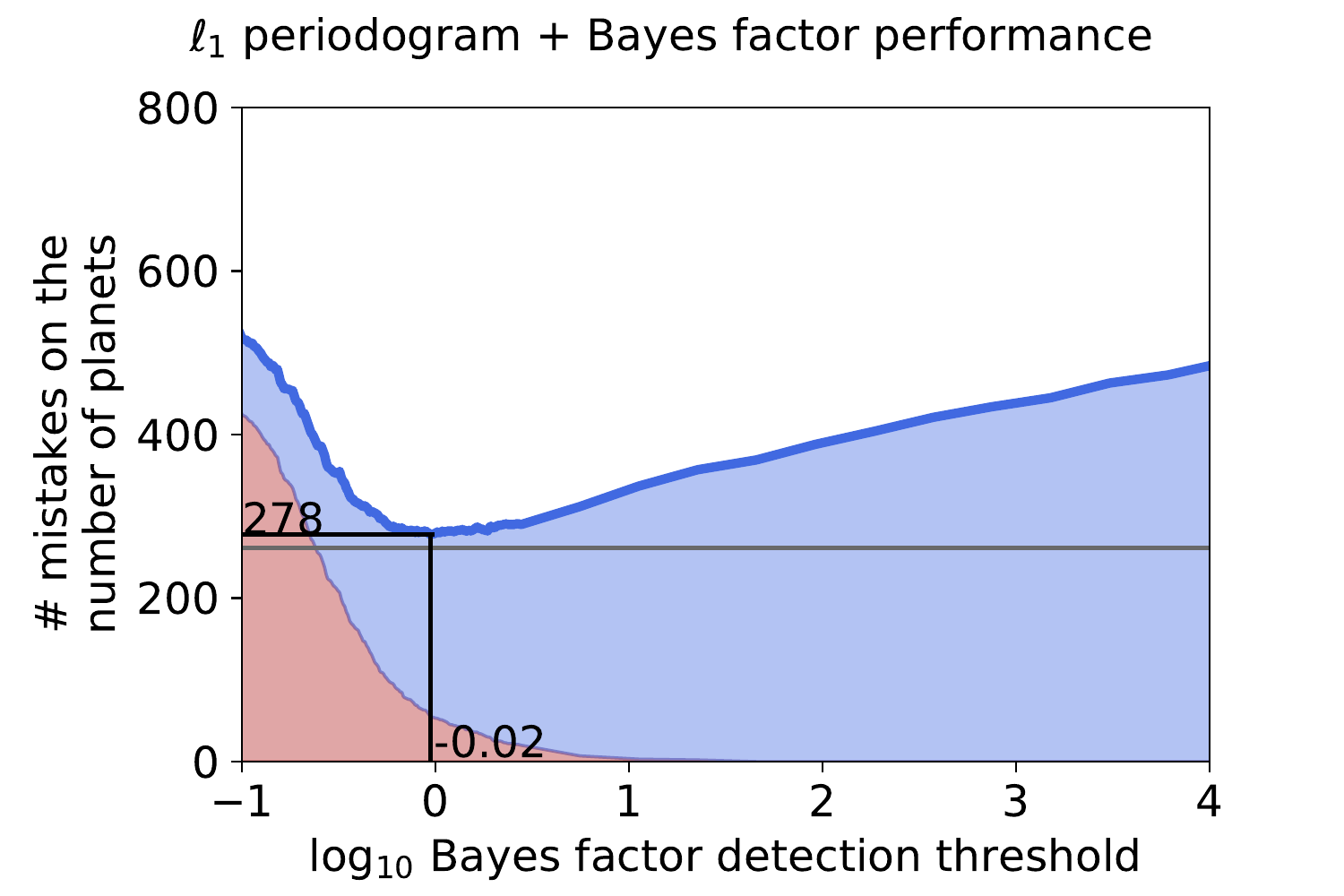}};
	\path (0.15,4) node[above right]{\large(d1)};
	\path (8,0) node[above right]{\includegraphics[width=0.36\linewidth]{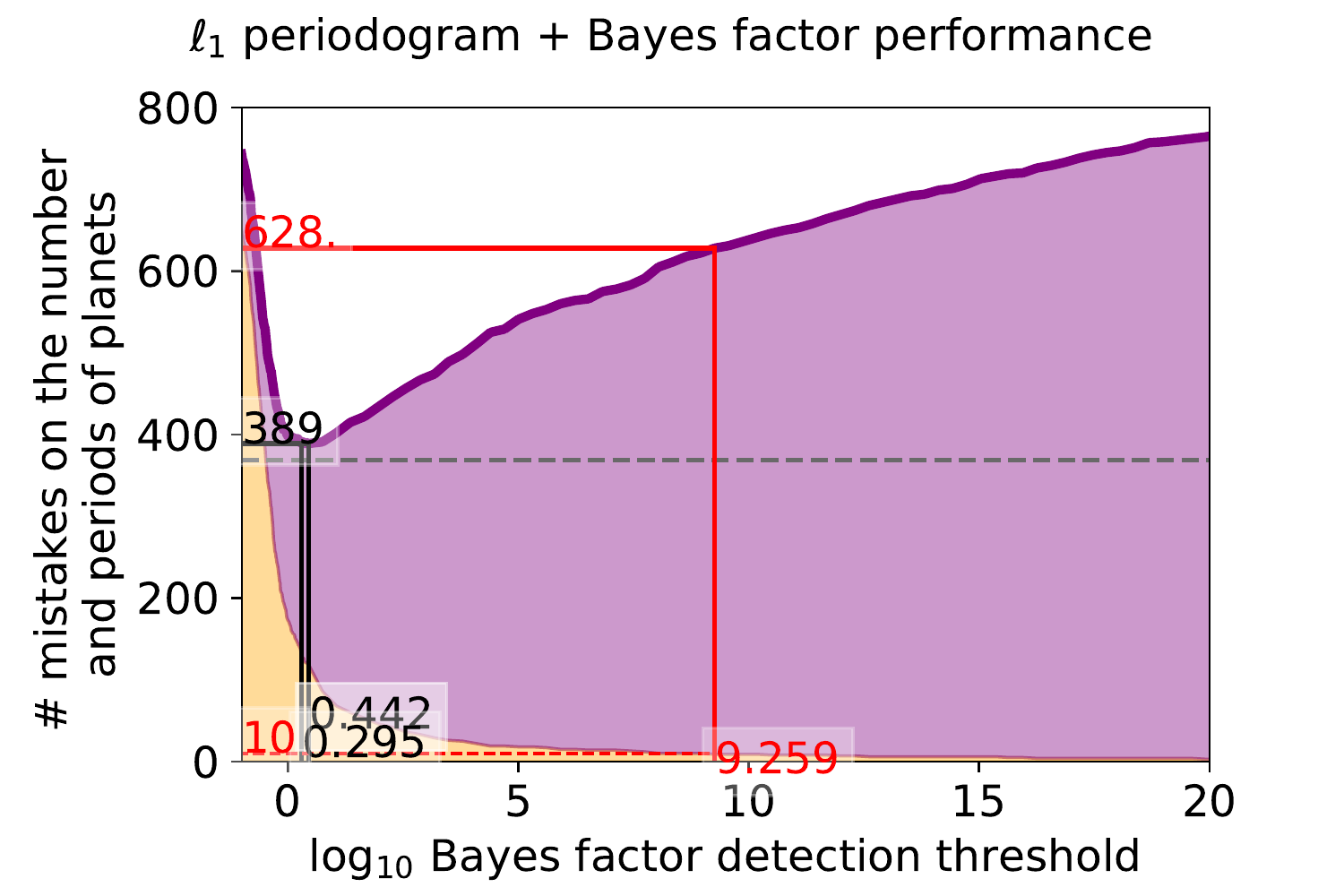}};
	\path (8.15,4) node[above right]{\large(d2)};
	\end{scope}
	\begin{scope}[yshift=-18cm]
	\path (0,0) node[above right]{\includegraphics[width=0.36\linewidth]{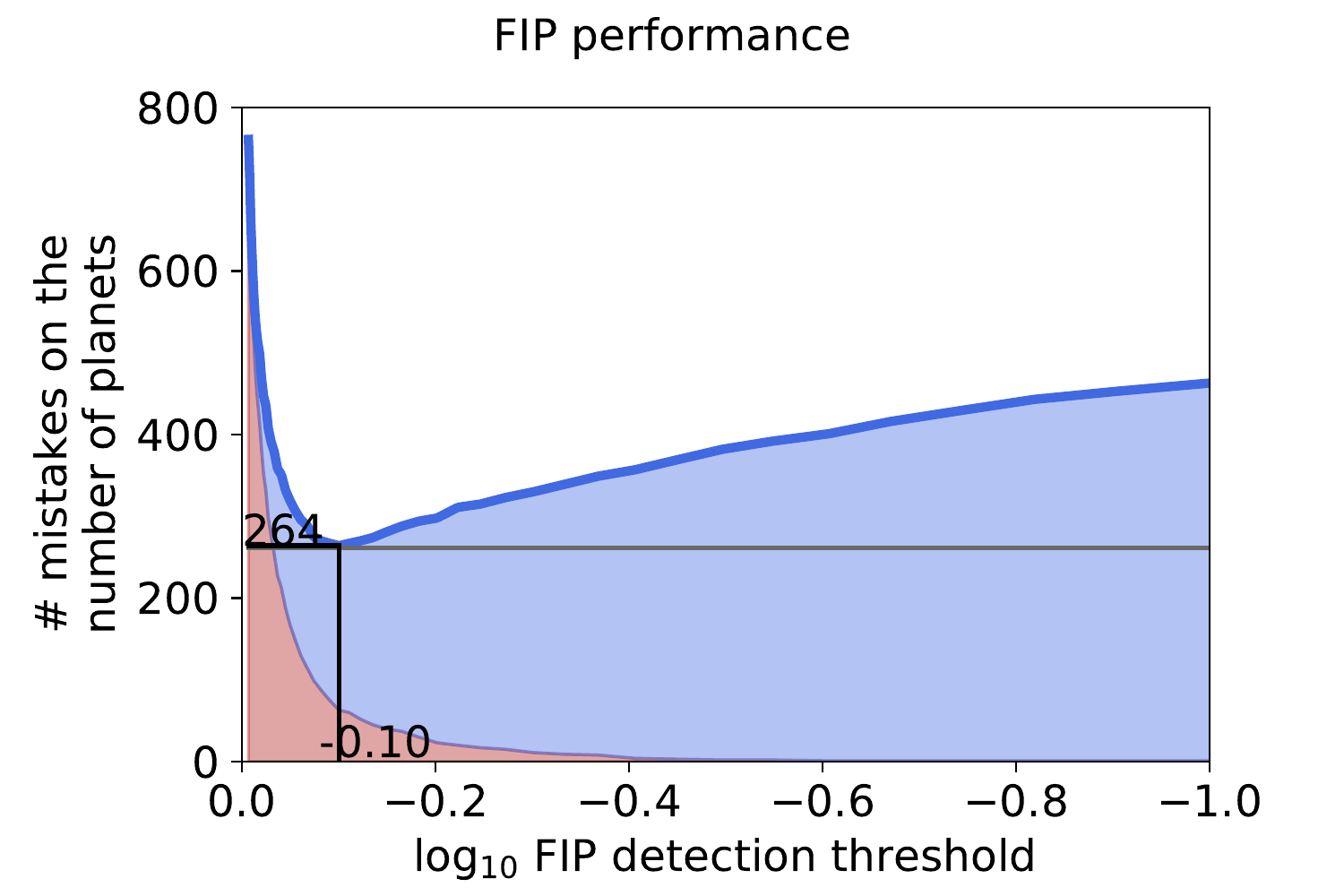}};
	\path (0.15,4) node[above right]{\large(e1)};
	\path (8,0) node[above right]{\includegraphics[width=0.36\linewidth]{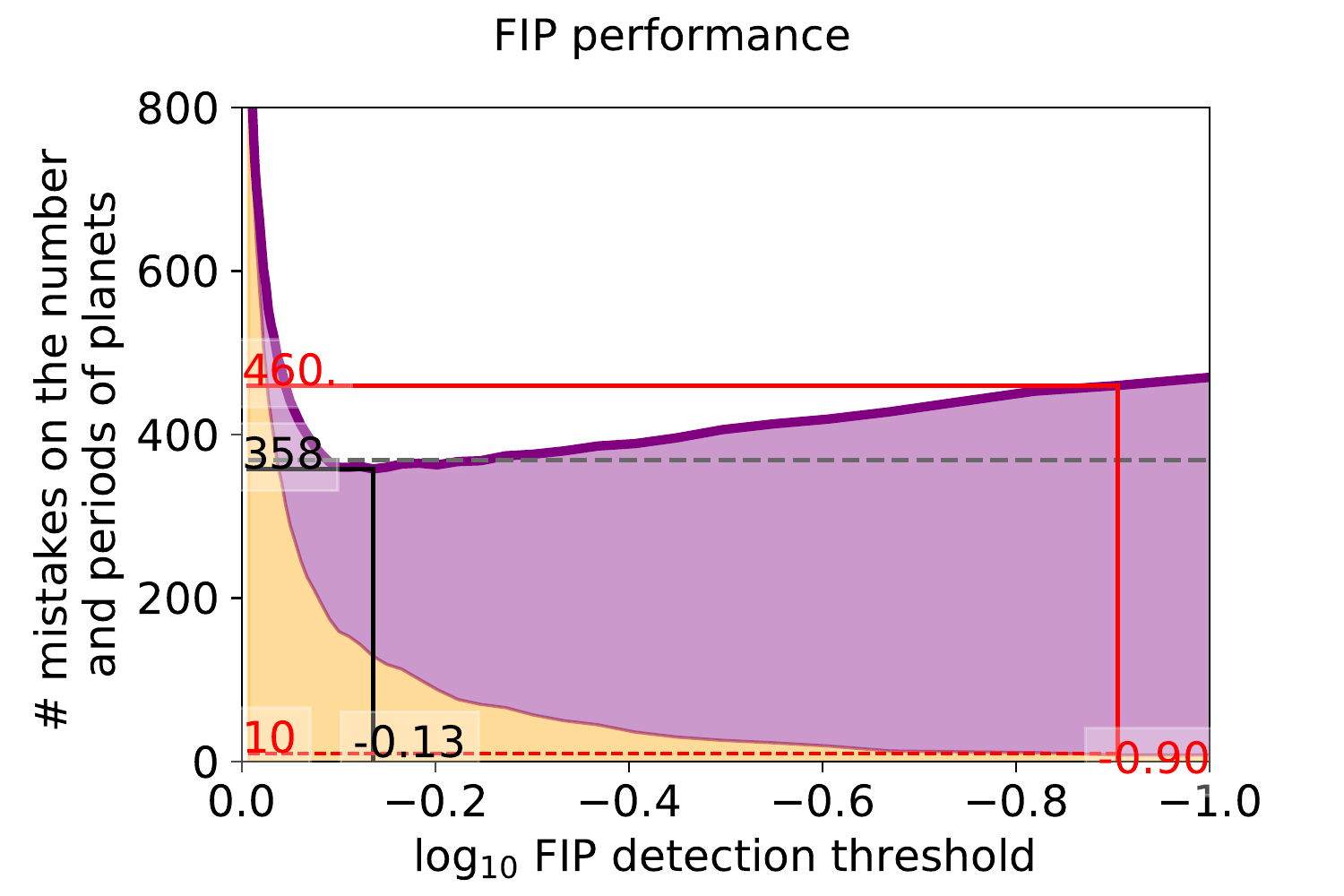}};
	\path (8.15,4) node[above right]{\large(e2)};
	\end{scope}
	\end{tikzpicture}
	}
		\vspace{-0.7cm}
	\caption{ Total number of mistakes on the number of planets (left) and number and period of planets (right) out of a thousand datasets \ch{with correlated noise (Simulation 2, described in Section~\ref{sec:simulation2})} as a function of the detection threshold for different statistical significance metrics. From top to bottom: Periodogram + FAP, Periodogram + Bayes factor, $\ell_1$ periodogram + FAP and FIP.  The total number of false positives and false negatives are represented in light red and blue \ch{shaded areas}. The false and missed detections are represented in orange and purple \ch{shaded areas}, respectively. The minimum and maximum values of the detection threshold corresponding to the minimum number of mistakes are marked with black lines. \ch{See the continuation of the figure below.} }
	 	\label{fig:rnoise}
\end{figure*}	

\begin{figure*}
	 \ContinuedFloat 
	 	\centering
	  \subfloat[][]{  	
  \hspace{-1cm}
	\begin{tikzpicture}
	\path (0,0) node[above right]{\includegraphics[width=0.36\linewidth]{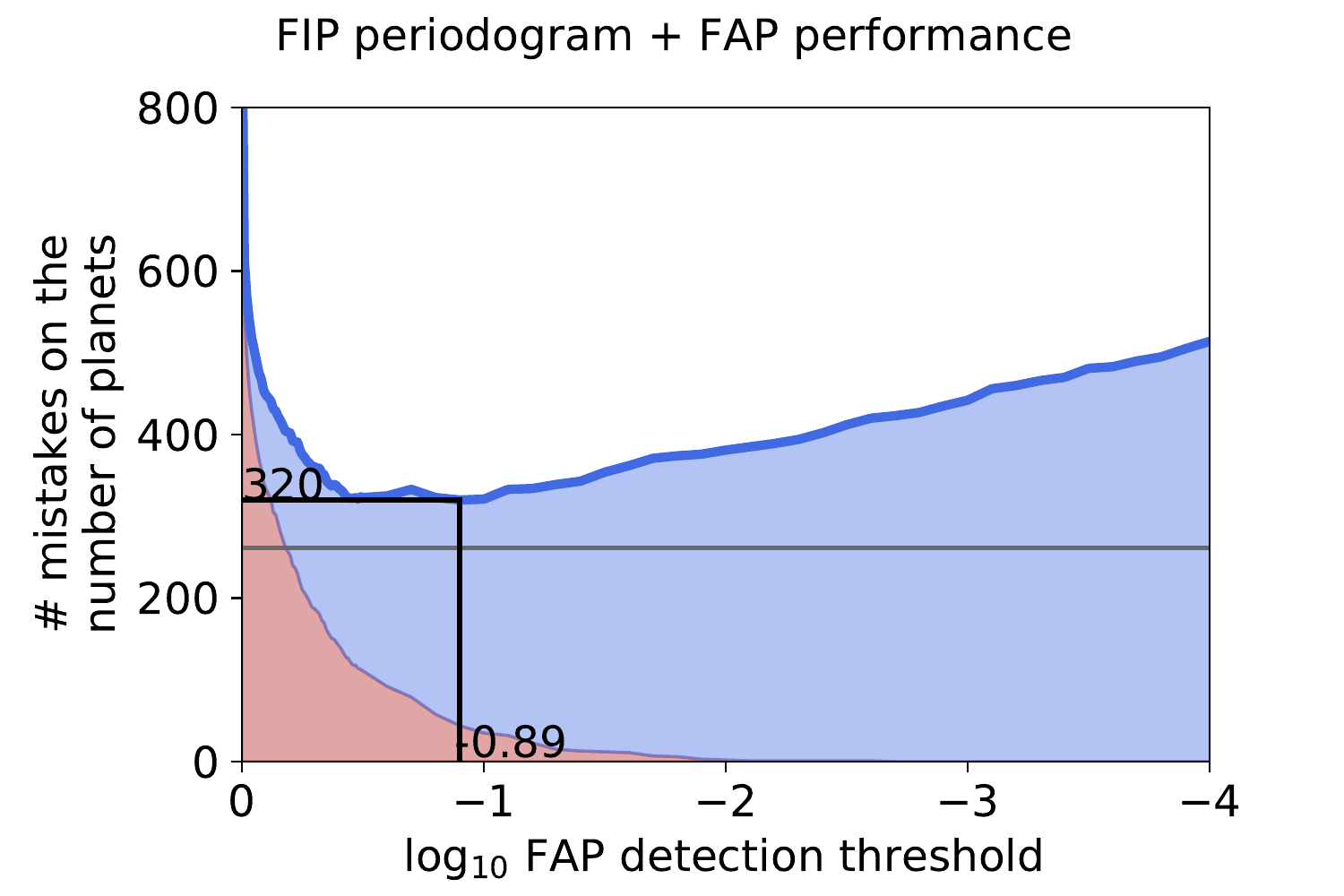}};
	\path (0.15,4) node[above right]{\large(f1)};
	\path (8,0) node[above right]{\includegraphics[width=0.36\linewidth]{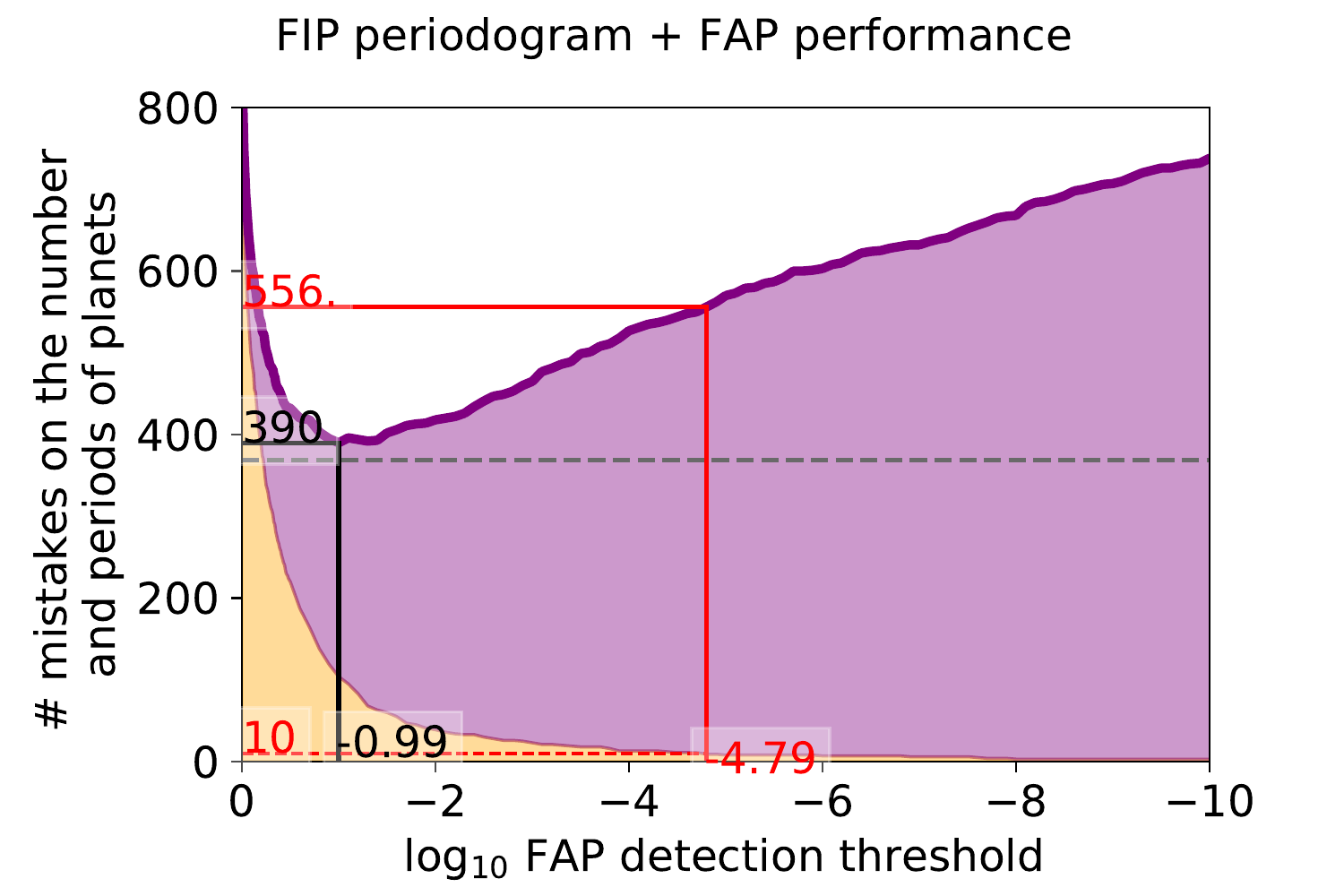}};
	\path (8.15,4) node[above right]{\large(f2)};
	\begin{scope}[yshift=-4.5cm]
	\path (0,0) node[above right]{\includegraphics[width=0.36\linewidth]{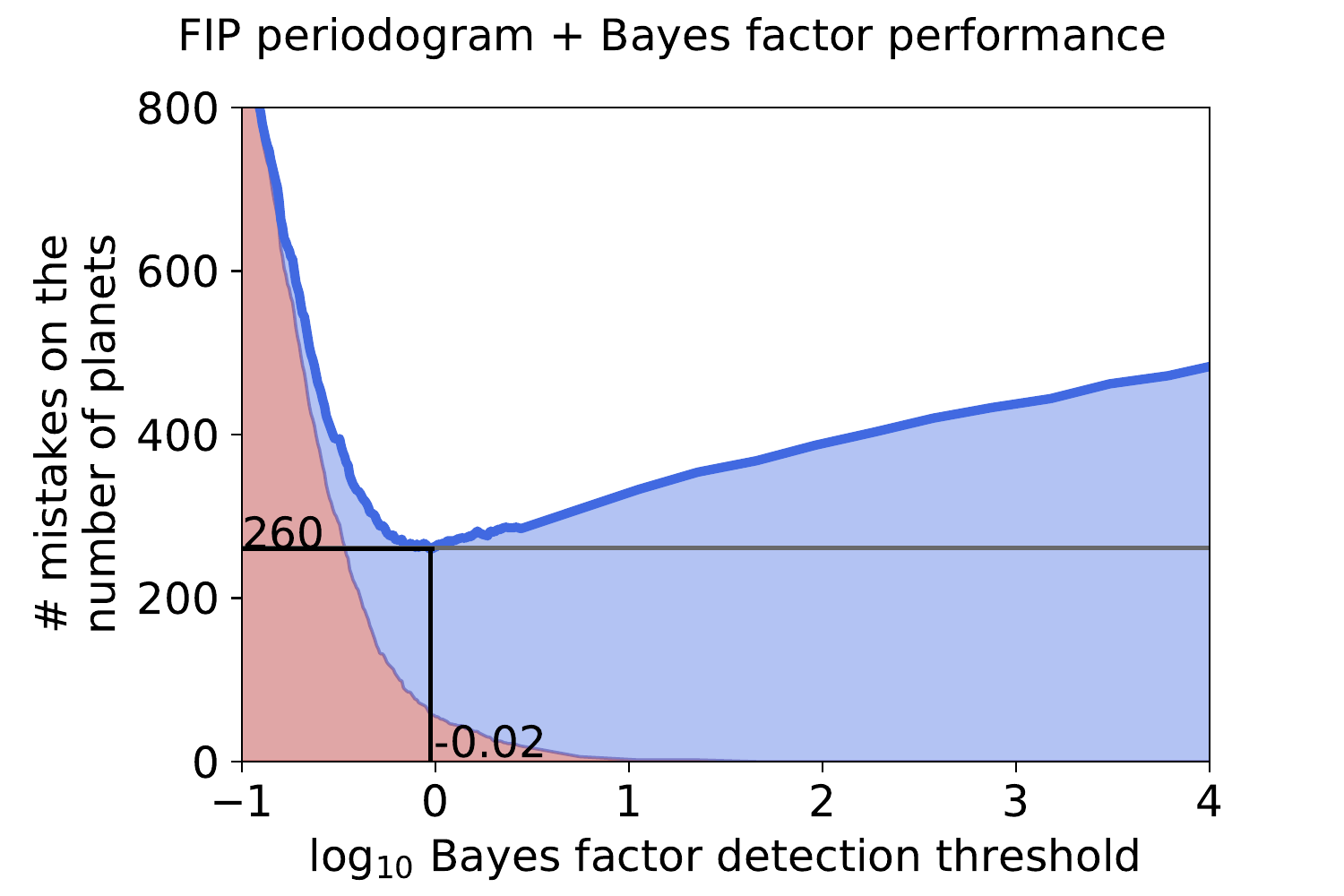}};
	\path (0.15,4) node[above right]{\large(g1)};
	\path (8,0) node[above right]{\includegraphics[width=0.36\linewidth]{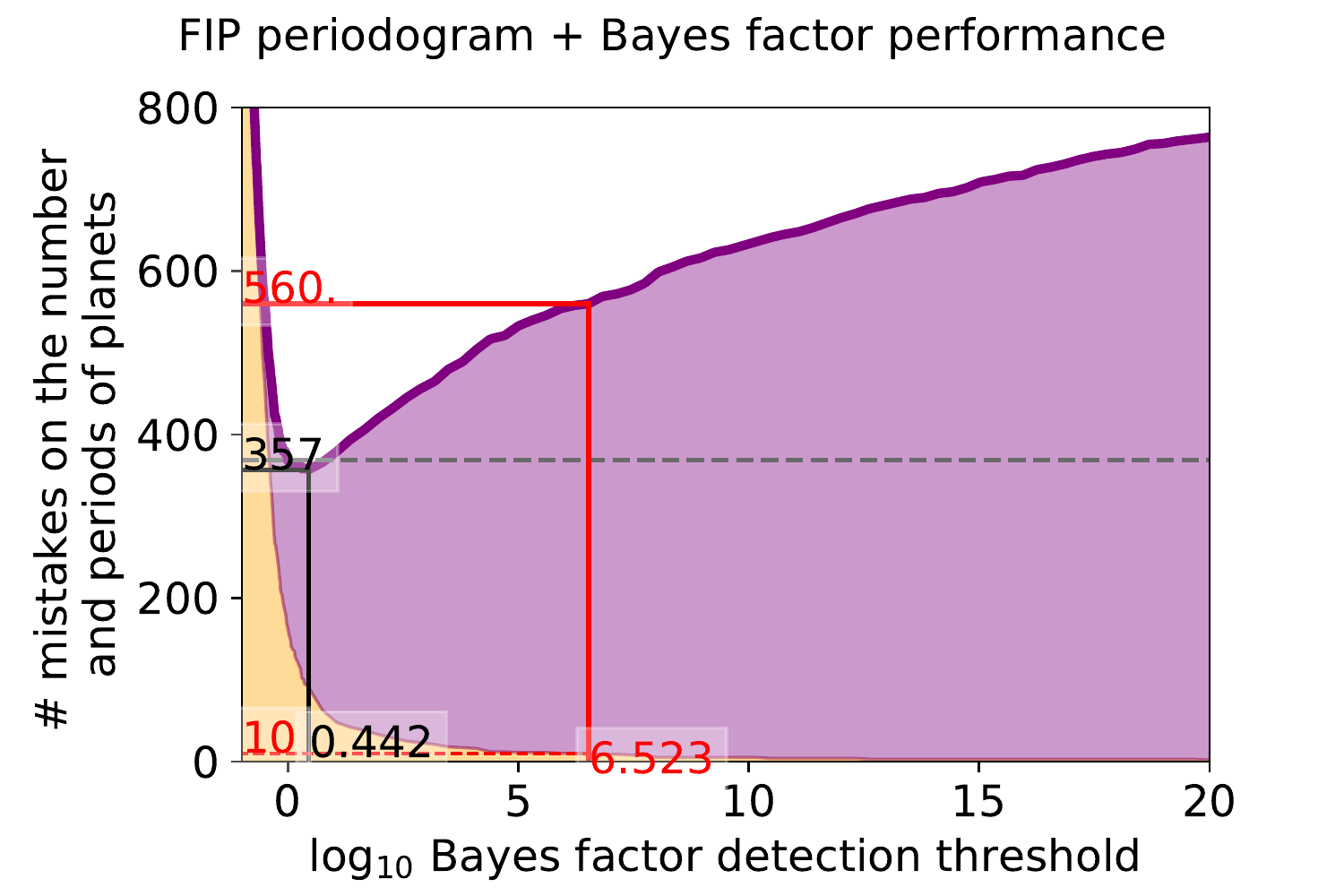}};
	\path (8.15,4) node[above right]{\large(g2)};
	\end{scope}
	\end{tikzpicture}
	}
	\vspace{-0.7cm}
	\caption[]{ \ch{Same quantities as above, for the detection criteria FIP periodogram + FAP (f1 and f2) and FIP periodogram + Bayes factor (g1 and g2). } }
	\label{fig:rnoise} 
\end{figure*}

	 \iffalse
	 \begin{figure}
	 	\centering
	 	
	 	\includegraphics[width=0.75\linewidth]{FAP_periods_circ2plalowSNR1000datasets_fpthres}
	 	
	 	\includegraphics[width=0.75\linewidth]{Periodogram_and_BF_periods_circ2plalowSNR1000datasets_fpthres}
	 	
	 	\includegraphics[width=0.75\linewidth]{l1_and_FAP_periods_circ2plalowSNR1000datasets_fpthres}
	 	
	 	\includegraphics[width=0.75\linewidth]{l1_and_BF_periods_circ2plalowSNR1000datasets_fpthres}
	 	
	 	\includegraphics[width=0.75\linewidth]{FIP_periods_circ2plalowSNR1000datasets_fpthres}
	 	
	 	\includegraphics[width=0.9\linewidth]{wronginf_circ2plalowSNR1000datasets_fpthres.pdf}
	 	\caption{Number of mistakes as a function of the number of false positives ($\log$ scale) for the different detection methods. }
	 	\label{fig:wfpthres}
	 \end{figure}
	 \fi
	 
\subsubsection{Summary}
\label{sec:threshold}

\paragraph{Threshold selection}
%---> The FIP threshold does not depend on the SNR

It appears in sections~\ref{sec:simulation1} and~\ref{sec:simulation2} that the optimal FIP threshold in the simulations is $10^{-0.13}$ = 0.74, which seems very permissive, and there are almost no false positives at FIP = 0.1. In the simulation 1,  since there are approximately 1000 planets truly in the data and there are 75 false negatives, at FIP = 0.1, one should expect approximately (1000 - 75)$\times$ 0.1 = 92 false positives at this level, but there are only two (which in fact, are close to the edge of the 1/$T_{obs}$ condition).  This seems at odds with the idea that 1 out of 10 peak with FIP = 0.1 should correspond to a missed detection. However, the reasoning above is faulty.  Indeed, in the FIP validation case the calculation is done conditioned on having events with same FIP $\alpha$. In the threshold selection case, there are no guarantees to find events that would be selected by taking the maximum of the FIP periodograms with threshold $\alpha$.

To see this, let us consider two limiting cases, in which signals are either extremely clear or very close to the noise level. The FIP incorporates the prior on the amplitude of the signal. In the first case, there is a clear cut separation between signals which are confidently detected and non detections, and a high  FIP threshold even of 90\% might well already provide a very low, potentially null number of false positives. If the prior signal amplitude is closer to the noise, then the FIPs of the maximum peaks will be concentrated towards lower values. Indeed, we see that between simulations 1 and 2, at the optimal threshold 0.74, there are many more false positives at low SNR (simulation 2) than in the high SNR case. 

The optimal FIP threshold should be of the order of 50\%. Indeed, if the FIP is below 50\%, it is more likely that there is a planet than not. It appears in both simulations that a FIP threshold of 1 - 10 \% is appropriate. In real cases, the number of planets is unknown and model errors might create spurious signals. We therefore consider 1\% as an appropriate threshold.  

\paragraph{Performances} \ch{In the two  simulations,  both the period selection and the level of significance play a role in the performance in the method. In Fig.~\ref{fig:wmistakes} and Fig.~\ref{fig:rmistakes}, the methods using the periodogram, $\ell_1$ periodogram and FIP periodogram perform increasingly better. The strong influence of the period selection method here comes from the relatively small number of observations, which results in aliasing.}

\ch{We find that for the periodogram and $\ell_1$ periodogram the FAP performs better as a detection threshold than the Bayes factor, but this is the reverse for the FIP periodogram. Overall, the FIP as a detection criterion offers the best performances in the low false positive regime.}

\ch{These results are obtained for priors corresponding to the distributions with which the data was generated. In the following sections we consider the influence of the prior and likelihood choices.  }
	 
	 \subsection{Sensitivity to the prior}
	 \label{sec:prior}

\begin{figure}
	\noindent
	\centering
	\hspace{-1cm}
	\begin{tikzpicture}
	\path (0,0) node[above right]{\includegraphics[width=\linewidth]{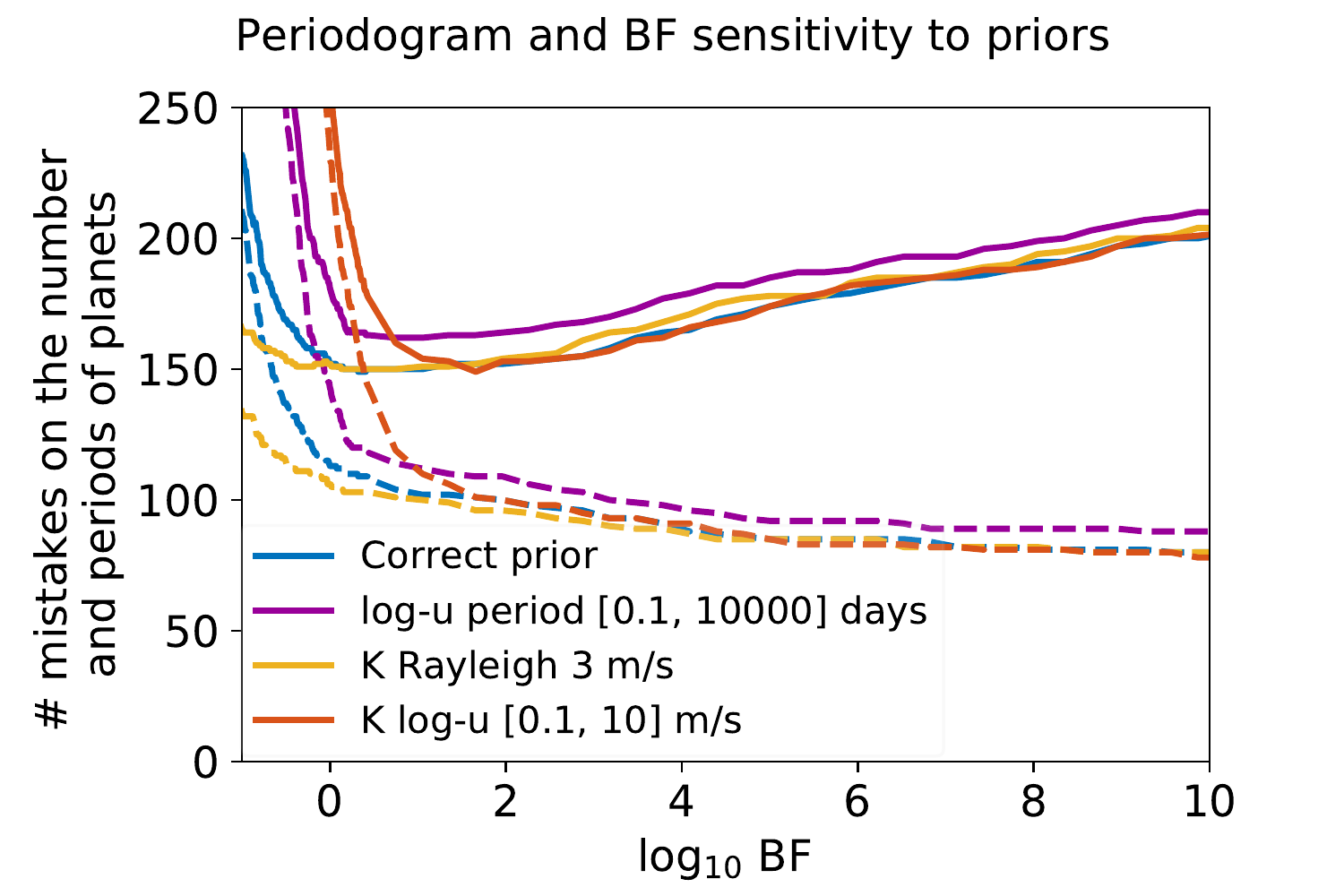}};
	\path (0.4,5.6) node[above right]{\large(a)};
	\begin{scope}[yshift=-6.2cm]
	\path (0,0) node[above right]{\includegraphics[width=\linewidth]{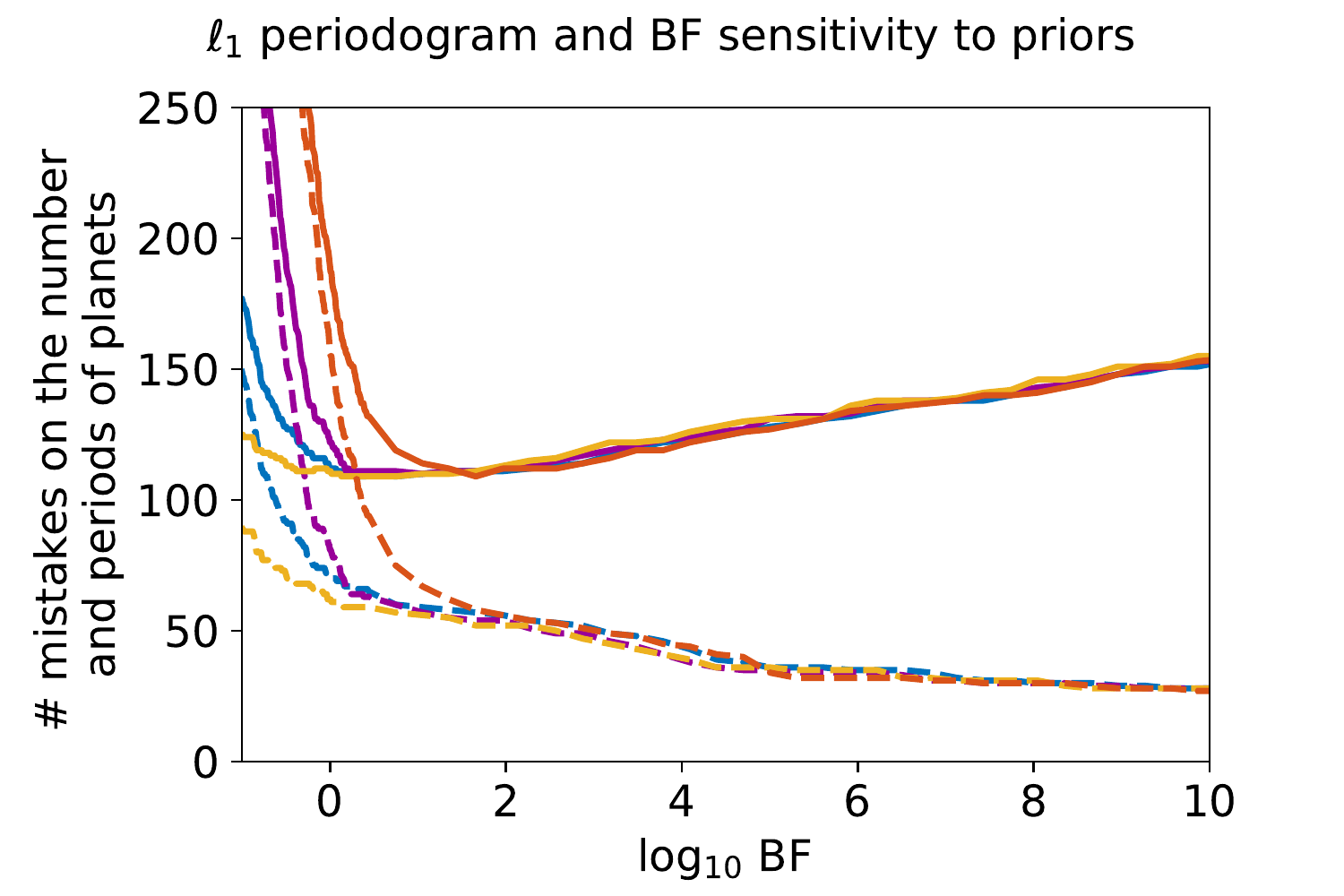}};
	\path (0.4,5.6) node[above right]{\large(b)};
	\end{scope}
	\begin{scope}[yshift=-12.4cm]
	\path (0,0) node[above right]{\includegraphics[width=\linewidth]{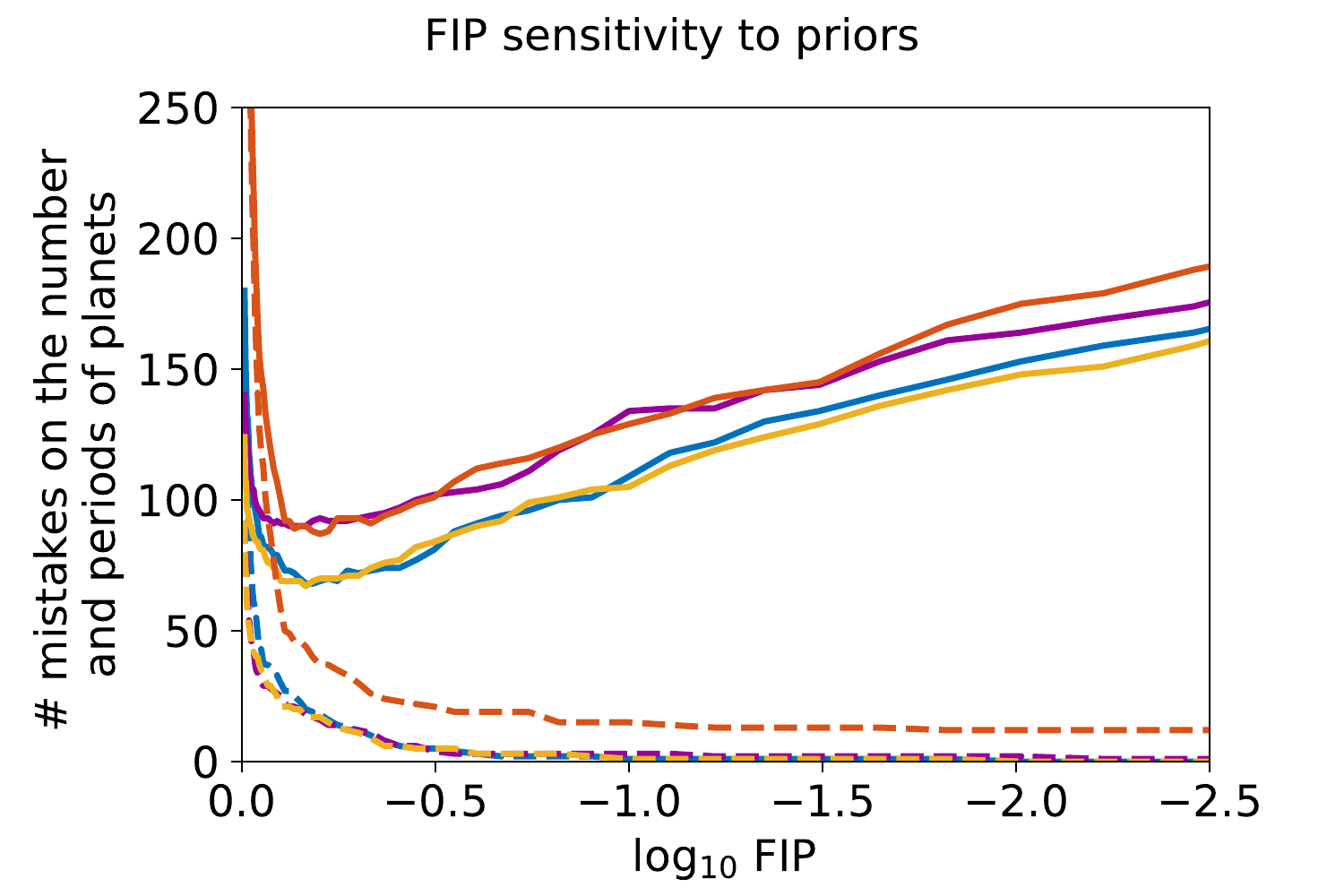}};
	\path (0.4,5.6) node[above right]{\large(c)};
	\end{scope}
	\end{tikzpicture}
	\caption{ Total number of mistakes (plain line) and total number of false detections (dashed lines) on a thousand simulated systems as a function of the threshold (from top to bottom: periodogram + BF, $\ell_1$ periodogram and BF, FIP), for different assumptions on the priors described in Section~\ref{sec:prior}. In blue: correct priors (1.5 - 100 days $\log$-uniform in period Rayleigh with $\sigma$ = 1.5 m/s on $K$), in purple: 1.5 - 10,000 days $\log$-uniform in period, in yellow: Rayleigh with $\sigma$ = 3 m/s on $K$ , in red: $\log$-uniform on [0.1, 10] m/s on $K$. Dashed lines represent the number of false detections as a function of the detection threshold and plain lines represent the total number of mistakes (false and missed detection) as a function of the threshold.   }
	 	\label{fig:priors1}
\end{figure}

	\begin{figure}
	    \centering
	    \includegraphics[width=\linewidth]{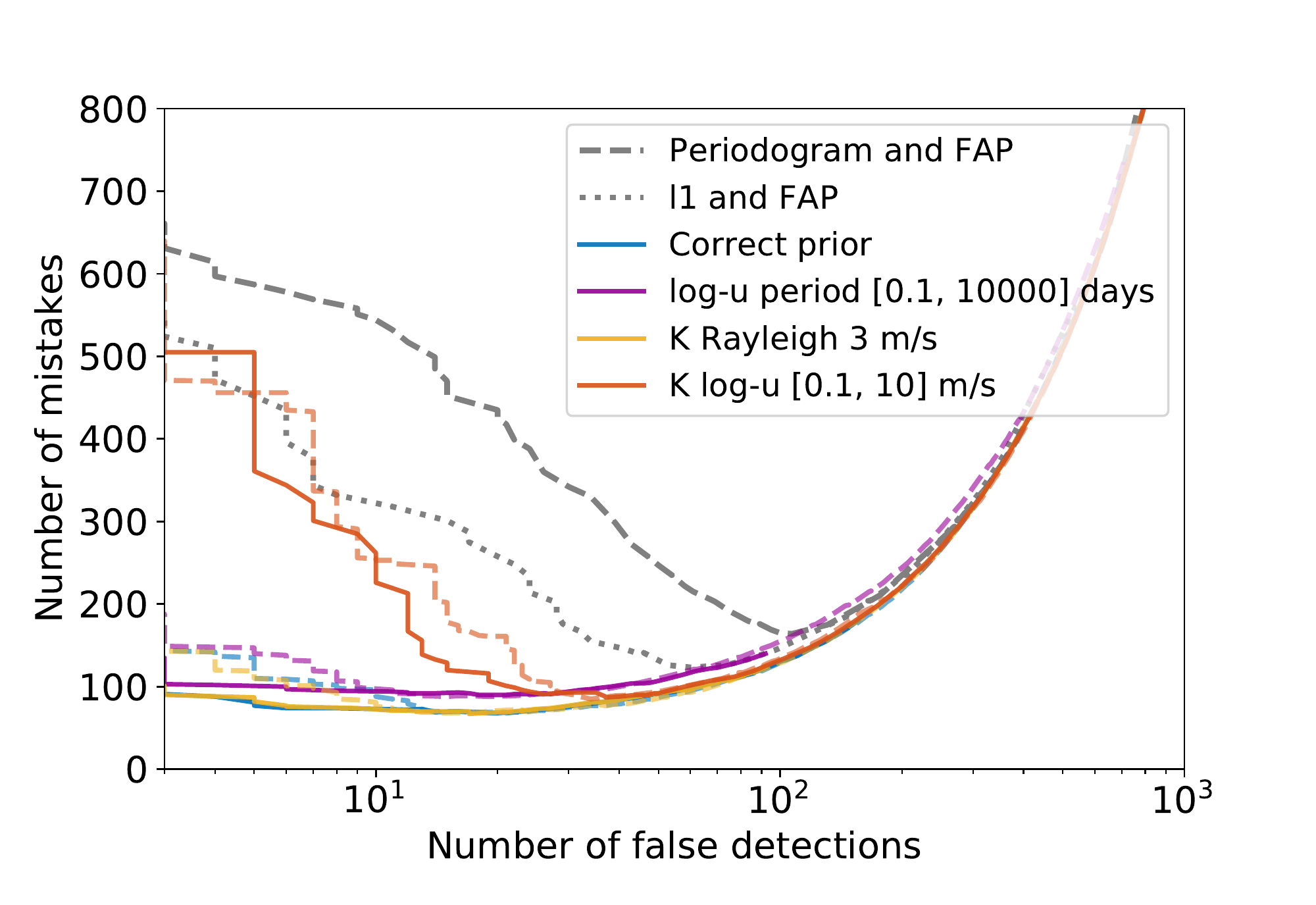}
	    \caption{\ch{Number of mistakes (false detection + missed detections) as a function of the number of false detections obtained with the FIP for different priors. Plain lines represent the detections with the FIP and dashed lines with FIP periodograms + Bayes factor.  }}
	    \label{fig:priors_FIP}
	\end{figure}
	
\begin{figure}
	 	\includegraphics[width=\linewidth]{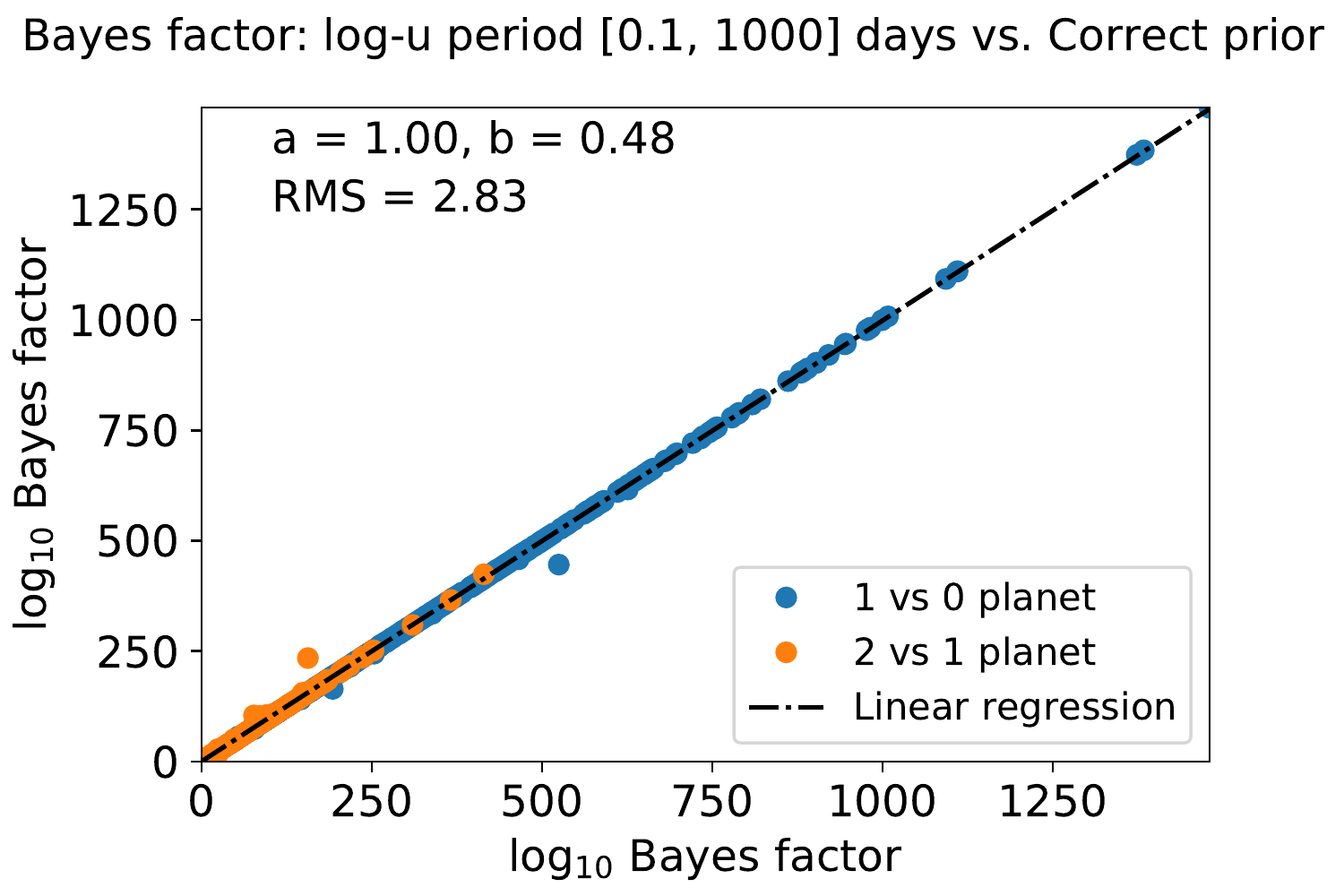}

	 	\caption{Logarithm (base 10) of Bayes factor comparing models with 1 vs. 0 (blue) and 2 vs. 1 (orange) planets computed on 1000 simulated RV time-series (black points). In abscissa, the Bayes factor is computed with correct priors. In ordinate, the Bayes factor is computed with identical priors except on period, where the prior width is $\log$-uniform on [1.5, 10000] days instead of [1.5, 100]. The red line represents a linear regression.   }
	 	\label{fig:bf}
    \centering

\end{figure}

	 All planet detection criteria depend on the underlying distribution of planets (the true distribution in nature of semi-major axes, mass eccentricities, etc.). In the cases of the BF, PNP and FIP one has to explicitly define a  prior distribution, which may or may not represent accurately the true population. The definition of the FAP does not involve a prior, but the detection performances of the FAP also depend on the true population, in particular on whether the event sought after is rare or common~\citep[e. g.][]{soric1989}. In this section we focus on the effect of the prior choice on the detection properties of the criteria that explicitly use a prior distribution. 
	 To do so, we perform a simulation. We here generate a thousand data sets with priors from Table~\ref{tab:priorscirc} and a beta distribution on eccentricity with $a$=0 and $b=15$. As in the previous sections,  in this simulation, the semi amplitude has a Rayleigh distribution with $\sigma$ = 1.5 m/s (see Section~\ref{sec:margin}), and the period has a $\log$-uniform distribution on 1.5 - 100 days.  We then analyse the data assuming
	 \begin{enumerate}
	 	\item The correct priors
	 	\item Correct priors except on periods, where it is assumed $\log$-uniform on $[1.5, 10000]$ days. In that case, periodograms and $\ell_1$ periodograms are computed on the period range $[1.5, 10000]$ days.
	 	\item Correct priors except on semi-amplitude, where it is a Rayleigh distribution with $\sigma$ = 3 m/s, that is twice the true $\sigma$. 
	 	\item Correct priors except on semi-amplitude, where it is assumed $\log$-uniform on $[0.1, 10]$ days.	 
%	 	\item Correct priors except on semi-amplitude, where it is assumed $\log$-uniform on $[0.1, 10000]$ days.	 	
	 \end{enumerate}
	 We focus the discussion on the effects of these wrong assumptions on the ability of the methods to retrieve both the correct number and period of planets. 
    As described in Section~\ref{sec:performances}, we consider that a planet is successfully recovered if it is significant and its frequency is retrieved to an accuracy $<1/T_{\mathrm{obs}}$ where $T_{\mathrm{obs}}$ is the observation time-span.  %If a planet is detected but there is no true planet within $\pm 1/T_{\mathrm{obs}}$ it is labelled as a false detection, if no planet is detected while there is one, it is labelled as a missed detection. We refer to the sum of missed and false detection as the number of mistakes. 
    
	%In Fig.~\ref{fig:priors1}, we represent the total number of mistakes (false detections and missed detections) as a function of the number of false detections for the different priors used. Each plot corresponds to a different metric. From top to bottom: Periodogram and BF, $\ell_1$ periodogram and BF and FIP. The results corresponding to assumptions 1 to 4 listed above are represented in cyan, orange, purple, blue and red. Besides, we represent in dotted line the results of the periodogram (blue) and $\ell_1$ periodogram + FAP (green) on a frequency range of 1.5 to 100 days, which do not depend on the prior definition, and  are thus identical in all cases. We find that the pattern of false and missed detection as a function of false detections depends very little on the priors chosen for the BF used in combination with the periodogram or $\ell_1$ periodogram. This is also true for the FIP, for assumptions on the priors 1,2 and 3. However, for $\log$-uniform prior on the semi-amplitude we see that the performance of the FIP deteriorates. 
	%The analysis above gives an indication on whether having an incorrect prior changes the performances of the detection criteria. However, it does not give information on whether, for a given criterion and detection threshold, changing the priors changes the detections made. 
	
	In Fig.~\ref{fig:priors1}, we represent the number of false detections (dashed lines) and total number of mistakes (plain lines)  on the thousand systems analysed as a function of the threshold adopted. These plots are identical to those of Fig.~\ref{fig:wnoise} and Fig.~\ref{fig:rnoise} except that we overplot the results obtained with different priors. Each plot corresponds to a detection method (from top to bottom: Periodogram + BF, $\ell_1$ periodogram + BF and FIP as described in Section~\ref{sec:analysis}). Colors blue, purple, yellow and red correspond to assumptions on the prior listed above 1, 2, 3 and 4 respectively.  We note that for a given detection threshold, the variation of the number of mistakes is at most 25\%  in the regions where the number of false positives is below 50 out of a thousand system, which is the region of interest. We find that the prior 2 (prior larger on period, purple curve) performs similarly or more poorly than other priors. This is explained by the fact that a larger prior on period offers chances to select a planet with period in the 100 - 10,000 days region, which would automatically be a false positives since planets are generated between 1.5 and 100 days. On the other hand, having a larger prior penalises the addition of a planet, such that viable candidates are not deemed significant with the larger prior. Finally, we note that for the FIP, the prior 4 exhibits the worst performances: the number of false positives (red, dashed line) decreases much more slowly than for the other priors. \ch{The method FIP periodogram + Bayes factor exhibits a pattern similar to the FIP. To further compare these methods, we plot in Fig.~\ref{fig:priors_FIP}  the total number of mistakes as a function of the number of false detections, as in Fig.~\ref{fig:wmistakes} and Fig.~\ref{fig:rmistakes}. Different colors correspond to different priors with the same conventions as Fig.~\ref{fig:priors1}, plain lines correspond to the FIP and dashed lines to FIP periodogram + Bayes factor. In both cases the performances degrade most when for the $\log$-uniform prior on $K$, and get closer to the $\ell_1$ periodogram + FAP (grey dotted line). }
	
	These observations are explained by the fact that changing the prior might change the level of significance (the Bayes factor) as well as the posterior distribution of periods and semi amplitude, and in turn the period selected for the planets. On the change of Bayes factor, we found in our simulations that changing the prior from case 1 (the correct one) to 2, 3, 4 only induces a multiplicative factor on the Bayes factor. Denoting by $BF_i$ the Bayes factors obtained with prior $i$, we perform a linear regression, $ \log_{10} BF_i = a_i  \log_{10} BF_1 +b_i$. The values of $BF_2$ as a function of $BF_1$ are represented in Fig.~\ref{fig:bf} (blue and orange points correspond respectively to Bayes factors of 1 vs 0 and 2 vs 1 planets) the  linear model is shown in black.
	We find  $a_2=1.00$, $b_2=0.48$ and a root mean square of the residuals (RMS)  $\text{RMS}_2=2.83$.    This means that when using prior 2 while the data was generated with prior 1, the Bayes factor is over-estimated on average by a factor $\approx 10^{0.48} = 3$  with typical variations of a factor $10^{2.83 }=676$ around this value. For $i=3$, we find $a_3=1.00$, $b_3=-0.55$, $\text{RMS}_3=1.16$, such that the dispersion around the linear model is smaller.  For $i=4$, $a_4=1.00$, $b_4=0.41$, $\text{RMS}_4=2.31$. In Appendix~\ref{app:priors}, we study analytically the effect of priors on semi-amplitude as they get wider. We show that once the prior encompasses the high likelihood region, as it widens it penalizes models with more planets.
	In terms of detection threshold, for instance in case 3, it means that having a Gaussian prior of 1.5 m/s and a detection threshold at BF = 100 is very similar to having a  Gaussian prior of 3 m/s and a detection threshold of BF = 100/3.5 = 28, provided the candidate periods of the planets are identical in both cases. 

	However, changing the priors might change the posterior distribution of periods and semi amplitudes, the convergence of numerical methods and in turn the peaks selected by the FIP periodogram. This indeed happens when using prior 4 ($\log$-uniform semi-amplitude) instead of 1,2 or 3. We investigate closely the 10 false positives with the lowest FIP when using prior 4. It appears that all of them happen in the same situation: there are two planets truly in the data, and the wrongly selected period appears at one of the two principal aliases of a true planet. Denoting by $\omega_0$ the frequency of the true planet, the spectral window is such that we expect aliases at $\pm \omega_0 + \Delta \omega_i$ where $i=1$ or 2, $\Delta \omega_1 = 1/0.997 $ days\textsuperscript{-1} and $\Delta \omega_2 = 1/31$  days\textsuperscript{-1}. 
	
\begin{figure}
	 		 	\centering
       \includegraphics[width=\linewidth]{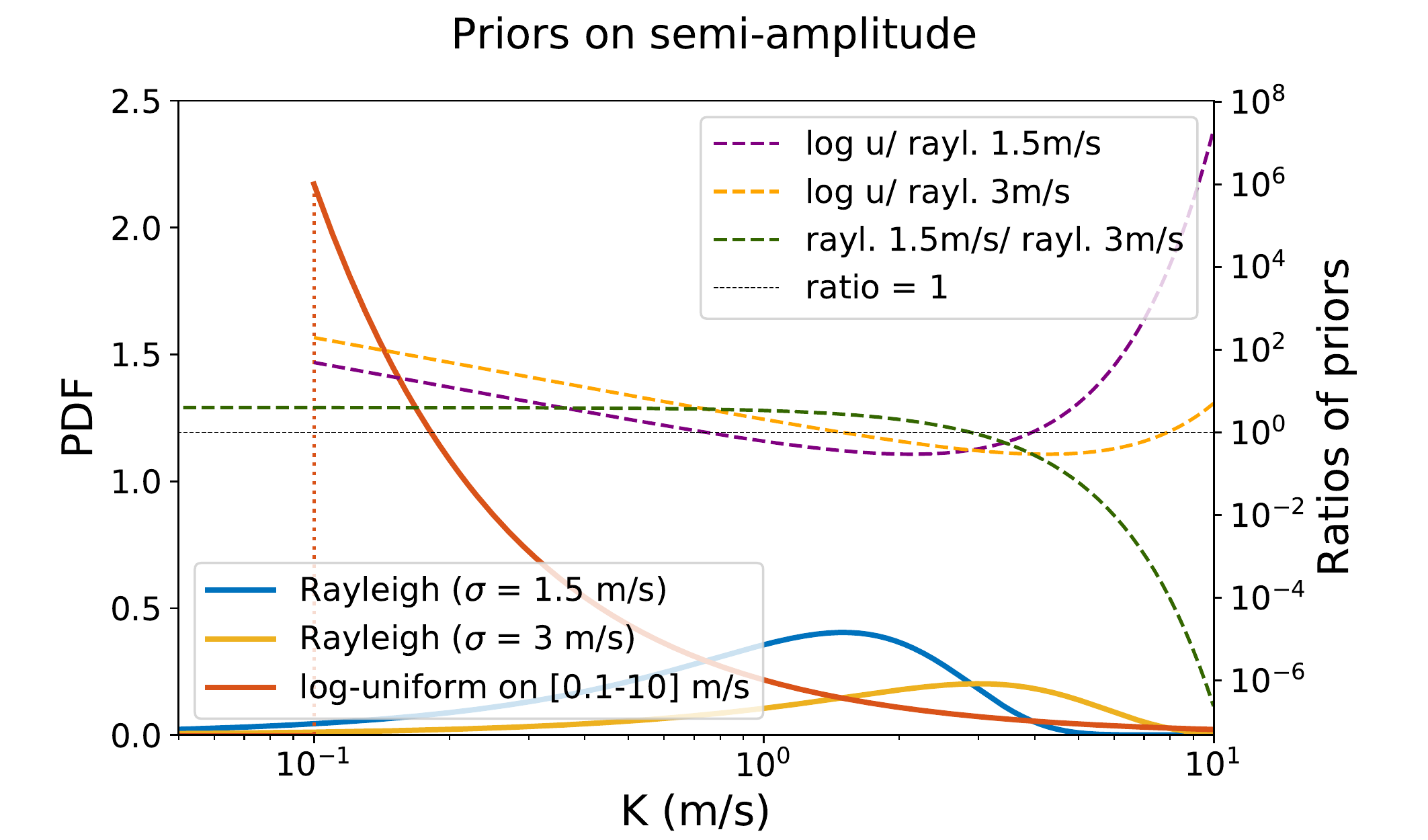}
	 	\caption{Priors on semi-amplitude. The red, yellow and blue curve correspond to Rayleigh priors with $\sigma$=1.5, $\sigma$ = 3 m/s and $\log$-uniform on [0.1 -10] m/s. The blue dotted line indicates the 0.1 m/s limit. We represent in dashed lines the ratio of log-uniform  and Rayleigh priors, as well as of the Rayleigh priors (scale is on the right $y$ axis). }
\label{fig:priorsK}
\end{figure}

	%Indeed, the convergence of the algorithm is harder to achieve with the $\log$-uniform K prior and  the behaviour of the prior close to small amplitudes might boost the significance of low amplitude signals. 	We have seen in Section~\ref{sec:prior} that when the data is generated with a Rayleigh prior with $sigma$ = 1.5 m/s, the performances of the FIP decrease when the data is  analysed with a  prior on semi amplitude ($K$) $\log$-uniform on [0.1, 10] m/s. This is particularly apparent on Fig.~\ref{fig:priorFIP}, showing the number of mistakes (false and missed detections) as a function of the false detections for the FIP for different assumptions on the prior. When using the  [0.1, 10] m/s $\log$-uniform prior on $K$, in the low false positive regime ($\lessapprox$ 20), the number of missed detection increases (red curve). We attribute this behaviour to either or both remaining numerical issues and the enhancement of low signal probability with the $\log-uniform$ prior. 
Our interpretation is that this is due to the different behaviours of Rayleigh and $\log$-uniform  priors close to small amplitudes, as well as potential remaining numerical errors.   
Indeed, the results for the priors 4 ($\log$-uniform on semi-amplitude, red curves in Fig.~\ref{fig:priors1}) are obtained with ten times as many live points as other simulations (increasing the number of live points leaves the behaviour of the other metrics unchanged). With the original number of live points the number of mistakes at in the low false detection rate regime was higher. The algorithm might spuriously select aliases.
Secondly, the behaviour of the prior close to small amplitudes  plays a role. In Fig.~\ref{fig:priorsK}, we represent in solid lines the priors on $K$ considered here (in blue, yellow and red the priors of assumptions 1, 3 and 4) and in dotted lines the ratios of priors. The orange dotted line represents the ratio of the $\log$-uniform and Rayleigh prior with $\sigma$=3m/s. It appears that the $\log$-uniform prior is 195 times higher than the Rayleigh prior ($\sigma$=3m/s) at $K = 0.1$ m/s, and on average 20 times higher on the interval 0.1 - 1 m/s, which  artificially enhances the significance of low amplitude signals, given that they are very rare in our simulation. It appears that in the case of prior 4, the wrongly selected aliases usually correspond to signal of smaller amplitudes (<0.5 m/s), which are boosted by the $\log$-uniform prior with respect to the true amplitude of signals, which is $>$1 m/s.

%\begin{figure}
%	 		 	\centering
%       \includegraphics[width=\linewidth]{figures/priors_FIP.pdf}
%	 	\caption{Number of mistakes (false and missed detections) as a function of the false detections for the FIP for different assumptions on the prior. }
%\label{fig:priorFIP}
%\end{figure}

	 We finally stress that changing the prior on the orbital elements should also be coupled to a change of the number of planets expected. Indeed, if the prior on the period, semi-amplitude or eccentricity is extended, then one expects a higher yield of planets. Let us consider the period case. The Titius-Bode law states that planets are typically uniformly spaced in $\log$ semi-major axis. If the search was initially performed with a $\log$-uniform prior on period between  1 day and $P$, then with a $\log$-uniform prior on periods between  1 and $P\times 2^{\frac{3}{2}}$ then one expects to find twice as many planets. The increase in the expected number of planets compensates the penalisation introduced by having a wider period prior.

	 %The detection criterion~\eqref{eq:criterion} necessitates to define priors on the orbital elements. The criterion might favour a different detections depending on the priors. We here show that the detections claimed based on~\eqref{eq:criterion} and Bayes factors strongly depend on the prior chosen on the semi-amplitude $K$. Loosely speaking, the broader the prior on $K$ is, the more the addition of a planet in the model is penalized.
	 
	 %	This can easily be seen in a simplified setting with the analytical formula~\eqref{eq:intlin}.

	 %In the limit of high $L$ is proportional to $1/L^{2k}$. In computing~\eqref{eq:criterion} as well as Factor comparison, we would compute 	\begin{align}	\frac{p(\bm{y} |\bm{\eta}^{k+1},k+1)}{p(\bm{y} |\bm{\eta}^{k},k)} \underset{L \rightarrow \infty}{\propto}   \frac{1}{L^{2}} 	\end{align}

	 %	\begin{align}	\frac{p(\bm{y} |\bm{\eta}^{k+1},k+1)}{p(\bm{y} |\bm{\eta}^{k},k)} \underset{L \rightarrow \infty}{\propto}   \frac{1}{L^{2}}.	\end{align} 

%\begin{figure}
%	 		 	\centering

%\includegraphics[width=\linewidth]{figures/priors_FIP and BF.pdf}
	
%	\includegraphics[width=\linewidth]{figures/priors_FIP and FAP.pdf}
	
%		 	\includegraphics[width=\linewidth]{figures/priors_PNP and FIP.pdf}
%	 	\caption{Number of mistakes (false detections and missed detections) for a given detection metric (from top to bottom: FIP, FIP and BF, FIP and FAP, FIP and PNP for different assumptions on the priors.}
%\label{fig:priors2}
%\end{figure}

	 \subsection{Sensitivity to the likelihood: averaging over noise models} 	 
	 \label{sec:likelihood}

	 \begin{figure}
	 	\centering
	 	\includegraphics[width=0.9\linewidth]{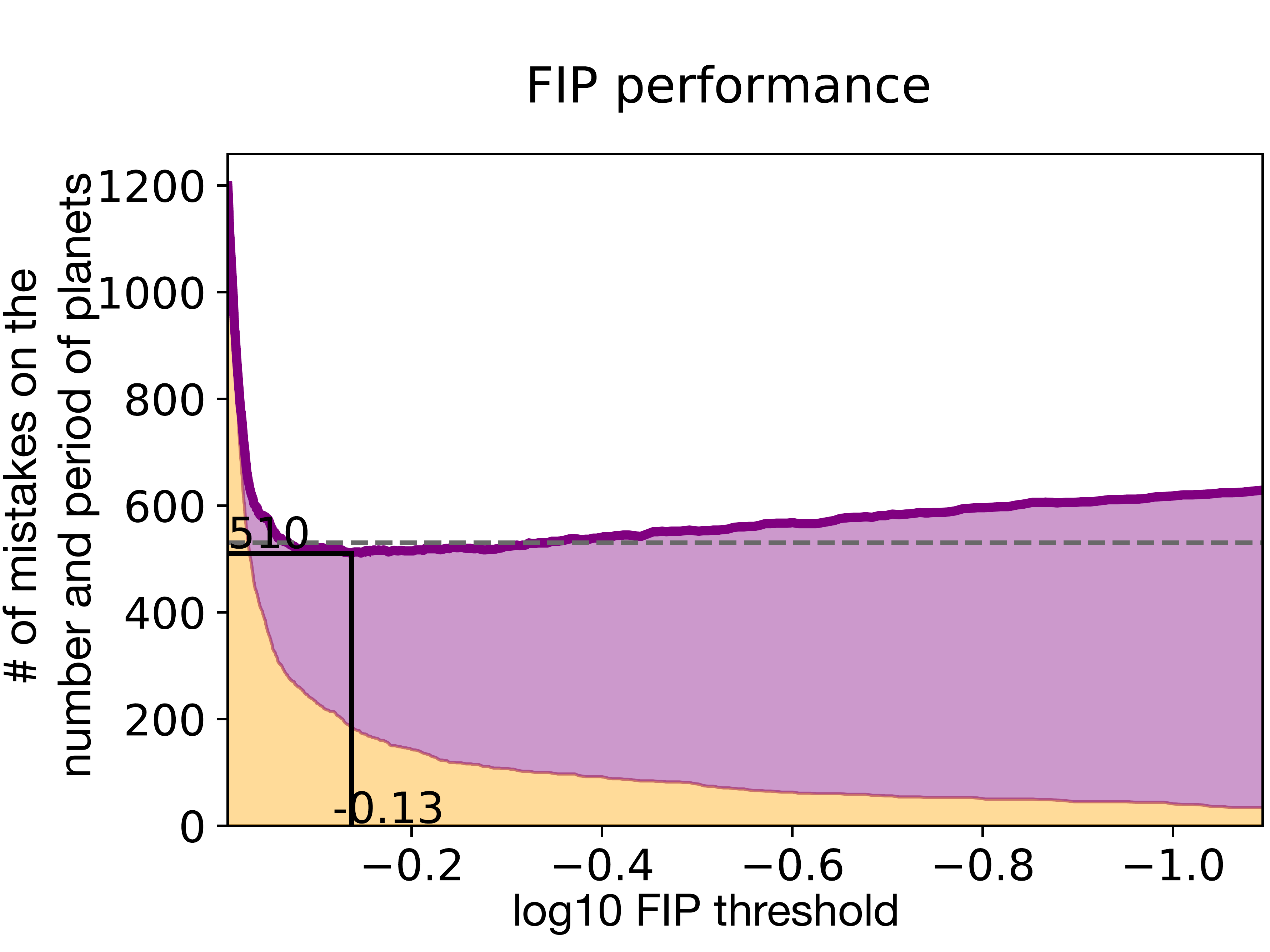}
	 	\caption{Total number of false and missed detections (yellow and purple shaded areas respectively) on a thousand system as a function of the FIP threshold when selecting the noise model with a fit of the ancillary indicators.}
	 	\label{fig:noiselev1}

	 	\centering
	 	\includegraphics[width=0.9\linewidth]{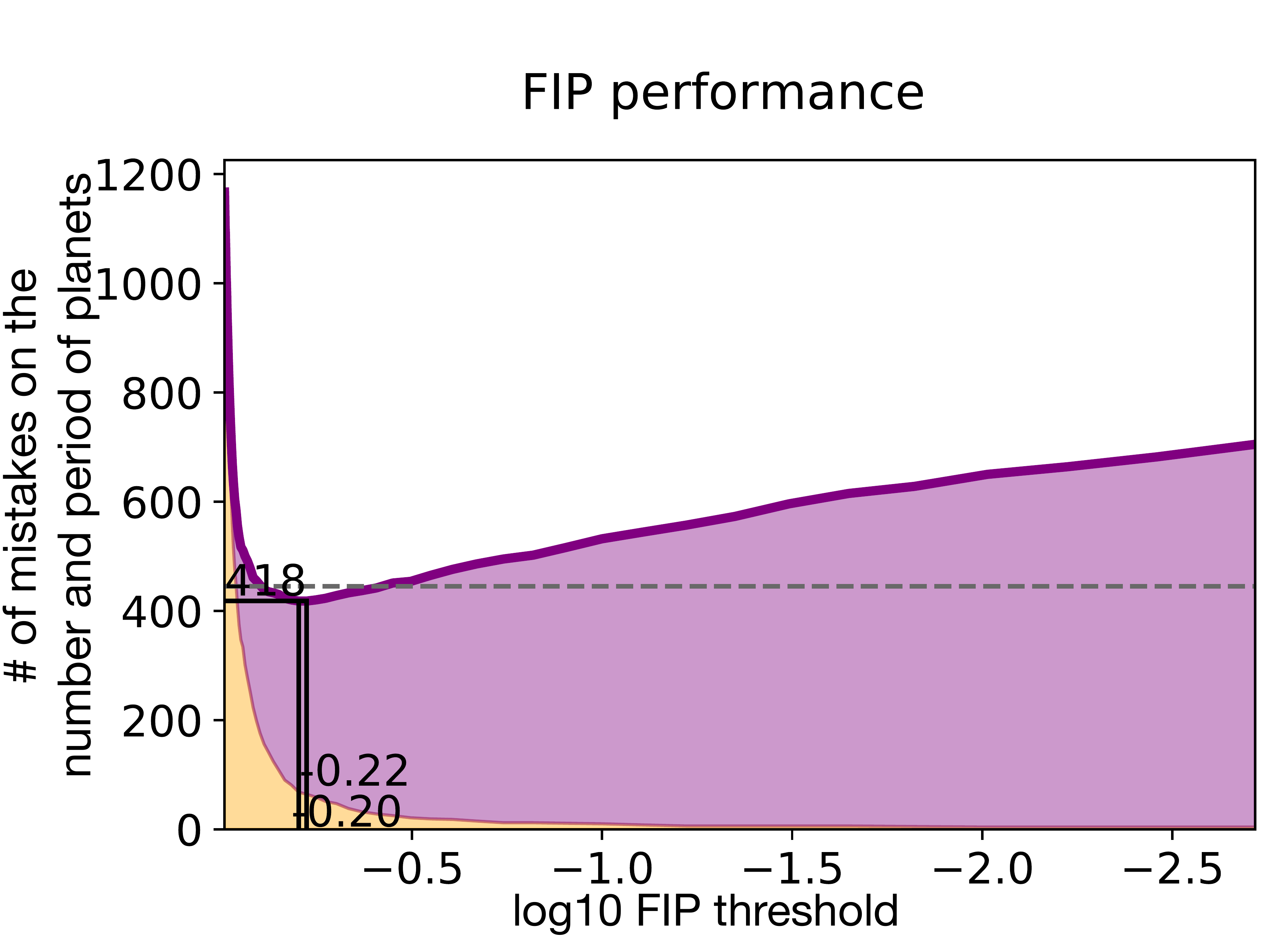}
	 	\caption{Total number of false and missed detections (yellow and purple shaded areas respectively) on a thousand systems as a function of the FIP threshold  when selecting the noise model with a fit of the ancillary indicators.}
	 	\label{fig:noiselev2}
	 \end{figure}

	 \begin{figure}
	 	\includegraphics[width=1.\linewidth]{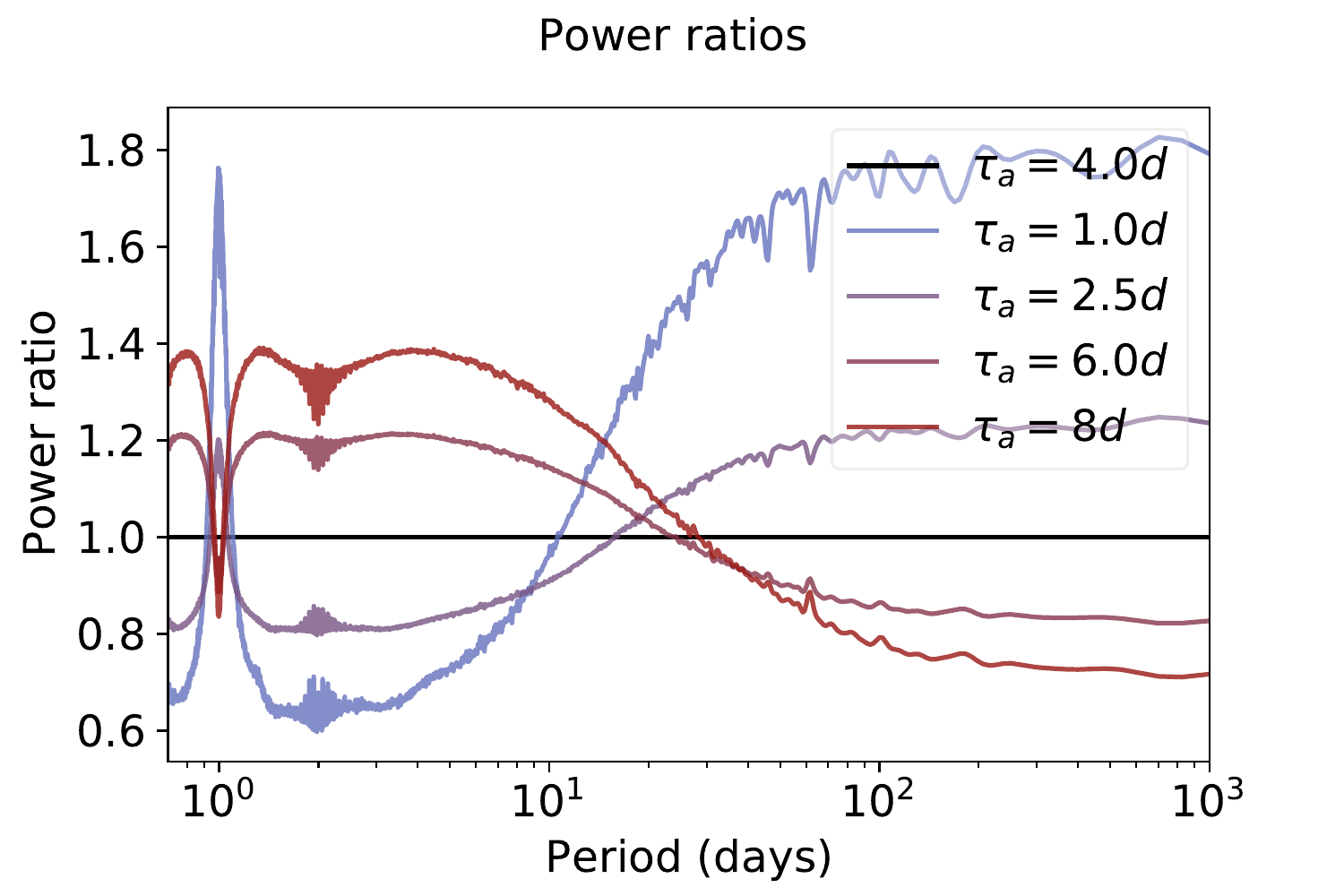}
	 	\caption{Expectancy of the ratio of periodograms computed with a noise with exponential covariance $\tau_a$ and the true timescale $\tau$ = 4 days. Each curve corresponds to a different value of $\tau_a$: 4 days (black) and 1, 2.5, 6 and 8 days (lavender to red).  }
	 	\label{fig:periodogram_ratios}
	 \end{figure}

	 Detections are sensitive to prior information, but also to the likelihood choice, which translates our assumptions on the noise. This is especially critical in the context of exoplanets detections, in which correlated stellar noises play a crucial role.  %The FIP~\eqref{eq:criterion} as well as Bayes factors have a desirable property, which is to average the detection criterion over the noise models. 
	 In this section, we study the sensitivity of detections as a function of the method used to estimate the noise parameters.  As in section~\ref{sec:prior}, we perform a simulation. We generate 1000 datasets whose parameters follow the distributions listed in Table~\ref{tab:priorscirc}. On each dataset, we generate an additional correlated Gaussian noise with an exponential kernel, a decay time-scale of 4 days and an amplitude of 1 m/s, like in Section~\ref{sec:simulation2}. Additionally, we generate  a time-series with exactly the same noise properties (same covariance). This is meant to emulate an activity indicator which could have the same covariance properties as the radial velocities. 
	 
	 These datasets are analysed with the priors and likelihood that were used to generate the data, except on the noise.% that we allow a correlated noise component with an exponential kernel, a time-scale $\tau$ with a uniform prior on 0 to 15 days and an amplitude with a uniform prior on 0 to 4 m/s. 
	 We analyse the data with the FIP with two different noise models. 
	 \begin{enumerate}
	 	%\item We perform a maximum likelihood fit of the noise parameters (amplitude and time-scale) onto the RV data, and use the fitted values as fixed noise parameters in the analysis of the RV. 
	 	\item Fitted noise: we perform a maximum likelihood fit of the noise parameters (amplitude and time-scale) onto the simulated activity indicator, and use the fitted values as fixed noise parameters in the analysis of the RV. 
	 	\item Free noise: we perform a RV analysis where the noise parameters are allowed to vary with a uniform prior on 0 to 15 days and an amplitude with a uniform prior on 0 to 4 m/s.
	 \end{enumerate}
	 
	 The results of these simulations in terms of false and missed detections are represented in Fig.~\ref{fig:noiselev1} (fitted noise) and Fig.~\ref{fig:noiselev2} (free noise), we show the number of false (yellow) and missed (purple) detections as a function of the FIP threshold. When using a free noise model, the minimum number of mistakes is 418 compared to 510 in the fitted noise case. 
	 
	 To provide some intuition on the role of the noise parameters in detection capabilities, we consider the calculation of a periodogram with a fixed noise model. We compute the average value of the periodogram when the true covariance of the noise is $V_t$ while the analysis is made with covariance $V_a$, and divide it by the average value of the periodogram when the noise is generated and analysed with $V_t$. This quantity expresses to which extent the periodogram is over or under-estimated under incorrect assumptions on the noise.  In~\cite{delisle2019a}, an analytical expression of this quantity is provided. To simplify the discussion, we assume that both true and assumed covariances consist of nominal error bars and a noise of covariance $\sigma_R^2 e^{-t/\tau}$. Let us denote the true time-scale and amplitude  of the noise by $\tau_t, \sigma_{R,t}$ and $\tau_a, \sigma_{R,a}$ the assumed time-scale and amplitude. 
	 We assume $ \sigma_{R,a} = \sigma_{R,t}$ and consider the expectancy of the ratio of periodograms obtained assuming $\tau_a$ and $\tau_t$, supposing the signal contains only noise with time-scale $\tau_t$. 
	  In Fig.~\ref{fig:periodogram_ratios}, we show this value computed with the time stamps used to generate the data, for $\tau_t$ = 4 days. If $\tau_a$ = $\tau_t$, the expected ratio of periodograms is one, which is represented in black. We represent the periodogram ratio for $\tau_a$ = 1, 2.5, 6 and 8 days, represented in colors from blue to red. 
	  We see that if the time-scale of the noise is under-estimated ($\tau_a$ = 1 or 2.5 days), then one over-estimates the power at low frequencies, resulting in too high a confidence in detections at these periods. Conversely, at lower period ($\approx 1 - 10$ days), one under-estimates the periodogram power. Because of aliasing, close to one day, the behaviour is closer to that of the low frequencies.  This situation is reversed when the noise time-scale is over-estimated. In that case, the power at low and high frequencies is respectively under and over-estimated. 
	  
	  This discussion provides some intuition as to why it is advantageous to average the detection metric over noise models. Indeed, because the noise parameters, in particular time-scales,  are allowed to be greater or lower than the true noise time-scale, such that the significance of a given period will be naturally balanced. On the contrary, if the noise is fitted and its time-scale is under or over-estimated, this acts like a frequency filter which enhances or reduces the significance of signals depending on their periods. % If the noise time-scale is fitted onto the RVs especially, the presence of planets introduces correlation which will lead to an over-estimation of $\tau$. 

	\section{Conclusion}
	\label{sec:conc}
	
The goal of the present work is to determine good practices for the analysis of radial velocity data, given that assumptions on the priors and the noise model (the likelihood) might be incorrect. 
%to define a statistical significance metric which is efficient to detect exoplanets, whose scale can be interpreted easily, and is robust to model misspecifications. 
We noted that the existing significance metrics (Bayes factor~\citep[e.g.][]{gregory2007}, false alarm probability (FAP)~\citep[e.g.][]{scargle, baluev2008} do not exactly define the information needed to claim a planet detection, which cannot be made without an estimate of the period of the planet. The computation of Bayes factor and FAP need to be coupled to an analysis of the period of the planets such as a periodogram~\citep[e.g.][]{baluev2008}, a $\ell_1$ periodogram~\citep[e.g.][]{hara2017} or the analysis of the marginal distribution of periods~\citep[e.g.][]{gregory2007}. Furthermore, Bayes factors and FAP are defined on scales which are difficult to interpret

To address this issue, we defined a new statistical significance metric: the true inclusion probability (TIP) which by definition is the probability to have a planet in a given period interval, and the false inclusion probability (FIP) as 1 - TIP. We suggested to compute the FIP on a sliding frequency interval of fixed length, to produce a periodogram-like figure, which we call FIP periodograms, and to select the planets based on peaks with FIP values below a certain threshold (see Section~\ref{sec:fip}).
We suggested two ways to compute the FIP: with algorithms sampling orbital elements for a given number of planets (see Section~\ref{sec:comp1}) and ``trans-dimensional'' algorithms, able to sample directly the joint posterior of the number of planets and orbital elements (see section~\ref{sec:comp2}). We defined two convergence tests: for a given maximum number of planets, one can run several times the calculation of FIP periodograms and ensure that the values are sufficiently close to each other in a user-defined sense. Secondly, to determine the maximum number of planets, one can stop once the difference of FIP periodograms obtained with an additional planet is below a threshold, here also user-defined (see section~\ref{sec:practicaluse}). We have highlighted the following properties of the FIP.
\begin{itemize}
	\item The FIP has a clear meaning. If the model used in the analysis is correct, on average, a fraction $1-\alpha$ of statistically independent detections made with FIP $= \alpha$ are correct. This property can be applied to check that  the signal models (likelihood and priors) are appropriate. Indeed, the FIP gives a prediction on the number of true and false detections which can be checked on a given RV catalogue (see Section~\ref{sec:fundprop}). %Doing so, one can diagnose model errors. 
	
	\item The FIP has a built-in period search, such that it offers clear diagnoses of aliasing. It  mitigates the false detections linked to combination of aliases leading to detections at spurious periods (see Section~\ref{sec:aliasing}).

\item The framework of the FIP allows not only to detect, but also to exclude planets within a certain model, since by definition it is the probability of having no planet in a certain range (see Section~\ref{sec:nondetec}). 

\end{itemize}
The FIP can be used in a broader context, it can be defined whenever a Bayes factor can be. It can be serve as a detection criterion in particular to detect planetary transits, any type of parametrized periodic variation in time series and more generally parametrized patterns in data. 

In Section~\ref{sec:discussion}, we studied the performance of the FIP as well as those of existing significance metrics. We discussed the best practices in three cases: assuming the model used in the analysis is the same as the one with which the data was generated, in the case where the prior does not correspond to the distributions of the generated elements, and finally in the cases where the true noise model is unknown. Our findings are summarised below. % We found the following properties
%The initial motivation behind this definition was theoretical. First, it has been shown that decisions based on the probability of the event of interest (here the presence of a planet) often proves optimal in a certain sense. Secondly

\begin{itemize}

	\item In our simulation, the FIP offers the lowest number of false detections (planets detected at the wrong period, or when no planet is present) and missed detections (no planet detecte while there is a planet). The difference with other methods is particularly important when the detection threshold corresponds to low false positives. In that regime, the FIP still offers a low number of missed detections. As a comment, in high false positive regimes, all statistical metrics seem to have the same behaviour.

\item Detections are sensitive to the priors chosen on semi amplitudes and periods. The behaviour of the prior close to low amplitudes is especially critical for the significance of low amplitude planets. A further study of the priors to be chosen in this regime would be valuable.

	\item It is better practice to let the noise parameters vary than fixing them to fitted values, for instance values fitted on ancillary spectroscopic indicators such as bisector span~\citep{queloz2001} or $\log R'_{HK}$~\citep{noyes1984}.

	\item The optimal FIP threshold  in terms of minimal number of false positive plus false negative is close to 50\%. A threshold of 1\% appears conservative and appropriate, but there is no need to define a clear cut threshold. 

\end{itemize}

We stress that all our simulations were intentionally simple, to isolate the effects of different assumptions. In the simulations, the maximum number of planets is known, and that the combinations of periods could be un-physical. Further comparisons on more realistic data such as in~\citep{pinamonti2017} would be valuable. 

\ch{The FIP as well as the Bayes factor are very sensitive to the prior on semi-amplitude. Depending on the prior, the significance might increase or decrease, and the period favoured might change from one alias to the other, even more so as the amplitude of the signal gets lower (see Section~\ref{sec:prior}).   For a robust detection of low amplitude signals, we suggest to check the robustness of the detection to a prior change.
Let us also note that the framework of the FIP is concerned with explicit alternatives: 1, 2, 3.. planets, specified noise models etc. If all those alternatives are faulty, the results might be unreliable. As a consequence, it is worth checking for systematic discrepancies between the data and the model as suggested in \citep{hara2019ecc}.}

We finally note that, as said in Section~\ref{sec:errorbars}, the FIP might be especially interesting to study populations. Usual population analysis methods one compares a forward model to detected planets, selected with a clear cut criterion~\citep[][]{gaudi2021}. The FIP provides a rigorous error bar on the detection, such that low signal to noise detection can be included rigorously in the analysis. For instance, one can compute the distribution of the number of true positives among a hundred detections made with FIP 50\%, which very clearly excludes 0. This aspect is left for future developments. 

%The analysis made so far is in very idealised cases, in order to understand precisely the differences between the methods. In practice, the relative and absolute planet parameters (period, mass, eccentricity, phase and argument of periastron) cannot take certain of the values they have here. We leave the systematic comparison of the different metrics in more realistic cases for future work. 

\begin{acknowledgements}
\ch{The authors thank Jo\~ao Faria for his review, which helped to improve our work.  N. C. H  thanks Roberto Trotta for his insightful suggestions. }
	N. C. H. and J.-B. D. acknowledge the financial support of the National Centre for Competence in Research PlanetS of the Swiss National Science Foundation (SNSF).
\end{acknowledgements}

\bibliographystyle{aa}
\bibliography{biblio}

\appendix

\section{Model details}
\label{app:model}
	%The sine and cosine of the true anomaly can be computed from equation~\eqref{eq:cosnu2bis},~\eqref{eq:sinnu2bis} and the Kepler equation~\eqref{eq:keplereq2bis}.%In the above notation, $\btheta = (e,K,P,\omega,M_0)$. 

%Our model 
%Our variables are the number of planets, $k$, the parameters of the deterministic model 
We assume a Gaussian noise model, such that the likelihood function for a given number of planets $k$ is
\begin{align}
p(\bm{y} | \bm{\theta}, \bm{\beta},k) &= \frac{1}{\sqrt{(2\pi)^N |\mathbf{V}(\bm{\beta}) | }} \e^{ -\frac{1}{2} (\bm{y} - \bm{f}_k(\bm{t} ,\bm{\theta}))^T  \mathbf{V}(\beta) ^{-1}  (\bm{y} - \bm{f}_k(\bm{t} ,\bm{\theta})) }  \label{eq:likelihood} \\
\bm{f}_k( \bm{\theta}) &= \vec g(\tilde{\bm{\theta}}) +  \sum\limits_{i=1}^k  \bm{f}(\bm{t},e_i,K_i,P_i,\omega_i,{M_0}_i) \label{eq:model}
\end{align}
where $\bm{f}(\bm{t},e,K,P,\omega,M_0)$ is defined as a Keplerian function evaluated at times $\bm{t}$. Such functions are defined as
\begin{align}
f(t,e,K,P,\omega,M_0) & = K(\cos\left(\omega + \nu(t,e,P,M_0)\right) + e \cos \omega) \label{eq:vexprbis}\\
\cos \nu & = \frac{\cos E - e}{1 - e\cos E} \label{eq:cosnu2bis}\\
\sin \nu & = \frac{\sqrt{1-e^2}\sin E}{1 - e\cos E} \label{eq:sinnu2bis}\\
E - e \sin E &= M_0 + \frac{2 \pi}{P} t \label{eq:keplereq2bis}.
\end{align}
The symbols $t,e,K,P,\omega,M_0$ designate respectively the measurement time, eccentricity, semi-amplitude, period, argument of periastron and mean anomaly at $t=0$. 

The function $\vec g(\tilde{\bm{\theta}})$ includes  some other model features (offset, trend, Gaussian process...), $\tilde{\bm{\theta}}$ denotes all the parameters that are not orbital elements of the planets.
The covariance matrix $ \mathbf{V}$ is parametrized by $\vec \beta$ and the suffix $T$ denotes the matrix transposition. 
We also define prior probabilities on $\vec \theta, \vec \beta$ and $k$. Their explicit expressions, as well as those of $\vec g(\tilde{\bm{\theta}})$  and $ \mathbf{V}(\bm{\beta})$ will be made precise when necessary.

\section{Datasets}
\label{app:datasets}

\begin{figure*}
    \centering
    \includegraphics[width=0.9\linewidth]{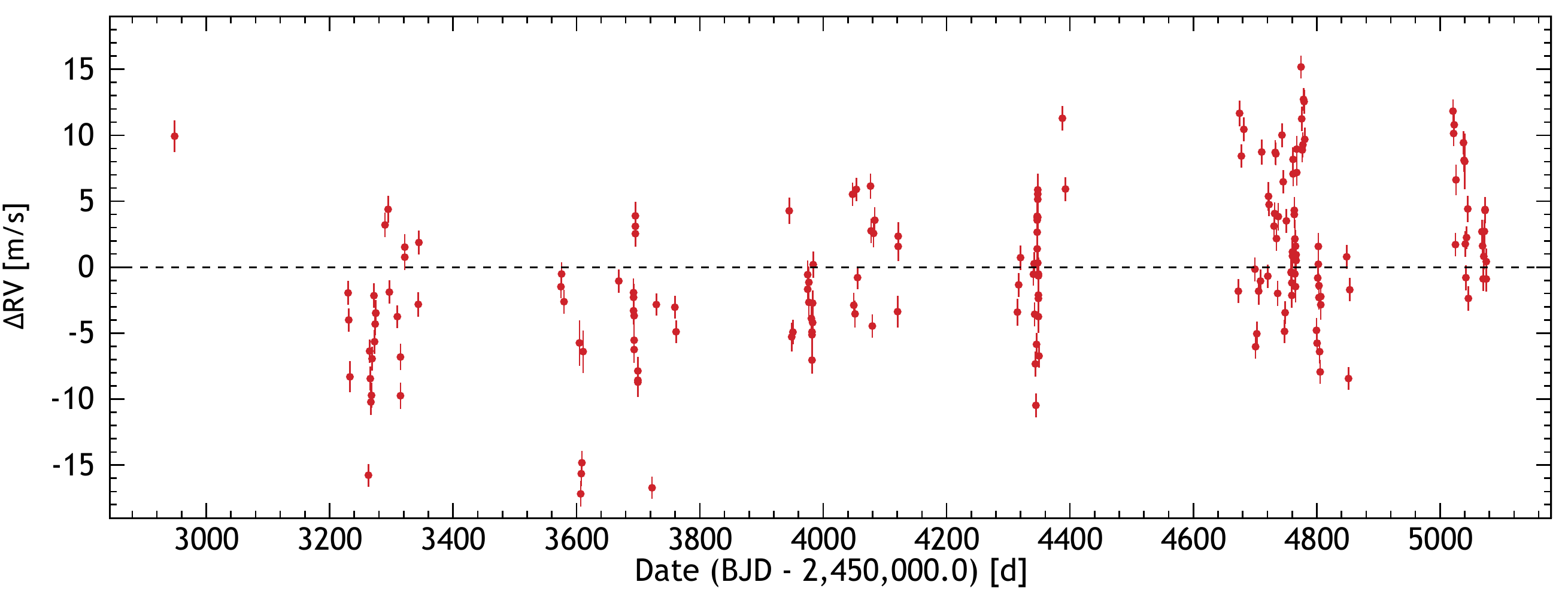}
    \caption{190 first HARPS radial velocity measurements of HD 10180. }
    \label{fig:hd10180_data}
\end{figure*}

\ch{In this appendix, we show the datasets used in this work. For the analysis of Section \ref{sec:application}, we analyse the 190 first points of the HARPS dataset of HD 10180 \citep{lovis2011}. The data are shown in Fig.~\ref{fig:hd10180_data}, it spans on 3.8 years, and the typical nominal uncertainty is 0.6 m/s. }

\begin{figure*}
    \centering
    \includegraphics[width=0.9\linewidth]{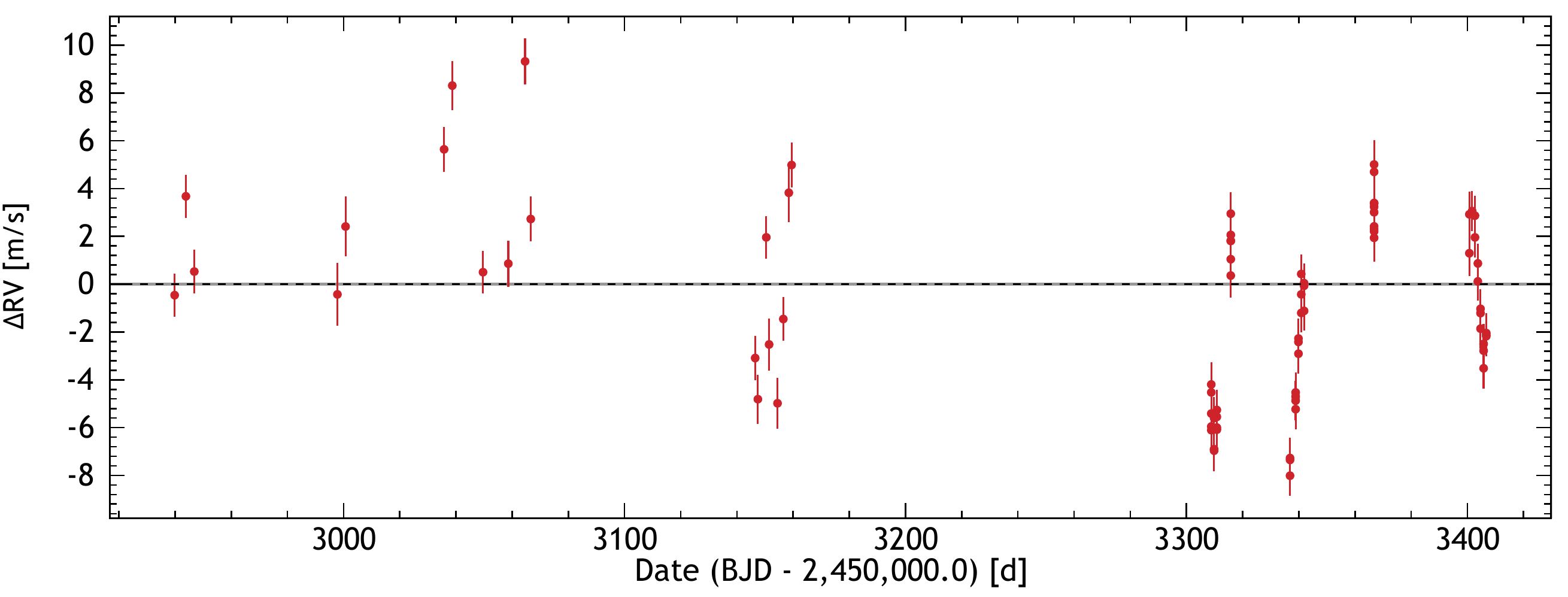}
    \caption{80 first HARPS radial velocity measurements of HD 69830. }
    \label{fig:hd69830_data}
\end{figure*}

\ch{The example of Section \ref{sec:practicaluse} and the simulations of Section \ref{sec:discussion} are performed using the time of measurements of HD 69830 \citep{lovis2006}. The data are shown in Fig.~\ref{fig:hd69830_data}, they span on 466 days. From BJD 2458330, there are several nights with several measurements per night separated typically by 4 minutes. The median nominal uncertainty is 0.45 m/s.  We stress that in Section \ref{sec:discussion}, we never use the measured radial velocities of  HD 69830, only the time of measurements and nominal error bars.   }


\section{Marginalisation over linear parameters}

\label{app:marg}

\subsection{Analytical expressions}
\label{app:marg_analytical}

Let us consider a model with $k$ planets. The likelihood function is then given by Eq.~\eqref{eq:likelihood} and~\eqref{eq:model}. The model~\eqref{eq:vexprbis} can be rewritten as 
\begin{align*}
f(t,e,K,P,\omega,M_0)  & = f(t, A ,B, e, P, M_0) \\ & = A\cos\left( \nu(t,e,P,M_0)\right)  \\ &+  B\sin\left( \nu(t,e,P,M_0)\right) + C. \label{eq:vexpab}
\end{align*}
We then re-write Eq.~\eqref{eq:model} in the form
\begin{align}
\vec f_k  =  \mat M_k(\vec  \eta) \vec x %+ h(\vec \eta)
\end{align}	
where $\vec \eta$ is the vector of non-linear parameters, $\vec x$ is the vector of linear parameters, and $\mat M_k$ is a matrix whose columns include $\cos \nu_i(\vec t)$ and $\sin \nu_i(\vec t)$ where $\nu_i$ is the true anomaly of planet $i$, as well as a column with all entries equal to one. The likelihood can then be written
\begin{align}
p(\bm{y} | \vec x, \bm{\theta}, \bm{\beta}, k) &= \frac{1}{\sqrt{(2\pi)^N |\mathbf{V}(\bm{\beta}) | }} \e^{ -\frac{1}{2} (\bm{y} - \mat M_k(\vec  \eta) \vec x))^T  \mathbf{V}(\beta) ^{-1}  (\bm{y} - \mat M_k(\vec  \eta) \vec x)) } .
\end{align}	
The evidence of a model with $k$ planets can be split in three integrals, over $\vec  x, \vec  \eta$ and $\vec  \beta$. When the prior distribution of $\vec x$ $p(\vec x)$ is Gaussian, the integral over $ \vec  x$ has an analytical expression. In the case  where $p(\vec x)$  has a null mean and a covariance matrix $\Sigma$, then denoting by  $N$ the number of measurements (the dimension of $\vec y$), 
\begin{align}
p(\bm{y} |\bm{\eta}, \bm{\beta},k) & = \int p(\bm{y} |\vec x,  \bm{\eta}, \bm{\beta},k)p(\vec x ) \dd \vec x \\ & = \frac{1}{\sqrt{2\pi}^N} \frac{\e^{-\frac{1}{2} \chi_\Sigma^2}}{\sqrt{|\mat V| |\mat \Sigma | |\mat Q|  }}\label{eq:intlin}
\end{align}
where  $|\mat X|$ designates the determinant of a matrix $\mat X$ and
\begin{align}
\mat Q &= \mat G + \mat \Sigma^{-1} \label{eq:q}\\
\mat G &=  \mat M^T  \mat V^{-1}  \mat M\label{eq:g}\\
\vec b & =\mat M^T \mat V^{-1} \vec y  \\
\chi_\Sigma^2 & = \vec y^T \mat V^{-1}  \vec y - \vec b^T \mat Q^{-1} \vec  b \\&= \vec y^T\left(\mat V^{-1} - \mat V^{-1} \mat M  (\mat M^T  \mat V^{-1}  \mat M + \mat \Sigma^{-1})^{-1} \mat M^T \mat V^{-1}\right) \vec y .
\label{eq:b}
\end{align}

A Gaussian prior might be overly restrictive. Fortunately, the analytical formula can be generalised to mixtures of Gaussians. Let us consider a collection of possible covariance matrices $(\mat \Sigma_i)_{i=1..M}$. We suppose that the linear elements  $\vec x$ follow a Gaussian of mean 0 and covariance $\mat \Sigma_i$ with probability $p_i$ and $\sum_i p_i = 1$. This prior on $\vec x$ can be seen as a two step process: (1) selecting  a covariance with probability $(p_i)_{i=1..M}$, (2) drawing $\vec x$ from a Gaussian with the selected covariance. The prior distribution on $\vec x$ can be written 
\begin{align}
p(\vec  x) = \sum_{i=1..M} p(\vec  x|i)p_i
\end{align}		
where $p(\vec x|i)$ is a Gaussian distribution of mean 0 and covariance $\mat \Sigma_i$. By linearity of the integral, 
\begin{align}
p(\bm{y} |\bm{\eta}, \bm{\beta},k) & = \sum\limits_{i=1}^N p(\bm{y} |\bm{\eta}, \bm{\beta},k, i) p_i
\label{eq:lingaussmixture}
\end{align}
where $p(\bm{y} |\bm{\eta}, \bm{\beta},k, i)$ is given by Eq.~\eqref{eq:b} with $\mat \Sigma = \mat \Sigma_{i}$ 

We specify this general formula to a case that is useful in practice. Let us suppose that for each planet, the parameters $A$ and $B$ of equation~\eqref{eq:vexpab} follow a Gaussian mixture model with distributions of standard deviation $(\sigma_j)_j={1..Q}$ with probabilities $q_j$, with $\sum_{j={1..Q}} q_j=1$. The $\sigma_j$ might correspond to different populations (super-Earths, mini-Neptunes, Neptunes, Jupiters etc). For a given number $k$ of planets, there are $Q^k$ combinations of different priors, which can be indexed by   $i_1 \in \{1..Q\}, i_2 \in  \{1..Q\} ... i_k \in \{1..Q\} $, that is any combination of $k$ indices with each a value between 1 and $Q$. For instance, let us suppose that we have four planet compact multi-planetary system where planets are \textit{a priori} either Super-Earth, mini-Neptunes. We have a two component Gaussian mixture model on semi-amplitude with standard deviations $\sigma_1$ and $\sigma_2$ and probabilities $q_1$ and $q_2=1-q_1$. Then the Gaussian mixture on the linear parameters $(A_i, B_i)_{i=1..4}$, where $A$ and $B$ are defined as in eq.~\eqref{eq:vexpab}, has $2^4 = 16$ components: all planets are Super Earth, planet 1 is a mini-Neptunes, all other Super Earth etc.
In the general case, eq.~\eqref{eq:lingaussmixture} becomes
\begin{align}
p(\bm{y} |\bm{\eta}, \bm{\beta},k) & = \sum\limits_{i_1,...i_k \in \{1..Q\} }   p(\bm{y} |\bm{\eta}, \bm{\beta},k,i_1, ....i_k) q_{i_1}...q_{i_k}
\label{eq:gaussmixt}
\end{align}
$p(\bm{y} |\bm{\eta}, \bm{\beta},k,i_1, ....i_k)$ is a Gaussian distribution of covariance matrix $\mat \Sigma^{i_1,...i_k}$, defined as follows. 
Ordering the components of $\vec x = (A_1, B_1,.... A_k, B_k, \vec x' )$ where $\vec x' $ are the linear parameters that do not correspond to a planet, then $\vec \sigma$
\begin{align}
\mat \Sigma^{i_1,...i_k} = 
 \mathrm{diag}(\sigma_{i_1}, \sigma_{i_1}, \sigma_{i_2}, \sigma_{i_2},...., \sigma_{i_k}, \sigma_{i_k}, \sigma_{\vec x'}). 
\end{align}

When utilising this formula, the number of analytic evaluations of integrals depends exponentially on the number of components in the Gaussian mixture models. However, it might be unnecessary to compute all the $Q^k$ components of the model. To illustrate thus, let us suppose that $Q = 2$ and correspond to Earth-sized and Jupiter-sized planets. If it appears that the likelihood completely excludes a mass above  20 $M_\oplus$ then the marginalised likelihood terms involving a Jupiter mass at this period can be neglected. 

%\textcolor{red}{This section seems to be incomplete, I guess you were describing the optimization where we avoid calculating the likelihood for all combinations. -Nico}

\subsection{Computational time}

We tested the advantages of the marginalisation in terms of computational time. We compute the FIP periodogram of the SOPHIE radial velocity data with 179 points spanning on 3.2 years,  with  different assumptions on the prior on $A$ and $B$ in eq.~\eqref{eq:vexpab} or the noise. 
In the base model the prior on $A$ and $B$ is Gaussian, and the noise model has a free jitter. The calculation is performed with the analytical marginalisation described in Section~\ref{app:marg_analytical}. We then consider a model including correlated noise with an exponential kernel, such that the noise model is $\sigma_W^2 + \sigma_R^2 \e^{-\frac{\Delta t}{\tau}}$, where $\sigma_W^2,\sigma_R^2$ and $\tau$ are free parameters. Secondly, we replace the prior on $K$ with a $\log$-uniform one and a uniform prior on phase. Finally, we use a three component Gaussian mixture model on $A$ and $B$  and use the analytical marginalisation. We compute as a function of the number of planets in the model how much time it takes to compute the posterior distribution of orbital elements. 

\begin{figure}
	\centering
	\includegraphics[width=\linewidth]{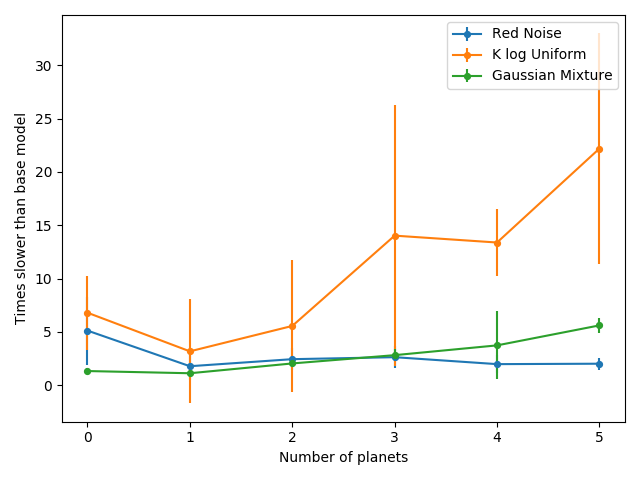}
	\caption{Runtime comparisons with different definition of the prior on semi-amplitude}
	\label{fig:speed_comp}
\end{figure}

In figure \ref{fig:speed_comp} we compare the time that it takes to compute the Bayesian evidence with $\textsc{polychord}$ for each configuration (Red noise: blue,  log Uniform prior on K: orange and Gaussian Mixture prior: green) compared to the base model.  The $y$-axis shows the ratio of the average time of the indicated configuration over the average time of the base model. We perform  five independent runs with each number of planets, the mean time is represented with plain lines and the standard deviation is represented as error bars.  
For the Gaussian Mixture model we used three populations, corresponding to super-Earths, sub-Neptunes and Jupiter-sized planets. We can see that the Gaussian Mixture prior and the Red Noise model are relatively stable between 2 and 5 times slower than the base model, and is systematically faster than a $\log$-uniform prior. 
The Gaussian Mixture model takes more time for higher number of planets because of the increasing $Q^k$ combinations (see eq.~\eqref{eq:gaussmixt}) that are necessary to compute, even when performing optimizations to remove some of those combinations. %\textcolor{red}{(Will these be described in the section before this? Mainly the one where we estimate the prior levels for each population and delete the combinations where the difference is large)}.
On the other hand, when we use a log Uniform prior for the semi-amplitude K we see a big increase in computational time which scales with the number of planets in the model. This is expected as each planet added to the model increases the number of free parameters by three (period, semi-amplitude and phase) instead of just one (period) when we marginalize over the linear parameters.

\color{black}

\section{Validation tests}
\label{app:validation}
\begin{figure}
        \caption{Fraction of events with probability $p_j$ where there actually was a planet injected as a function of $p_j$ for the case with 250 measurements instead of 80.}
        \label{fig:successalpha-250dpoints}
        \centering
        \includegraphics[width=8cm]{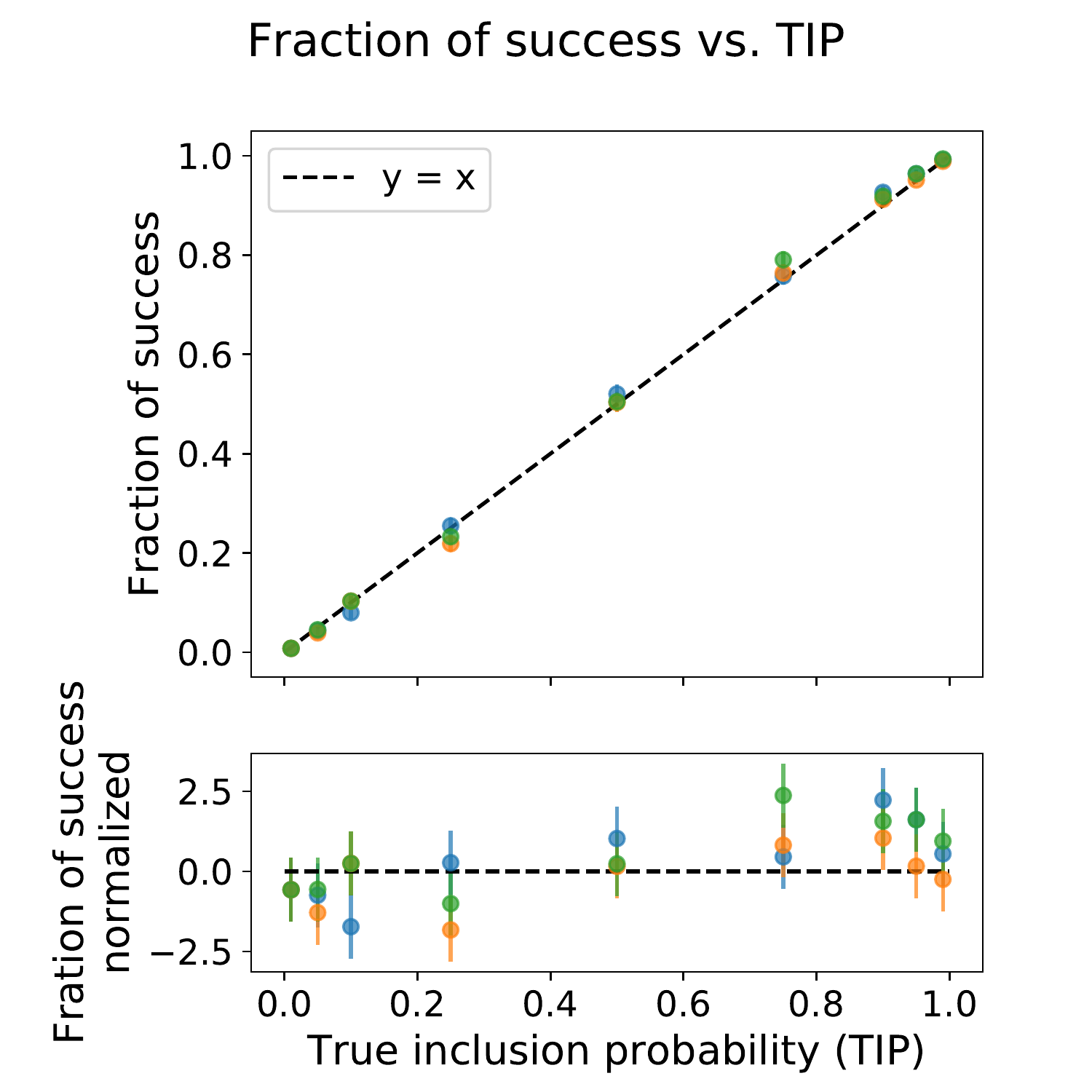}
    \end{figure}

\begin{figure}
        \caption{Fraction of events with probability $p_j$ where there actually was a planet injected as a function of $p_j$ for the case with up to 4 planets.}
        \label{fig:successalpha-4pla}
        \centering
        \includegraphics[width=8cm]{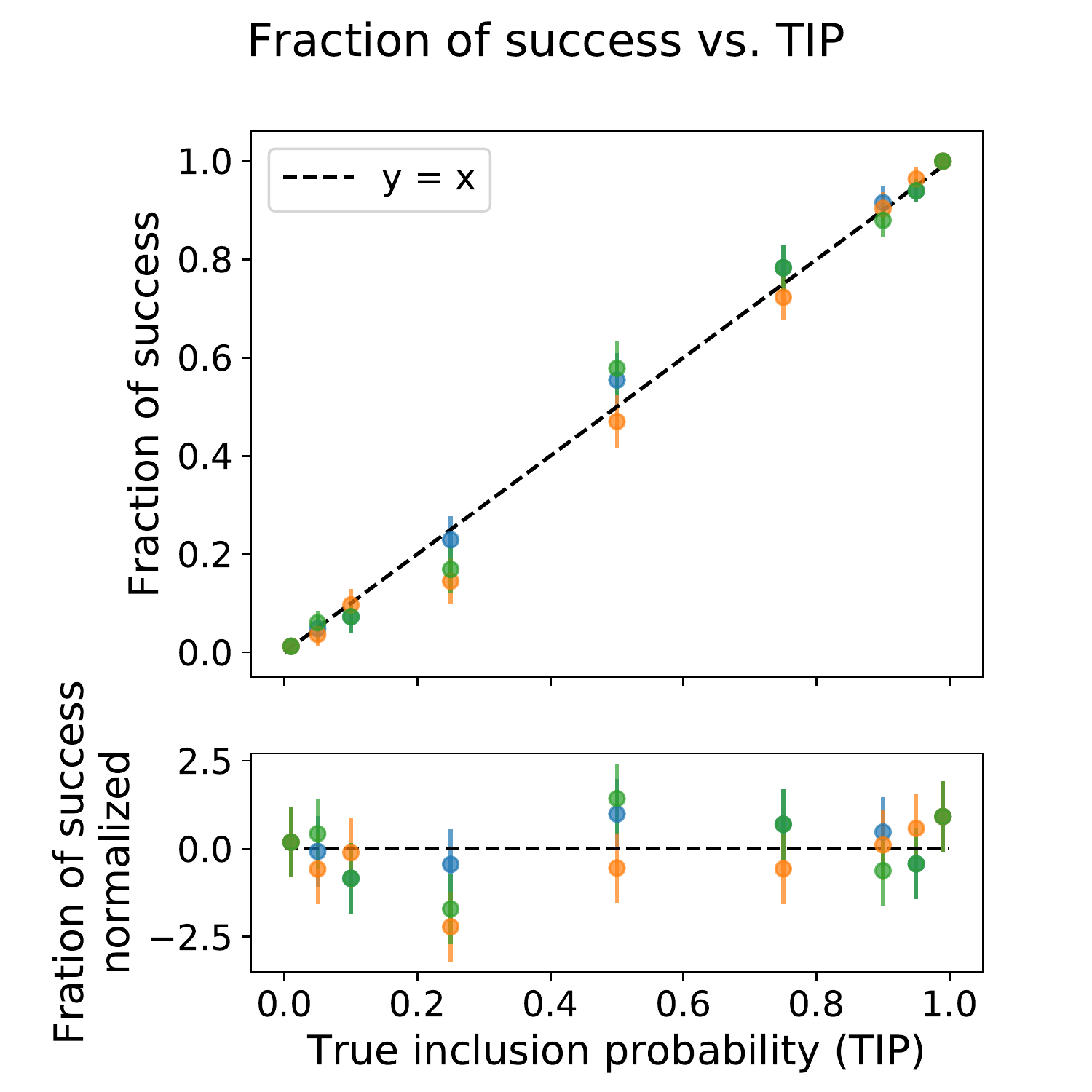}
    \end{figure}

\begin{figure}
        \caption{Fraction of events with probability $p_j$ where there actually was a planet injected as a function of $p_j$ for the case with correlated noise.}
        \label{fig:successalpha-rednoise}
        \centering
        \includegraphics[width=8cm]{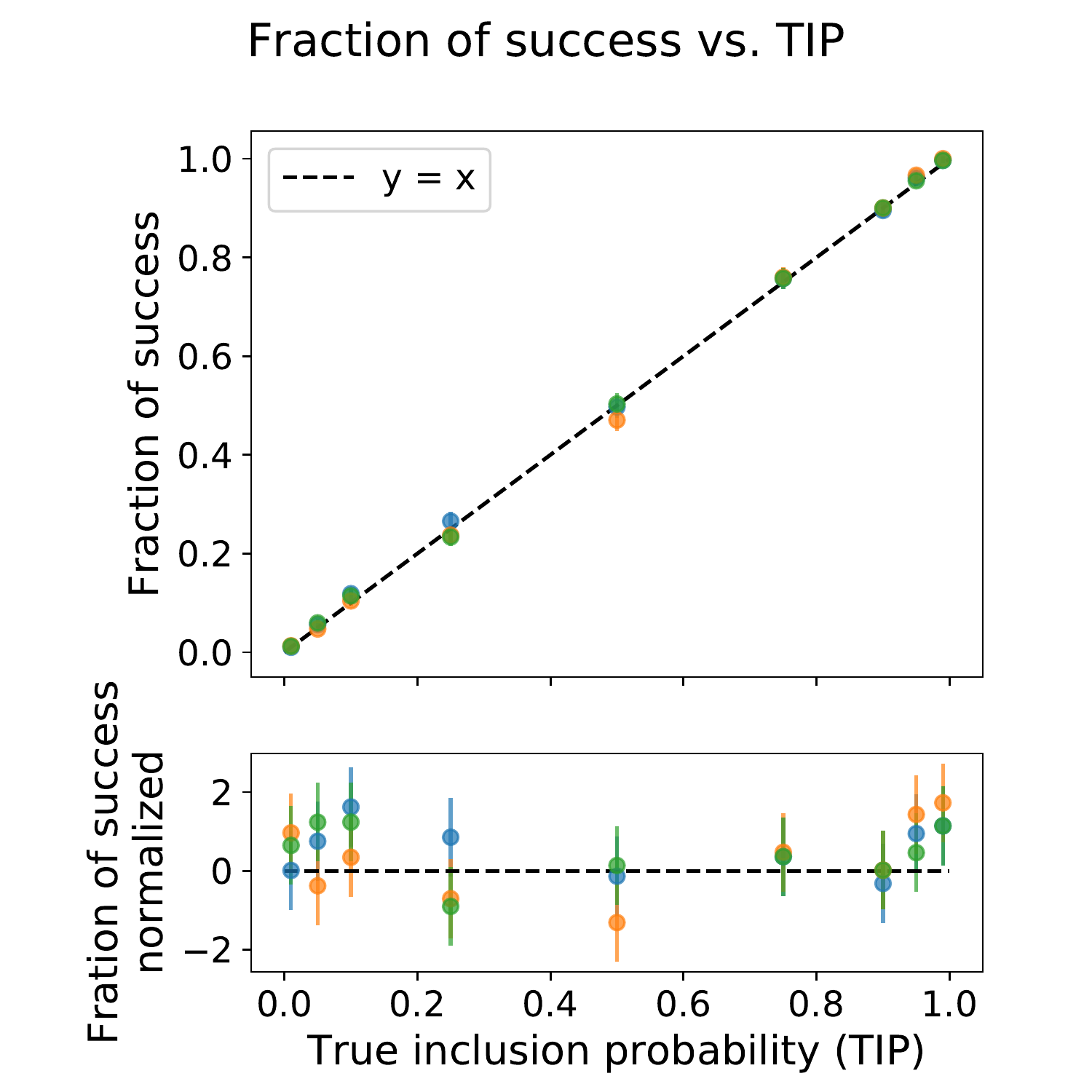}
    \end{figure}

\begin{figure}
        \caption{Fraction of events with probability $p_j$ where there actually was a planet injected as a function of $p_j$ for the case with eccentric orbits and $e$ generated from a Beta distribution with $a=0$ and $b=15$.}
        \label{fig:successalpha-kepsmecc}
        \centering
        \includegraphics[width=8cm]{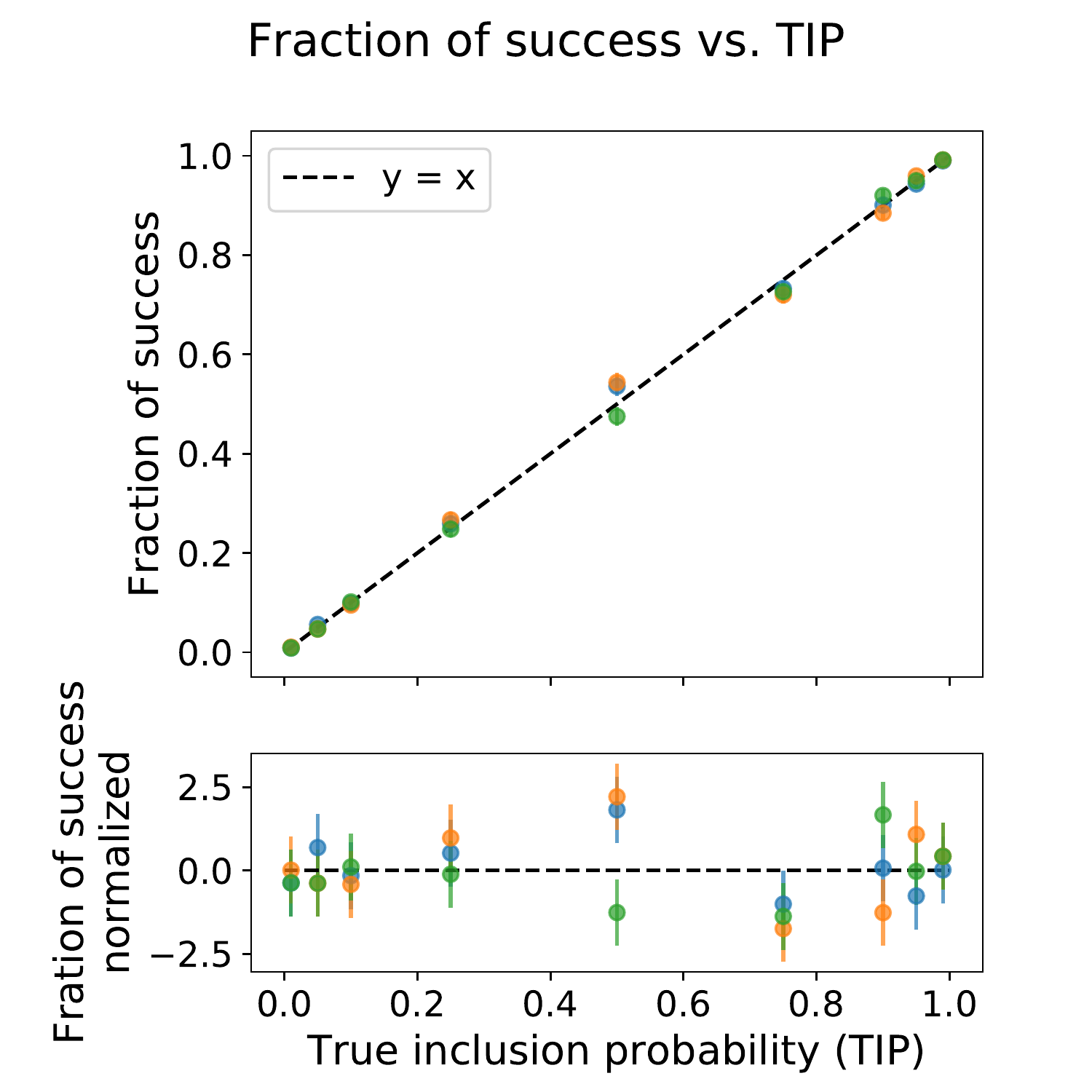}
    \end{figure}

\begin{figure}
        \caption{Fraction of events with probability $p_j$ where there actually was a planet injected as a function of $p_j$ for the case with eccentric orbits and $e$ generated from a Beta distribution with $a=0.867$ and $b=3.03$ as in~\cite{kipping2014}.}
        \label{fig:successalpha-kep}
        \centering
        \includegraphics[width=8cm]{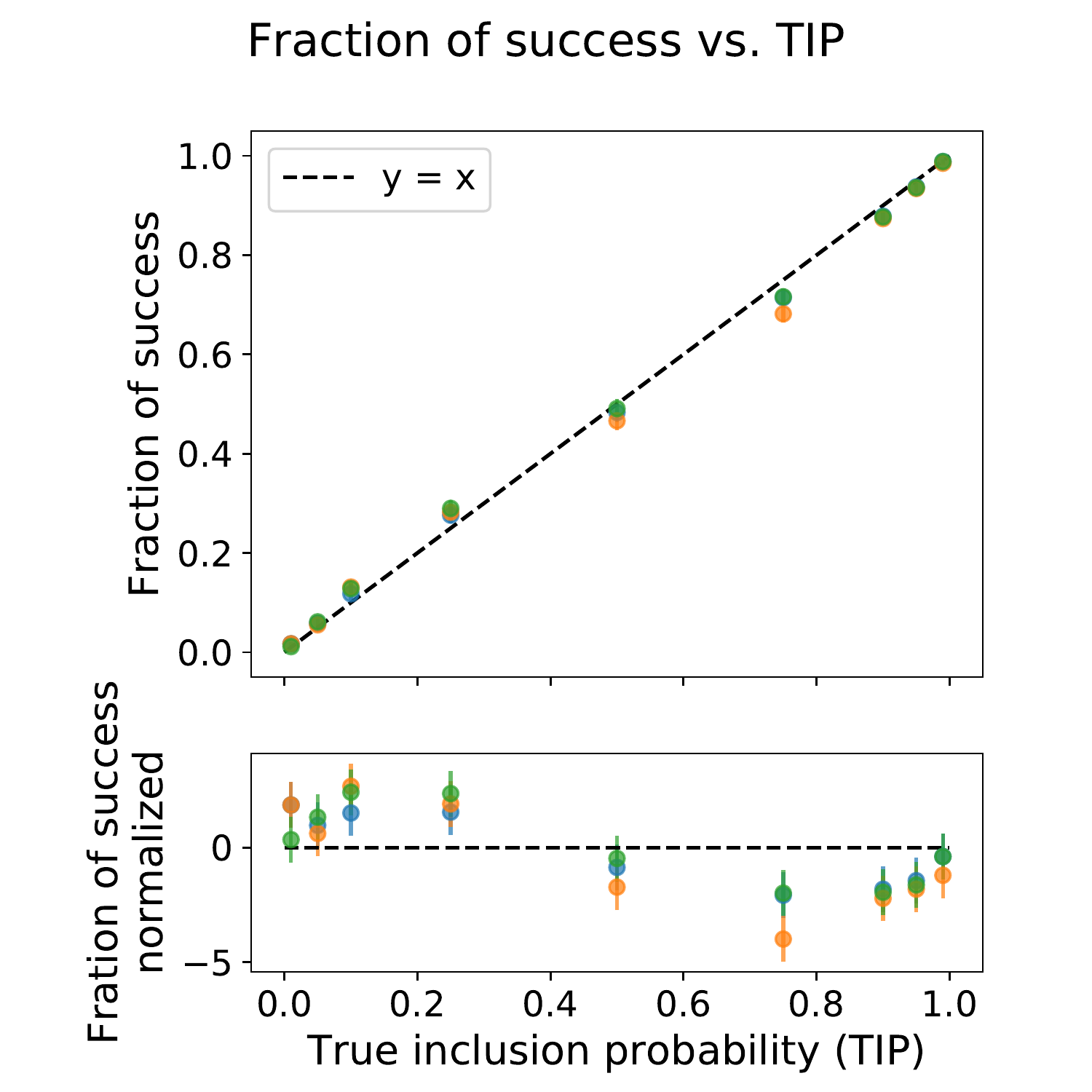}
    \end{figure}

In this appendix, we list the tests performed to validate our numerical methods. These are built exactly on the same principle as the test performed in~\ref{sec:validation} with different input models, listed below. 
	 \begin{itemize}
	 	\item Same priors as Table~\ref{tab:priorscirc}. We generate data on the 250 first HARPS measurement times of HD 69830 instead of the 80 first ones. See Fig. \ref{fig:successalpha-250dpoints}.
	 	\item  	Same priors as Table~\ref{tab:priorscirc} except that systems are generated with 0 to 4 planets See Fig. \ref{fig:successalpha-4pla}.
	 	\item  Same priors as Table~\ref{tab:priorscirc}, except that we add a correlated noise with exponential kernel, 1 m/s amplitude  and  a decay time-scale of 4 days. See Fig. \ref{fig:successalpha-rednoise}.
	 	\item Same priors as Table~\ref{tab:priorscirc}, except that orbits are eccentric (uniform argument of periastron and $e$ generated from a Beta distribution with $a=0$ and $b=15$. See Fig. \ref{fig:successalpha-kepsmecc}.
	 	\item Same priors as Table~\ref{tab:priorscirc}, except that orbits are eccentric (uniform argument of periastron and $e$ generated from a Beta distribution with $a=0.867$ and $b=3.03$ as in~\cite{kipping2014}. See Fig. \ref{fig:successalpha-kep}.
	 \end{itemize}
	 
\section{Prior influences}
\label{app:priors}

	The detection criterion~\eqref{eq:criterion} necessitates to define priors on the orbital elements. The criterion might favour a different detections depending on the priors. We here show that the detections claimed based on~\eqref{eq:criterion} and Bayes factors  depend on the prior chosen on the semi-amplitude $K$. We find that, loosely speaking, once the prior encompasses the $K$ with high likelihood, the broader the prior on $K$ is, the more the addition of a planet in the model is penalized.

This can easily be seen in a simplified setting with the analytical formula~\eqref{eq:intlin}.
	Let us consider two models with no other components but $k$ and $k+1$ planets, and a fixed covariance matrix of the noise $\mat V$. We further suppose that the models with $k$ and $k+1$ planets have specified period, $\bm{\eta}^k = (P_1,...,P_k)$ and $\bm{\eta}^{k+1} = (P_1,...,P_{k+1})$. 
%In that case the only non-linee $\vec \eta = (P_1,...,P_k)$. We suppose that the posterior distributions of the models with $k$ and $k+1$ planets are peaked in the neighbourhoods of $(P_1,...,P_k)$ and $(P_1,...,P_{k+1})$ and negligible elsewhere.
Assuming that the prior on the linear parameters $A$ and $B$ of all the planets (see eq.~\eqref{eq:vexpab}) is Gaussian of null mean and variance $L^2$, then $\mat \Sigma = L^2 \mat I$ where $I$ is the $2k$ identity matrix. Then~\eqref{eq:intlin} can be re-written
\begin{align}
p(\bm{y} |\bm{\eta}^k,k) & = \int p(\bm{y} |\vec x,  \bm{\eta},k)p(\vec x ) \dd \vec x \\ & = \frac{1}{\sqrt{2\pi}^N} \frac{\e^{-\frac{1}{2} (\vec y^T \mat V^{-1}  \vec y - \vec b^T \mat(\mat G +\frac{1}{L^2} \mat I)^{-1} \vec  b)}}{L^{2k}\sqrt{|\mat V| \left|\mat G + \frac{1}{L^2} \mat I \right| }}  \\ & \underset{L \rightarrow \infty}{\propto}   \frac{1}{L^{2k}}  \label{eq:a}
\end{align}

If we assume that the noise is white of standard deviation $\sigma$, denoting by $\chi^2_k$ the $\chi^2$ of the least square estimate fit of $M_k$ with a linear model $ \mat M$ and a Gaussian prior on the parameters of mean 0 and covariance $\mat \Sigma$,  it can be shown that 
\begin{align}
B_{k+1}  \approx \frac{\e^{\frac{1}{2} ( \chi^2_{k} -  \chi^2_{k+1} ) }}{ 1 +  L^2 \frac{N}{2\sigma^2} }.
\label{eq:bfapprox}
\end{align}
The Bayes factor, and thus the strength of the detection  increases as the exponential of the $\chi^2$ difference of the  $k+1$  and $k$ model. On the other hand, it is penalized by the denominator term, which accounts for the fact that the evidence of the model with $k+1$ planets has a larger parameter space.  In the limit of high $L$ which is asymptotically proportional to $ \frac{1}{L^{2}}$. 	

Similarly, considering a flat prior in $A$ and $B$ on $[-L, L]$,  one can easily show that as $L$ grows, the Bayes factor is also proportional to $ \frac{1}{L^{2}}$. 	For a flat prior in $K = \sqrt{A^2+B^2}$ on $[0, L]$, the limit is $\frac{1}{L}$.
and for a log-uniform prior on $K$ on $[e^{-L_m}, e^{L_M}]$,   $ \frac{1}{\ln L_M - \ln L_m}. $. This behaviour is also apparent in the analysis of~\cite{sinharay2002}, once the prior encompasses the high likelihood region, as it gets wider the Bayes factor favours more and more the simple model.

\end{document}